\documentclass[a4paper,fleqn,externalfonts]{cas-sc}

\usepackage[authoryear]{natbib}
\usepackage{fontspec}
\usepackage{unicode-math}
\defaultfontfeatures{Ligatures=TeX}
\usepackage[section]{placeins}
\usepackage{float}

\usepackage{booktabs}
\graphicspath{{figures/pairwise_redesign/}}

\hypersetup{
  pdftitle={Stochastic Shockwave Dynamics in First-Order Traffic Flow},
  pdfauthor={Yanlin Zhang and Alireza Talebpour},
  pdfsubject={Stochastic first-order traffic flow and congestion-front statistics},
  pdfkeywords={stochastic traffic flow, kinematic waves, shock clustering,
    Markov processes, Lax pair, queue spillback}
}

\newcommand{\rhoj}{\rho_{\mathrm j}}
\newcommand{\A}{\mathcal A}
\newcommand{\B}{\mathcal B}
\newcommand{\Qtr}{\mathcal Q_{\mathrm{tr}}}
\newcommand{\Cp}{\mathcal C_{p}}
\newcommand{\dd}{\,\mathrm d}
\newcommand{\RH}[2]{[q]_{#1,#2}}

\newcommand{\argmaxp}{\operatorname*{arg^+\,max}}

\newcommand{\diag}{\operatorname{diag}}
\newdefinition{theorem}{Theorem}
\newdefinition{lemma}[theorem]{Lemma}
\newdefinition{proposition}[theorem]{Proposition}
\newdefinition{corollary}[theorem]{Corollary}
\newdefinition{assumption}[theorem]{Assumption}
\newdefinition{definition}[theorem]{Definition}
\newdefinition{remark}[theorem]{Remark}
\newproof{proof}{Proof}
\newproof{proofsketch}{Proof sketch}

\begin{document}
\let\WriteBookmarks\relax
\renewcommand{\topfraction}{.90}
\renewcommand{\bottomfraction}{.85}
\renewcommand{\textfraction}{.08}
\renewcommand{\floatpagefraction}{.60}
\raggedbottom
\makeatletter
\setlength{\@fptop}{0pt}
\setlength{\@fpsep}{18pt}
\makeatother
\setcounter{topnumber}{3}
\setcounter{bottomnumber}{2}
\setcounter{totalnumber}{5}

\shorttitle{Stochastic Shockwave Dynamics in First-Order Traffic Flow}
\shortauthors{Zhang and Talebpour}

\title[mode=title]{Stochastic Shockwave Dynamics in First-Order Traffic Flow}

\author[1]{Yanlin Zhang}
\ead{yanlinz4@illinois.edu}

\author[1]{Alireza Talebpour}
\cormark[1]
\ead{ataleb@illinois.edu}

\affiliation[1]{organization={Department of Civil and Environmental Engineering, University of Illinois Urbana-Champaign},
                addressline={205 North Mathews Avenue},
                city={Urbana},
                postcode={61801},
                state={Illinois},
                country={United States}}

\cortext[1]{Corresponding author}

\begin{abstract}
Moving congestion fronts vary across realizations even when the traffic-flow
law is fixed. This paper develops a generator-based formulation of the
Lighthill--Whitham--Richards (LWR) model with random initial density and a
deterministic fundamental diagram. Randomness enters through the initial
spatial density field; every realization follows the same conservation law.
For a smooth strictly concave fundamental diagram, we adapt established
one-sided spatial Markov closure to traffic variables under the stated
spatial-process and regularity assumptions, and derive a kinetic balance for
physically normalized front intensities. Evolving the spatial generator
provides density laws and congestion-front statistics without solving a
road-grid problem for every realization. The formulation is continuous in
physical space and time, while its numerical implementation approximates
density state and integrates physical time. Four experiments distinguish deterministic consistency from stochastic
comparison. Riemann and signal examples recover independently constructed
exact LWR solutions. Smooth strictly concave approximations approach the
deterministic triangular-diagram signal solution over the tested finite
horizons; this does not establish stochastic closure at contacts or the
capacity kink. A matched Monte Carlo Godunov ensemble evaluates front counts,
sizes, mergers, and density distributions. The evolved generator also yields
an instantaneous travel-time distribution for a frozen density field,
compared with the same functional of the matched Godunov profiles. This
quantity is not experienced travel time through evolving traffic. Resolution
checks identify remaining normalization and small-front-loss effects. The
contribution is direct evolution and interpretation of traffic-state and
front distributions under a fixed first-order traffic law.
\end{abstract}

\begin{keywords}
stochastic traffic flow \sep kinematic waves \sep shock clustering \sep
Markov processes \sep Godunov method \sep travel-time reliability
\end{keywords}

\maketitle

\section{Introduction and Background}
\label{sec:introduction}

Congestion is experienced and managed through moving traffic boundaries. A
queue-forming front connects an upstream and a downstream density state; its
speed determines when congestion reaches a detector, bottleneck, or upstream
intersection, while its amplitude indicates the contrast between the two
traffic states. When several fronts are present, their relative speeds
determine whether they separate, weaken, or merge. These quantities vary
across days even under similar demand conditions, so average density or speed
alone cannot describe queue-spillback risk and travel-time reliability. The
transportation question addressed in this paper is therefore how the
population of congestion fronts evolves: how many fronts occur per unit road
length, which density-state pairs they connect, how their speeds and
amplitudes are distributed, and how frequently their interactions remove or
create state transitions.

This question sits within the broader view of traffic as a driven,
self-organizing many-particle system. Driver--vehicle units respond to one
another and to roadway constraints, and small disturbances can organize into
waves, localized clusters, and transitions between traffic states
\citep{helbing2001}. Empirical freeway data likewise exhibit unstable waves,
moving localized clusters, persistent wide jams, and the ``boomerang'' effect
\citep{schonhof-helbing2007}. Microscopic car-following and
cellular-automaton models explain such patterns through vehicle-level rules,
whereas mesoscopic models evolve distributions of vehicle speeds and
macroscopic models evolve aggregate fields such as density and flow
\citep{helbing2001}. The present paper belongs to the first-order macroscopic
branch. It does not propose a new behavioral law for individual drivers.
Instead, it asks which statistical description of a random macroscopic
traffic field is preserved by kinematic-wave dynamics and can be evolved
without following every realization separately.

The statistical-physics lineage also includes the foundational contributions
of Robert Herman and Ilya Prigogine. Herman and his collaborators analyzed
the propagation of disturbances and local and asymptotic stability in
car-following chains \citep{herman1959}. Prigogine and Andrews developed a
Boltzmann-like equation for the vehicle-speed distribution, including passing
and the formation of clusters of vehicles moving at the same speed
\citep{prigogine-andrews1960}. Prigogine and Herman's
\emph{Kinetic Theory of Vehicular Traffic} subsequently presented a kinetic
description of traffic through evolving vehicle-speed distributions
\citep{prigogine-herman1971}. This work is an important antecedent to treating
traffic statistically. The kinetic description used here has a different
unknown: it evolves the spatial transition rates between density states at
LWR congestion fronts, rather than a distribution of individual vehicle
speeds. Its front-merging term is therefore not the vehicle-interaction term
of the Prigogine--Herman theory.

The deterministic foundation, retained for every realization of the random
initial density field, is the Lighthill--Whitham--Richards (LWR) conservation
law
\[
  \partial_t \rho+\partial_x q(\rho)=0,
\]
where \(\rho\) is traffic density and \(q(\rho)\) is the fundamental diagram
\citep{lighthill1955,richards1956}. In traffic terms, the derivative
\(q'(\rho)\) is the speed at which information travels through a smooth
density region, and the secant slope of the fundamental diagram between two
states is the speed of the front separating them. Newell's simplified
kinematic-wave theory, the cell transmission model (CTM), and Godunov-type
schemes use this wave structure to propagate queues and load road networks
\citep{newell1993,daganzo1994,lebacque1996}. Variational and
Hamilton--Jacobi formulations express the same traffic-flow surface through
cumulative counts, passage times, or vehicle positions and make the roles of
initial data, boundary data, and entropy selection explicit
\citep{daganzo2006,laval-leclercq2013}. We follow the same traffic-translation
principle: each stochastic object introduced below is tied to a feature of a
road density profile or to a measurable population of traffic fronts.

A deterministic LWR or CTM calculation evolves one density profile for one
set of initial and boundary data. Repeated calculations under day-to-day
variation produce an ensemble that can be summarized by mean density,
exceedance probabilities, or selected sample paths. The unknown proposed
here is different. Imagine traversing a road at a fixed time. Density changes
continuously along recovery regions and jumps at congestion fronts. The
continuous slope describes how density relaxes between fronts; the frequency
of a transition from a lower state \(\rho\) to a higher state \(\eta\)
describes how often that type of front is encountered per unit distance. A
spatial Markov infinitesimal generator combines these two traffic quantities: its drift
records continuous recovery and its jump kernel records the state-pair
intensity of congestion fronts. Evolving this generator is therefore a way
to evolve the front population rather than only a collection of density
sample paths.

Existing stochastic traffic models address complementary parts of the
uncertainty problem. Stochastic CTM formulations propagate uncertainty in
density, demand, and supply \citep{sumalee2011}; headway-based models and
Gaussian approximations connect microscopic randomness to macroscopic state
estimation \citep{jabari-liu2012,jabari-liu2013}; stochastic kinematic-wave
and fundamental-diagram models characterize congestion and flow
distributions \citep{laval-chilukuri2014,qu2017}; and conservative jump
processes provide mesoscopic supply--demand descriptions
\citep{tordeux2014}. These approaches can generate fronts in individual
realizations, and Monte Carlo ensembles can estimate queue and spillback
probabilities. The unresolved traffic-facing question is how an inherited
spatial closure can be translated, physically normalized, and evaluated so
that front counts, adjacent traffic states, and gains and losses caused by
front mergers become direct operational outputs.

Traffic-clustering studies make this front-population question natural.
Faster vehicles joining slower clusters produce coarsening behavior analogous
to shock aggregation \citep{ben-naim1994}, while reduced traffic models link
density-wave formation to Burgers-type dynamics in different stability
regimes \citep{nagatani1998,nagatani2000}. These studies show that wave
formation and aggregation are recurring traffic mechanisms, but they do not
evolve the state-pair intensity of entropy fronts in a stochastic first-order
density field.

A mathematical starting point is provided by stochastic scalar conservation
laws. Classical results establish the Hopf representation, entropy-shock
selection, and conservation-law well-posedness
\citep{burgers1974,hopf1950,lax1957,dafermos2016}; subsequent work shows that
important spatial stochastic structures can persist under inviscid Burgers
dynamics \citep{carraro-duchon1998,bertoin1998}. Chabanol and Duchon's
generator evolution is a formal precursor \citep{chabanol-duchon2004},
while later work establishes one-sided Markov closure and develops kinetic
and Lax descriptions of shock clustering
\citep{menon-srinivasan2010,menon2011}.
Those results provide the broad closure and operator provenance used here.
The traffic formulation requires a concave fundamental diagram, upward
density jumps at admissible congestion fronts, traffic-consistent signs for
characteristic and Rankine--Hugoniot motion, a physically normalized measure
of fronts per unit road length, and mappings from that measure to front
speeds, amplitudes, and traffic-state distributions. The derivations needed
for these traffic variables are given within the paper and its appendices.

Related traffic research further distinguishes the sources of randomness.
\citet{chiabaut2010heterogeneous} connect heterogeneous parameters in a
stochastic Newell car-following model to macroscopic congestion patterns and
an equivalent deterministic representation under their aggregation
conditions. Their randomness originates in differences between vehicles.
Here, the already aggregated initial density field is random; its spatial
generator is not derived from individual driver-parameter distributions. A
micro-to-macro connection is therefore a potential extension, not a
relationship estimated in this paper.

\citet{leclercq2016capacity} incorporate distributions of insertion headways
and vehicle characteristics into analytical merge-capacity calculations.
That setting distinguishes mean-capacity approximations from upstream
dynamics affected by heterogeneity. In contrast, the homogeneous generator
law studied here contains no ramp-insertion or moving-bottleneck mechanism.
It describes randomness in the initial road field rather than continuing
stochastic vehicle insertions.

\citet{laval2010mechanism} explain the formation and propagation of
stop-and-go waves through driver behavior. This distinguishes a mechanism
that generates oscillations from the evolution of an initially perturbed
traffic field. The present model concerns the propagation and coalescence of
admissible disturbances under classical LWR; it adds no behavioral mechanism
for spontaneous instability.

The deterministic foundation must also be distinguished from stochastic
closure. \citet{herbin2011entropy} clarify that nondifferentiability of the
fundamental diagram does not, under the relevant assumptions, invalidate
existence and uniqueness of the entropy solution. Together with the
Hamilton--Jacobi traffic representations of \citet{laval-leclercq2013}, this
supports using a triangular diagram in deterministic LWR without implying
that the smooth stochastic-generator derivation extends through its capacity
kink. Section~\ref{subsec:discussion-flux} addresses that separate issue.

The theory and experiments are organized around four progressively harder
questions:
\begin{enumerate}
\item[\textbf{RQ1.}] Does the front representation reduce to the classical
LWR entropy solution for a deterministic Riemann shock?
\item[\textbf{RQ2.}] Which traffic calculations require a smooth strictly
concave fundamental diagram, and which remain valid for a continuous
piecewise-linear concave triangular diagram?
\item[\textbf{RQ3.}] How closely do directly evolved shock counts,
jump-size distributions, merger rates, and density statistics agree with a
matched Godunov ensemble for a stochastically initialized
many-front field?
\item[\textbf{RQ4.}] How accurately does the evolved infinitesimal generator
reproduce the distribution and reliability of instantaneous travel time
compared with a matched Monte Carlo Godunov ensemble? Here instantaneous
travel time uses the road's speed profile at one specified time, without
accounting for subsequent changes during the trip \citep{fhwa2012}.
\end{enumerate}
The first two locate the framework inside established traffic-flow
theory by comparison with independently constructed exact solutions. The
latter two test the statistical outputs that motivate the generator
description by comparison with matched Monte Carlo Godunov ensembles. These
two validation modes are kept separate: an exact solution is never used to
grade a road-grid method, and no exact reference is inserted into a stochastic
Godunov comparison.

Together, these questions move from classical consistency to the setting in
which the proposed representation is most useful: direct evolution and
evaluation of a random kinematic-wave law. Exact Riemann solvers and wave constructions are preferable for a
small deterministic wave system; Newell--Lax--Hopf methods and CTM are
especially effective for triangular traffic flow; and Monte Carlo Godunov or
CTM can estimate distributional quantities by resolving many realizations.
The proposed framework contributes a particular chain: a stochastic spatial
density process is represented by a Markov generator, compatible generator
dynamics yield a kinetic front balance, and that balance is mapped to shock
and reliability distributions. The implementation avoids a finite-volume
mesh in physical road space and time while still approximating density state
and integrating finite-dimensional evolution equations numerically.

The analytical scope is intentionally specific. The pathwise closure assumes
a smooth, strictly concave fundamental diagram and a restricted class of
spatial Markov, bounded-variation density profiles with upward jumps. The
continuum front-frequency equations additionally require Feller regularity,
a common invariant operator core, finite front rates, and spatial
stationarity on a homogeneous, unforced road section. Ramps, signals,
controlled boundaries, and general bottleneck forcing are not yet closed at
the generator level. The triangular diagram is treated using its exact deterministic traffic
solution and a family of smooth strictly concave approximations. Agreement
over the tested finite horizons does not establish stochastic closure at
contacts or the capacity kink, or uniform accuracy over longer times and
distances. The travel-time result
describes instantaneous travel time, not the travel time experienced by a
vehicle moving through the evolving field \(\rho(x,t)\). The continuum operator compatibility remains
conditional on the stated regularity assumptions. Acceleration, reaction
time, lane changing, capacity drop, hysteresis, and endogenous growth of
stop-and-go instability remain outside this first-order macroscopic scope.

The paper makes three methodological contributions. First, it states the
conditions under which a random traffic-density profile retains its spatial
Markov structure and upward-front orientation under LWR evolution with a
fixed, smooth, strictly concave fundamental diagram.
Second, it derives a closed evolution for continuous recovery slopes and
front state-pair intensities. The resulting transport and merger terms
describe changes in front counts, amplitudes, and speeds in traffic variables;
space--time consistency also yields the associated operator and finite-state
Lax representations. Third, it maps the evolved generator to one- and
multi-point density laws, front distributions, and an instantaneous
travel-time distribution. The numerical realization discretizes
density state rather than physical road cells. The four experiments then
separate exact classical consistency, fundamental-diagram regularity,
proposed-versus-Godunov shock statistics, and proposed-versus-Godunov
reliability. The novelty is the closed traffic-facing description of a front
population and its direct distributional outputs, not Rankine--Hugoniot
motion itself or an exclusive ability to estimate uncertainty by simulation.

The remainder of the paper is organized as follows. Section
\ref{sec:problem-description} defines the problem setting, the admissible
random density profiles, and the front-state representation. Section
\ref{sec:solution-approach} develops the front balance, traffic outputs, and
computational realization; the appendices contain the operator derivations
and the limited three-state algebraic check. Section
\ref{sec:numerical-results} answers RQ1--RQ4 in order. Section
\ref{sec:discussion} compares capabilities and limits, and Section
\ref{sec:conclusion} summarizes the transportation implications and next
steps.

\section{Problem setting}
\label{sec:problem-description}

Consider repeated observations of the same road section at the same time of
day. One realization may contain a front from 30 to 140 veh/km/lane 200 m
upstream of an observation point; another may contain several weaker fronts at different
locations. LWR determines how each observed front moves. The additional
question in this paper is how to describe the population across realizations:
the density found at a typical road location, the frequency of each adjacent
state pair, and the total number of fronts per kilometer.

Only three statistical quantities are needed to follow the main argument.
The marginal density law \(p\) describes conditions at a randomly selected
location, the conditional rate \(\lambda\) describes which state follows a
given state as one moves downstream, and \(F=p\lambda\) converts that
conditional transition into an unconditional front count. Table
\ref{tab:traffic-translation} gives their traffic meanings together with the
wave quantities used to evolve them. The remainder of the section supplies
the kinematic-wave conditions that make this representation consistent. The
technical Hopf construction justifies closure of the stochastic class; it is
not an additional behavioral traffic model.

\begin{table}[H]
\caption{Traffic interpretation of the principal model quantities.}
\label{tab:traffic-translation}
\centering
\small
\begin{tabular}{p{0.14\linewidth}p{0.50\linewidth}p{0.24\linewidth}}
\hline
Quantity & Traffic meaning & Units or normalization \\
\hline
\(\rho\) & Local traffic density & vehicles/km/lane \\
\(q(\rho)\) & Equilibrium flow (fundamental diagram) & vehicles/time/lane \\
\(q'(\rho)\) & Characteristic speed & km/time \\
\(\RH{\rho}{\eta}\) & Speed of a front joining states \(\rho\) and \(\eta\) & km/time \\
\(a(\rho,t)\) & Magnitude of a decreasing spatial density gradient & vehicles/km\(^2\)/lane \\
\(p(\rho,t)\) & Density distribution at a randomly selected road location & integrates to one \\
\(\lambda(\rho,\dd\eta,t)\) & Conditional rate of fronts from \(\rho\) into \(\dd\eta\) & expected fronts/km \\
\(F\dd\rho\dd\eta\) & Unconditional count in two adjacent-state bins & expected fronts/km \\
\(\Gamma(t)\) & Total congestion-front intensity & expected fronts/km \\
\hline
\end{tabular}
\end{table}

\subsection{Traffic waves}
\label{subsec:lwr-model}

The road coordinate \(x\) increases downstream, and
\(\rho(x,t;\omega)\in[0,\rhoj]\) denotes the density in realization
\(\omega\). The model is the Lighthill--Whitham--Richards conservation law
with random initial conditions \citep{lighthill1955,richards1956}:
\begin{equation}
\label{eq:lwr}
\begin{aligned}
 \partial_t\rho(x,t;\omega)
 +\partial_x q\bigl(\rho(x,t;\omega)\bigr)&=0,\\
 \rho(x,0;\omega)&=\rho_0(x;\omega).
\end{aligned}
\end{equation}
The same fixed fundamental diagram \(q\) applies to every realization.
Randomness enters through the initial spatial density field, not through a
random fundamental diagram, a stochastic forcing term, or temporal noise
injection. Once an initial realization and admissible deterministic data are
specified, its entropy evolution is deterministic. Density, wave speeds, and
front configurations nevertheless have distributions across realizations.

The stochastic object studied here is therefore the law induced by LWR with
random initial density. The generator representation evolves that law; it
does not introduce a new constitutive relation or identify the underlying
behavioral source of the initial variability. The smoothness and
spatial-process assumptions below specify when this statistical evolution
closes. We suppress \(\omega\) in subsequent expressions when no ambiguity
arises. The characteristic information speed remains
\(c(\rho)=q'(\rho)\).
Capacity always means the maximum of the fundamental diagram:
\[
 q_{\max}=\max_{0\leq r\leq\rhoj}q(r),\qquad
 \rho_c\in\operatorname*{arg\,max}_{0\leq r\leq\rhoj}q(r).
\]
For the Greenshields model \citep{greenshields1935}, \(\rho_c=\rhoj/2\); for the triangular diagram \citep{newell1993},
\(\rho_c=w\rhoj/(v_f+w)\). A congested state with \(\rho>\rho_c\)
has flow below capacity and must not be identified with the capacity state.

The stochastic closure used later was developed for a convex flow with
downward jumps, whereas traffic convention uses a concave flow and upward
queue fronts. We therefore make the change of variables explicit rather
than importing signs by analogy:
\begin{equation}
\label{eq:traffic-convex-transform}
  u=-\rho,\qquad f(u)=-q(-u),\qquad
  f'(u)=q'(-u),\qquad f''(u)=-q''(-u)>0.
\end{equation}
It follows directly that
\(u_t+f(u)_x=-(\rho_t+q(\rho)_x)=0\). An upward traffic jump
\(\rho_L<\rho_R\) becomes the downward jump
\(u_L=-\rho_L>u_R=-\rho_R\), while its Rankine--Hugoniot speed \citep{lighthill1955,richards1956} is unchanged:
\begin{equation}
\label{eq:transform-speed}
 \frac{f(u_R)-f(u_L)}{u_R-u_L}
 =\frac{q(\rho_R)-q(\rho_L)}{\rho_R-\rho_L}.
\end{equation}
Consequently, the convex-flow entropy inequality translates to
\(q'(\rho_L)>s>q'(\rho_R)\). This calculation fixes the jump orientation
and every secant-speed sign used below. The Markov closure and operator
framework are inherited from the convex-flow shock-clustering theory cited
with the closure statement in Section \ref{subsec:traffic-closure-theorem};
the traffic-variable transformation and its physical interpretation are
stated here.

This is the classical kinematic-wave model of \citet{lighthill1955} and
\citet{richards1956}. The next assumption selects the smooth concave regime
used for the analytical results. Strict concavity is standard in that regime
\citep{daganzo2006,dafermos2016}; the \(C^3\) clause and the restriction of
the triangular diagram to numerical comparison are assumptions imposed here.

\begin{assumption}[Smooth strictly concave fundamental diagram]
\label{ass:fd}
The physical fundamental diagram \(q:[0,\rhoj]\to\mathbb R_+\) is \(C^2\),
satisfies \(q(0)=q(\rhoj)=0\), and is strictly concave:
\[
  q''(\rho)<0,\qquad 0<\rho<\rhoj .
\]
For the differentiated kinetic equations, \(q\in C^3((0,\rhoj))\), with the
one-sided bounds needed for the integrals below. A triangular fundamental
diagram is used only in the numerical section as a nonsmooth comparison (or
as a smooth-flow limit); it is outside the smooth strict-concavity theorems
stated here.
\end{assumption}

Variational traffic-flow theory backtracks an observation \((x,t)\) to the
part of the initial road profile that generated it
\citep{newell1993,daganzo2006,laval-leclercq2013}. Under the conditions below,
this Hamilton--Jacobi construction produces the LWR entropy solution and
preserves the traffic ordering of its discontinuities. Its equation, concave
dual, density selector, and monotonicity proof are collected in Appendix
\ref{app:hopf-orientation}; the main text needs only the resulting front
orientation.

The next assumption adopts the one-sided spatial-Markov initial class used by
that stochastic conservation-law closure. The bounded traffic state interval
and upward-density orientation are imposed here.

\begin{assumption}[Admissible random initial field]
\label{ass:initial-field}
The two-sided initial density \(x\mapsto\rho_0(x)\) is a spatial Markov
process with locally bounded-variation paths and no downward jumps. Its
potential
\[
  U_0(s)=U_0(0)+\int_0^s\rho_0(r)\dd r
\]
is assumed to make the variational maximum finite. The precise right-process,
coercivity, and filtration conditions used in the closure proof are stated in
Appendix \ref{app:hopf-orientation}.
\end{assumption}

Under Assumptions \ref{ass:fd}--\ref{ass:initial-field}, the standard concave
Hopf--Lax solution \citep{hopf1950,lax1957,newell1993,daganzo2006} has an
important traffic consequence: its selected positive-time discontinuities
are upward density fronts. Appendix \ref{app:hopf-orientation} states the
formula and proves the required monotonicity of the backtracked road origin.
Thus strict flow concavity fixes the traffic-wave orientation before the
stochastic front model is introduced.

The variational solution also fixes the ordering of discontinuities. We now
record the Rankine--Hugoniot and Lax relations that will determine every
transport and collision sign in the generator equations.

For \(\rho<\eta\), define the Rankine--Hugoniot speed
\begin{equation}
\label{eq:rh-speed}
  \RH{\rho}{\eta}=\frac{q(\eta)-q(\rho)}{\eta-\rho}.
\end{equation}

This jump speed and its entropy ordering are classical
\citep{lighthill1955,richards1956,lax1957,dafermos2016}. Strict concavity
places the front speed between the two characteristic speeds and makes the
upstream member of an ordered adjacent-front pair faster than the downstream
member. This is the kinematic reason adjacent fronts can merge. Appendix
\ref{app:collision-positivity} records the complete inequalities and their
short proof.

\subsection{Front-state model}
\label{subsec:traffic-closure-theorem}

We now translate a random density profile into the three quantities previewed
in Table \ref{tab:traffic-translation}. Between fronts, the coefficient
\(a\) records the magnitude of a decreasing recovery gradient. At a front,
\(\lambda(\rho,\dd\eta,t)\) records how often a road currently in state
\(\rho\) jumps into a denser state near \(\eta\) per unit distance. The
result below ensures that this one-sided road description is preserved by LWR
before any coefficient equation is imposed.

The stochastic conservation-law literature establishes spatial Markov
closure and an associated operator description for suitable one-sided random
initial data \citep{menon-srinivasan2010,menon2011}. We restate the required
equations and derivations in traffic-density variables. The contributions
developed here are the concave-flow/upward-front orientation, the explicit
front-frequency normalization, the traffic performance mappings, and the
scope conditions needed for transportation use.

In stochastic-process terminology, a profile with no downward jumps is
spectrally positive. Here the term simply records that congestion fronts are
upward density transitions; it does not add a new traffic assumption. The
closure statement below specializes the established conservation-law result to a concave traffic flow under
\(u=-\rho\) and \(f(u)=-q(-u)\).

\begin{theorem}[Traffic-density Markov closure]
\label{thm:traffic-closure}
Under Assumptions \ref{ass:fd} and \ref{ass:initial-field}, for every \(t>0\)
the cadlag entropy representative \(x\mapsto\rho(x,t)\) is a possibly
spatially-inhomogeneous Markov process with no downward jumps.
\end{theorem}

The proof is given in Appendix \ref{app:markov-closure}. Operationally, the
theorem says that the same type of road-wise transition model can be used at
later times; LWR does not force the analyst to change stochastic state
descriptions as fronts move and merge. If the initial ensemble is spatially
stationary, translation invariance of LWR also preserves that stationarity.
The theorem alone does not guarantee smooth coefficient densities, so the
calculation below uses the additional regularity stated next.

Closure identifies the stochastic class but does not yet provide smooth
coefficients or a jump density. To pose the kinetic problem, we now impose the
additional regularity needed to write the spatial generator and its
front-number measure explicitly.

To obtain finite front counts and differentiable coefficient equations, we
work in the following regular regime. The precise integrability and operator
conditions are recorded in Appendix \ref{app:generator-adjoints}.

\begin{assumption}[Regular front-transition process]
\label{ass:generator-regularity}
For \(t\in[t_0,T]\), the spatial density field is a conservative
bounded-variation Feller process on \([0,\rhoj]\). Its continuous drift is
\(-a(\rho,t)\), with \(a\geq0\), and its finite-activity jump kernel
\(\lambda(\rho,\dd\eta,t)\) is supported on \(\eta>\rho\). Coefficients are
regular enough for the derivatives and integrals in Section
\ref{sec:solution-approach}, and boundary conditions keep density inside the
physical interval.
\end{assumption}

The resulting operator has the standard uncompensated bounded-variation
drift--jump form used for Feller processes
\citep{applebaum2009,ethier-kurtz1986}.
For a traffic reader, it is simply a compact bookkeeping rule: the first term
tracks movement through smooth recovery states, while the integral tracks
crossings of upward congestion fronts.

Under Assumption \ref{ass:generator-regularity}, the road transition operator
has the uncompensated bounded-variation form
\begin{equation}
\label{eq:spatial-generator}
  (\A_t\varphi)(\rho)
  =-a(\rho,t)\varphi'(\rho)
  +\int_\rho^{\rhoj}
   \bigl(\varphi(\eta)-\varphi(\rho)\bigr)
   \lambda(\rho,\dd\eta,t).
\end{equation}
There is no diffusion term because the paths have bounded variation and no
killing term because the process is conservative.

Assume in addition that the spatial law is stationary and has a one-point
density \(p(\rho,t)\). If
\(\lambda(\rho,\dd\eta,t)=\lambda(\rho,\eta,t)\dd\eta\), define the joint
shock-number density
\begin{equation}
\label{eq:F-definition}
  F(\rho,\eta,t)=p(\rho,t)\lambda(\rho,\eta,t),
  \qquad 0\leq\rho<\eta\leq\rhoj.
\end{equation}
Then \(F\dd\rho\dd\eta\) is the expected number, per unit road length, of
upward fronts whose left and right states fall in the indicated density
elements.

Section \ref{sec:solution-approach} therefore starts from three traffic-facing
state variables: \(p\) for density uncertainty at a location, \(a\) for
continuous recovery gradients, and \(F\) for the number and composition of
fronts per unit road length. The Markov closure is a pathwise result under
Assumptions \ref{ass:fd} and \ref{ass:initial-field}; the coefficient
evolution additionally uses the regular regime of Assumption
\ref{ass:generator-regularity}. The finite-state calculation is placed in
Appendix \ref{app:spectral-integrability}, where it serves as an optional
numerical consistency check rather than part of the traffic problem
definition. Section \ref{sec:solution-approach} now turns these three state
variables into evolution equations and observable traffic measures.

\section{Generator-based solution approach}
\label{sec:solution-approach}

\begin{figure}[tbp]
\centering
\includegraphics[width=\textwidth]{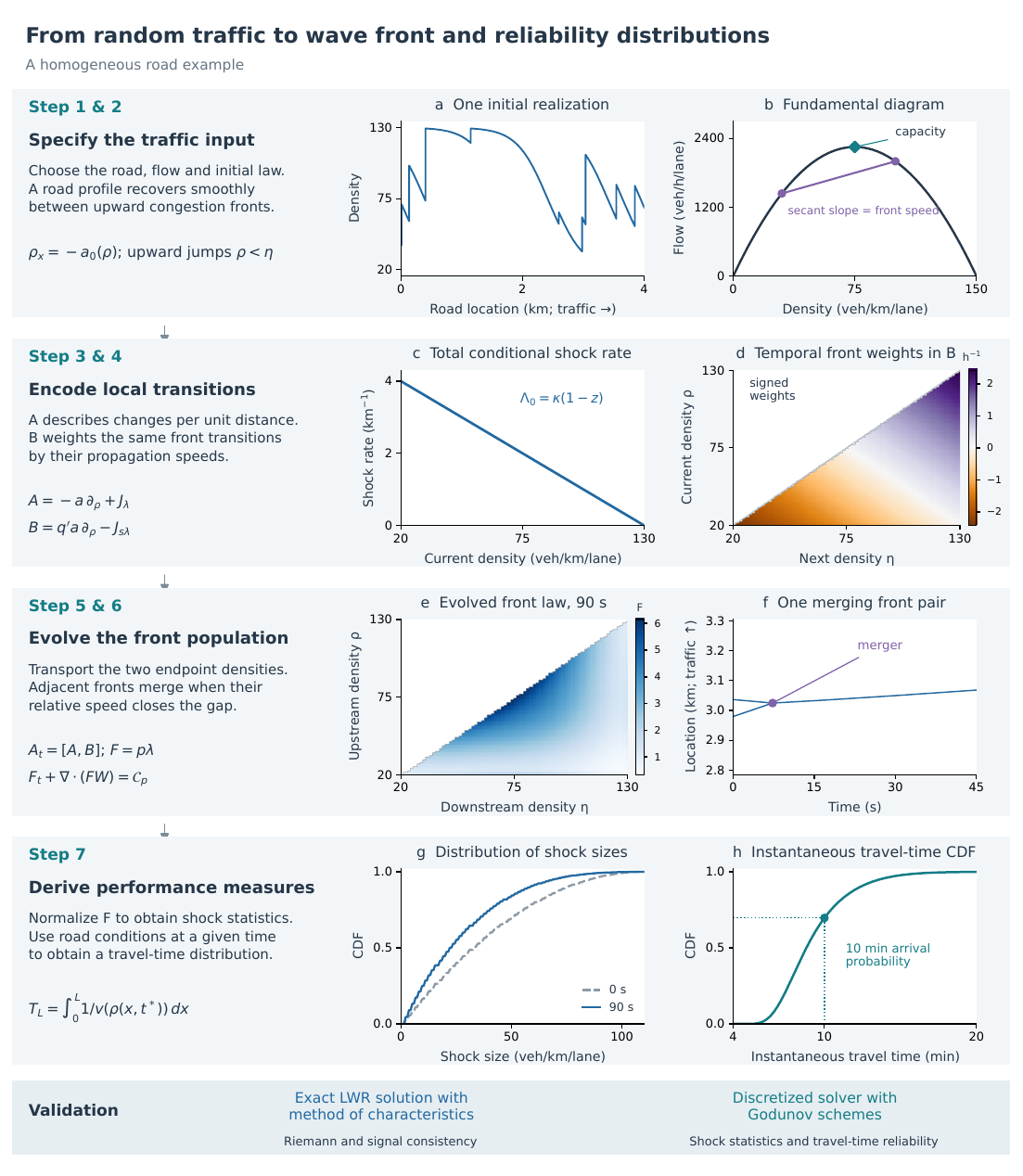}
\caption{Worked framework: specify a random traffic field, encode its
transitions, evolve the front population, and derive performance measures.
The insets use the homogeneous process of Experiments 3--4
(\(v_f=60\) km/h, \(\rhoj=150\) veh/km/lane; capacity 2250 veh/h/lane).
Panel (e) uses normalized density coordinates. The labeled paths in (f)
illustrate one realization and are not uniquely recovered from \(F\).
Validation follows the comparator assigned to each experiment.}
\label{fig:method-flow}
\end{figure}

The problem setting in Section \ref{sec:problem-description} supplies a
fundamental diagram and a random road profile; the solution approach turns
those inputs into an evolving law of congestion fronts. The calculation is
organized around four operations in Figure \ref{fig:method-flow}: specify a
random traffic field, encode its local transitions, evolve the front
population, and reconstruct traffic distributions. We first explain the
continuum equations and the outputs they determine, then describe their
numerical realization. The experiment section assigns an independent
comparator to each output.

Figure \ref{fig:method-flow}(c) displays the initial total conditional shock
rate \(\Lambda_0(\rho)=\kappa(1-z)\), where \(z=(\rho-20)/110\) and
\(\kappa=4\) km\(^{-1}\). Although the rate density is uniform over allowed
higher states, the available destination range shrinks as current density
increases. Panel (d) weights destination-bin rates by propagation speed,
and (e) shows the resulting nonuniform front law after evolution. Panel (f)
illustrates one merging pair; (g,h) translate the evolved law into shock-size
and travel-time distributions.

\subsection{Inputs and generator pair}
\label{subsec:generator-pair}

The first four steps translate familiar traffic-wave data into two operators.
The spatial generator \(\A\) describes how density changes when moving along
the road at a fixed time; the signed temporal operator \(\B\) describes how
the same smooth segments and fronts pass a fixed observation point. Writing
both operators in traffic variables makes the characteristic and
Rankine--Hugoniot speeds explicit.

\noindent\textbf{Step 1: traffic input.}
Specify a fixed fundamental diagram \(q(\rho)\), the road orientation, and
the probability law of the initial spatial density field
\(\rho_0(x;\omega)\). The continuum closure below concerns a spatially
homogeneous, unforced road. Deterministic initial and boundary data used in
the consistency tests are specified separately. A finite Riemann datum or a
signal boundary can still be represented by a deterministic atomic wave
measure and advanced with the same characteristic and secant-speed rules,
but that use does not establish closure under stochastic boundary forcing.

\noindent\textbf{Step 2: stochastic representation.}
Represent a traffic realization as a spatial Markov process with decreasing
recovery segments and upward density jumps. In the notation of Section
\ref{subsec:traffic-closure-theorem}, the continuous spatial drift is
\(b(\rho,t)=-a(\rho,t)\), where \(a\geq0\), and the jump kernel is
\(n(\rho,\dd\eta,t)=\lambda(\rho,\dd\eta,t)\), supported on \(\eta>\rho\).
This is the traffic reflection of the one-sided Markov class introduced in
Section \ref{subsec:traffic-closure-theorem}; the upward-jump convention is
fixed by the transformation in
\eqref{eq:traffic-convex-transform}. A deterministic finite collection of waves has an atomic front measure.
This singular consistency case is treated separately from a regular Feller generator.

\noindent\textbf{Step 3: initial spatial generator.}
The initial drift and jump kernel define
\begin{equation}
\label{eq:initial-spatial-generator}
 (\A_0\varphi)(\rho)
 =-a_0(\rho)\varphi'(\rho)
 +\int_\rho^{\rhoj}
  \{\varphi(\eta)-\varphi(\rho)\}
  \lambda_0(\rho,\dd\eta).
\end{equation}
At later times the same expression, with \(a_0,\lambda_0\) replaced by
\(a,\lambda\), is \(\A_t\) in \eqref{eq:spatial-generator}. For a stationary
road ensemble, \(\A_t^\dagger p_t=0\), and
\(F(\rho,\eta,t)=p(\rho,t)\lambda(\rho,\eta,t)\) is the unconditional number
of fronts per unit road length and pair of density-state intervals. A finite
nonstationary Riemann problem is instead embedded through its atomic jump
measure
\(M_0(\dd x,\dd\rho,\dd\eta)
=\delta_{x_0}(\dd x)\delta_{\rho_L}(\dd\rho)
 \delta_{\rho_R}(\dd\eta)\). The finite transition at a prescribed road location is not an ordinary
finite-rate Feller generator. Its front measure instead satisfies the weak
transport identity
\[
 \frac{\dd}{\dd t}\langle M_t,\psi\rangle
 =\langle M_t,s\,\partial_x\psi\rangle,
 \qquad s=\RH{\rho_L}{\rho_R},
\]
obtained from the distributional LWR balance. This identity agrees with the
signed front-speed weighting used in \(\B\); it does not define products of
singular generators in \([\A,\B]\). The deterministic experiments check
this front-measure specialization.

\noindent\textbf{Step 4: temporal operator.}
The fundamental diagram converts \(\A_t\) into
\begin{equation}
\label{eq:temporal-operator-main}
 (\B_t\varphi)(\rho)
 =q'(\rho)a(\rho,t)\varphi'(\rho)
 -\int_\rho^{\rhoj}
  \RH{\rho}{\eta}
  \{\varphi(\eta)-\varphi(\rho)\}
  \lambda(\rho,\dd\eta,t).
\end{equation}
The first coefficient is the characteristic speed \(q'(\rho)\) multiplied by
the recovery-gradient magnitude. The second coefficient uses the
Rankine--Hugoniot secant speed
\(\RH{\rho}{\eta}=[q(\eta)-q(\rho)]/(\eta-\rho)\) and the frequency of the
corresponding front. The minus sign records that a moving spatial front
crosses a fixed detector in the opposite orientation. Thus \(\B_t\) is a
signed compatibility operator, not necessarily a Markov generator in time.
Its adjoint evolves the one-point density law; the full forward expression is
given in Appendix \ref{app:generator-adjoints}.

For a smooth flow function, both coefficients in \eqref{eq:temporal-operator-main} are
unambiguous. For a continuous piecewise-linear concave flow function, the secant speed
remains defined for distinct states and \(q'\) is constant within each linear
branch. At the critical-density kink, however, \(q'\) is not unique. A direct
construction there requires a declared one-sided derivative and state split,
or a specified smooth-flow limit. Such a construction is outside the smooth
closure theorem and is treated as a scope test in Experiment 2B.

\subsection{Generator and front dynamics}
\label{subsec:generator-dynamics}

The next two steps enforce consistency between the road-wise and detector-wise
descriptions and then extract equations for traffic quantities. The operator
identity is a compact way to retain the coupled movement and merger of an
entire front population; the front-number equation below is the form used for
transportation interpretation and computation.

\noindent\textbf{Step 5: generator evolution.}
Compatibility of the two views gives the zero-curvature equation inherited
from the stochastic conservation-law framework
\citep{menon-srinivasan2010}:
\begin{equation}
\label{eq:generator-zero-curvature-main}
 \partial_t\A-\partial_x\B=[\A,\B].
\end{equation}
Only for a spatially homogeneous coefficient law does
\(\partial_x\B=0\), reducing \eqref{eq:generator-zero-curvature-main} to
\(\partial_t\A=[\A,\B]\). A signal, ramp, or spatial bottleneck generally
retains the \(\partial_x\B\) term and may also require source or boundary
operators. The stationary relation \(\A_t^\dagger p_t=0\) describes the
road-wise marginal at each physical time; it does not imply
\(\partial_t p_t=0\).

\begin{assumption}[Kinetic regularity]
\label{ass:kinetic-regularity}
In addition to Assumption \ref{ass:generator-regularity}, the continuum
front law is spatially homogeneous, \(p(\rho,t)>0\) on the active density
range, and \(a\), \(p\), and \(\lambda\) have the differentiability and
integrability required by the displayed derivatives and collision integrals.
The operators \(\A_t\) and \(\B_t\) preserve a common core admitting an
operator-determining family of compatible backward observables as specified
in Appendix \ref{app:zero-curvature}. Mixed derivatives are valid there,
binary mergers have
finite rate, simultaneous triple mergers have probability \(o(\dd t)\), and
no front-number flow enters through the density-state boundary.
\end{assumption}

The common-core, boundary, and adjoint conditions behind this assumption are
listed in Appendices \ref{app:generator-adjoints} and
\ref{app:zero-curvature}. They are conditions of the present continuum
result, not consequences claimed for every random traffic field.

\noindent\textbf{Step 6: kinetic front equation.}
Expanding the commutator first gives the continuous-recovery equation
\begin{equation}
\label{eq:a-drift-equation}
 \partial_ta(\rho,t)=q''(\rho)a(\rho,t)^2\leq0.
\end{equation}
For a strictly concave fundamental diagram, a decreasing recovery gradient therefore
flattens. The two density states attached to an isolated front are transported
with velocities
\begin{align}
\label{eq:phase-velocities}
 W_\rho(\rho,\eta,t)
 &=\bigl(q'(\rho)-\RH{\rho}{\eta}\bigr)a(\rho,t),\\
 W_\eta(\rho,\eta,t)
 &=\bigl(q'(\eta)-\RH{\rho}{\eta}\bigr)a(\eta,t).\notag
\end{align}
Strict concavity gives \(W_\rho\geq0\) and \(W_\eta\leq0\), so an isolated
front weakens in density-state space even when its physical trajectory moves
upstream. At the conditional-rate level, the commutator yields
\[
 \partial_t\lambda
 +\partial_\rho(\lambda W_\rho)
 +\partial_\eta(\lambda W_\eta)
 =\Qtr(\lambda,\lambda)
 +\lambda\left[
   \bigl(q'(\rho)-\RH{\rho}{\eta}\bigr)\partial_\rho a
   +q''(\rho)a
 \right],
\]
where \(\Qtr\) collects the three ordered two-front interactions. Appendix
\ref{app:commutator-expansion} gives the term-by-term operator derivation.

For traffic use, the conditional rate is converted to the physically
normalized number density \(F=p\lambda\). The resulting balance is the
upper-density-jump version of the shock-clustering kinetics associated with
the cited generator evolution; the explicit \(F/p\) normalization makes its
output a road frequency rather than a conditional transition rate.

\begin{proposition}[Front-frequency balance]
\label{prop:F-transport}
Under Assumption \ref{ass:kinetic-regularity}, the front-number density
\(F=p\lambda\) satisfies
\begin{equation}
\label{eq:F-transport}
 \partial_tF
 +\partial_\rho(FW_\rho)
 +\partial_\eta(FW_\eta)
 =\Cp(F,F).
\end{equation}
\end{proposition}

For \(\rho<\eta\), the collision operator is
\begin{align}
\label{eq:collision-operator-main}
 \Cp(F,F)(\rho,\eta)
 &=\int_\rho^\eta
  \bigl(\RH{\rho}{m}-\RH{m}{\eta}\bigr)
  \frac{F(\rho,m)F(m,\eta)}{p(m)}\dd m \notag\\
 &\quad-F(\rho,\eta)\int_\eta^{\rhoj}
  \bigl(\RH{\rho}{\eta}-\RH{\eta}{m}\bigr)
  \frac{F(\eta,m)}{p(\eta)}\dd m \notag\\
 &\quad-F(\rho,\eta)\int_0^\rho
  \bigl(\RH{m}{\rho}-\RH{\rho}{\eta}\bigr)
  \frac{F(m,\rho)}{p(\rho)}\dd m .
\end{align}
The first integral creates a composite \(\rho\to\eta\) front when an
intermediate state \(m\) disappears. The second and third remove the selected
front when it merges with a downstream or upstream neighbor. Strict
concavity makes every displayed relative speed nonnegative. Each binary
event removes two parents and creates one composite front, so the total front
count falls by one even though a gain appears at a new state pair. Appendix
\ref{app:number-density} derives \eqref{eq:collision-operator-main} from
adjacent-state probabilities and checks its units and signs.

\subsection{Traffic performance measures}
\label{sec:reconstruction-validation}
\label{subsec:mapping-back}

Step 7 converts \(\A_t\), \(p_t\), and \(F_t\) into quantities that can be
read as traffic performance measures. A deterministic atomic front is mapped
to a continuous trajectory; a stochastic front law is mapped to counts,
sizes, speeds, spacing, and density probabilities; and the spatial generator
at a specified time is mapped to an instantaneous travel-time distribution.

\noindent\textbf{Step 7: derive performance measures.}
For an atomic front joining \(\rho^-\) to \(\rho^+\), the jump part of
\(\B\) supplies the secant speed and hence
\[
 x_s(t)=x_s(0)+\int_0^t
 \RH{\rho^-(u)}{\rho^+(u)}\dd u.
\]
The states are constant for a Riemann shock and follow the phase transport in
\eqref{eq:phase-velocities} when adjacent smooth recovery regions change
them. Placing the reconstructed discontinuities in their continuous physical
coordinates, and filling the intervening regions with the characteristic or
recovery solution, gives a density realization. The signal examples apply
the same operation event by event to queue formation, discharge, interaction,
and clearance waves.

For a stationary stochastic road, the one-point law is \(p_t\). If
\(P_h(t)\) is the downstream transition law over distance \(h\), the Markov
product construction gives joint congestion probabilities at several
locations \citep{ethier-kurtz1986}. Its complete expression is recorded in
Appendix \ref{app:generator-adjoints}; in the finite-state realization,
\(P_h(t)=\exp(hA_N(t))\).

The total front intensity is
\begin{equation}
\label{eq:shock-intensity}
 \Gamma(t)=\int_0^{\rhoj}\int_\rho^{\rhoj}
 F(\rho,\eta,t)\dd\eta\dd\rho.
\end{equation}
When \(0<\Gamma(t)<\infty\), selecting a front in proportion to its road
frequency gives the normalized adjacent-state law
\begin{equation}
\label{eq:shock-palm}
 \mathbb P_{\rm shock}
 \{(\rho^-,\rho^+)\in\dd\rho\dd\eta\}
 =\frac{F(\rho,\eta,t)}{\Gamma(t)}\dd\rho\dd\eta.
\end{equation}
This is the standard Palm normalization for a stationary point process
\citep{daley-verejones2003}. Pushing this law through the maps in Table
\ref{tab:front-output-mappings} yields the shock-size and shock-speed
distributions. The spatial semigroup additionally yields the front-count law
on a finite link and the inter-front spacing law; these are distributional
outputs of the generator rather than statistics from one selected road
realization.

Table \ref{tab:front-output-mappings} relates each mathematical output to a
road-level quantity. Shock intensity measures how frequently drivers encounter
abrupt density increases; shock size measures their severity, and shock speed
describes how congestion boundaries move, not how fast vehicles travel.
Counts and gaps describe the spatial organization of congestion, while
travel-time reliability summarizes variation across entire road profiles.

\begin{table}[H]
\caption{Traffic performance measures and their reconstruction.}
\label{tab:front-output-mappings}
\centering
\small
\begin{tabular}{p{0.25\linewidth}p{0.33\linewidth}p{0.30\linewidth}}
\hline
Traffic measure & Meaning on the road & Reconstruction \\
\hline
Mean shocks per kilometre & Frequency of abrupt congestion boundaries & \(\Gamma(t)\) \\
Shock size & Density increase across a boundary & Distribution of \(\eta-\rho\) under \(F/\Gamma\) \\
Shock-wave speed & Speed and direction of a congestion boundary & Distribution of \(\RH{\rho}{\eta}\) under \(F/\Gamma\) \\
Congestion probability & Chance of high density at one or several locations & \(p_t\) and spatial transition semigroup \\
Shock count and spacing & Number on a link and gaps between neighbors & Marked and killed spatial-generator semigroups \\
Queue-boundary trajectory & Position of an identifiable queue boundary over time & Time integral of its secant speed \\
Instantaneous travel-time reliability & Variation in estimated link travel time at a specified time & Markov-reward law at that time \\
\hline
\end{tabular}
\end{table}

Instantaneous travel time is the distance integral of inverse vehicle speed
using conditions along the road at one specified time \(t^*\)
\citep{fhwa2012}. It estimates the time needed to traverse the link if these
conditions persist, rather than following a vehicle through changing
conditions. For its distribution, let \(Y_x\) be the spatial chain with row generator \(A_N\), stationary initial law \(\pi\), and density
states \(\rho_i\). Define
\[
 c_i=\frac{1}{v(\rho_i)},\qquad
 v(\rho)=\frac{q(\rho)}{\rho},\qquad
 C=\operatorname{diag}(c_0,\ldots,c_{N-1}).
\]
All retained states are below jam density and the free-flow limit is used at
zero density. The travel time accumulated over link length \(L\) is
\(T_L=\int_0^L c(Y_x)\dd x\). The finite-state Feynman--Kac formula
\citep{fitzsimmons-pitman1999} gives
\begin{equation}
\label{eq:feynman-kac-travel}
 \Phi(\theta,L)
 =\mathbb E_\pi[e^{-\theta T_L}]
 =\pi\exp\{L(A_N-\theta C)\}\mathbf1.
\end{equation}
The complete distribution is also computed from the conservative
Markov-reward balance
\begin{equation}
\label{eq:reward-transport}
 \partial_x g+\partial_\tau(gC)=gA_N.
\end{equation}
With \(g(0,\dd\tau)=\pi\delta_0(\dd\tau)\), its marginal over density
states gives the reward law. The first two conditional moments can be
computed independently from
\[
 m_0(x)=\mathbf1,\qquad
 m_1'=A_Nm_1+C\mathbf1,\qquad
 m_2'=A_Nm_2+2Cm_1,\qquad m_1(0)=m_2(0)=0.
\]
Thus \(\mathbb E_\pi T_L=\pi m_1(L)=L\pi c\) under stationarity and
\(\mathbb E_\pi T_L^2=\pi m_2(L)\). These column moment equations and the row
distribution equation use the same row-generator convention.
The reward law returns the mean, standard deviation, quantiles, buffer time, buffer index,
and planning-time index. This is a distribution across spatial density
fields. It is not a vehicle trajectory through a density field that continues
to evolve during the trip.

\subsection{Numerical realization}
\label{subsec:state-space-computation}

Equations \eqref{eq:a-drift-equation} and \eqref{eq:F-transport} are posed in
continuous physical coordinates, but their numerical evaluation still
requires approximation. We discretize possible density states and integrate
the resulting evolution equations in physical time. We do not divide the
road into finite-volume cells. A Godunov or CTM comparator, by contrast,
discretizes the physical road--time domain and is run realization by
realization.

\noindent\textbf{Direct front-law solver.}
\phantomsection\label{subsubsec:direct-front-solver}
Experiment 3 uses
\(z=(\rho-\rho_{\min})/(\rho_{\max}-\rho_{\min})\in(0,1)\) and the stationary
piecewise-deterministic input
\begin{equation}
\label{eq:pdmp-generator}
 (\A_0\varphi)(z)
 =-\kappa z(1-z)\varphi'(z)
 +\kappa\int_z^1\{\varphi(w)-\varphi(z)\}\dd w .
\end{equation}
It has recovery law \(z_x=-\kappa z(1-z)\), uniform one-point density
\(p_0(z)=1\), and initial front density \(F_0(z,w)=\kappa\) for \(z<w\).
This process is a controlled traffic-admissible experiment design, not an
empirically calibrated stop-and-go model; its piecewise-deterministic form is
standard \citep{davis1993}.

For the normalized flow function
\(\widetilde q(z)=q(\rho_{\min}+Dz)/D\), where
\(D=\rho_{\max}-\rho_{\min}\), the recovery equation has the pointwise
solution
\begin{equation}
\label{eq:a-numerical-explicit}
 a(z,t)=\frac{\kappa z(1-z)}
 {1-\widetilde q''(z)\kappa z(1-z)t}.
\end{equation}
The triangle \(0<z<w<1\) is divided into density-state bins. Conservative
upwinding treats the two state-transport terms, midpoint quadrature treats the
three merger terms, and a strong-stability-preserving Runge--Kutta method
advances time. At each stage, the discrete stationary adjoint reconstructs
\(p\). Positivity, state-space CFL, and collision gain--loss balance are checked.
The reconstructed marginal is normalized explicitly. At finite resolution,
this normalization does not also prove the continuum adjoint identity:
its raw mass defect is reported before normalization. State transport can
also carry unresolved small fronts into the diagonal \(z=w\), producing a
numerical count loss distinct from physical binary mergers. The audit records
both effects and their state-resolution sensitivity. The output
is an evolving law of front pairs, not an Eulerian road-cell density movie.

\noindent\textbf{Generator at a specified time.}
\phantomsection\label{subsubsec:snapshot-generator}
Multi-location and travel-time calculations use a conservative finite-state
representation of \(\A_t\) at a selected physical time. Let
\(z_0<\cdots<z_{N-1}\), let \(\Delta z\) be the bin width, and normalize
\(\pi_i\propto p_i\Delta z\). The bin-integrated upward rates are
\begin{equation}
\label{eq:snapshot-upward-rates}
 J_{ij}=\frac{F_{ij}}{p_i}\Delta z,
 \qquad j>i.
\end{equation}
To preserve the stationary law at finite resolution, the neighboring
downward recovery rates are obtained from a face balance rather than by
sampling \(a_i/\Delta z\). Define
\begin{equation}
\label{eq:snapshot-face-balance}
 U_i=\pi_i\sum_{j>i}J_{ij},\qquad
 I_i=\sum_{k<i}\pi_kJ_{ki},\qquad
 d_0=0,\quad d_{i+1}=d_i+U_i-I_i.
\end{equation}
Here \(d_i\) approximates the probability flow \(ap\) through density-state
face \(i\). The row generator at that time is
\begin{equation}
\label{eq:snapshot-generator-discrete}
 (A_N)_{ij}=J_{ij}\ (j>i),\qquad
 (A_N)_{i,i-1}=\frac{d_i}{\pi_i}\ (i\geq1),\qquad
 (A_N)_{ii}=-\sum_{j\ne i}(A_N)_{ij}.
\end{equation}
Under the discrete stationary balance, \(d_i\geq0\), the terminal face flow
vanishes, \(A_N\mathbf1=0\), and \(\pi A_N=0\) to roundoff. This matrix is a
density-state approximation to the spatial generator at the selected time. It is neither a
physical road mesh nor the upper-triangular pure-shock Lax matrix retained as
an algebraic check in Appendix \ref{app:spectral-integrability}.

The continuum formulation therefore avoids finite-volume discretization of
the physical road space--time domain and represents front locations and
interactions continuously in physical coordinates. Its implementation is not
discretization-free: it uses density-state bins, quadrature, and numerical
integration in physical time. This distinction defines the computational
claim tested against road-grid Godunov ensembles in Section
\ref{sec:numerical-results}. The homogeneous smooth-flow closure remains
conditional on Assumptions \ref{ass:generator-regularity} and
\ref{ass:kinetic-regularity}; deterministic boundary forcing and the
triangular-flow limit are clearly labeled extensions rather than additional
closure theorems.

\section{Experiments}
\label{sec:numerical-results}

The experiments progress from classical deterministic consistency to the
distributional questions for which the generator formulation is intended.
Experiments 1 and 2 use analytical validation against one independently coded
traffic-wave solution. Experiments 3 and 4 use numerical comparison against
one matched Monte Carlo Godunov ensemble. Keeping these tracks separate
separates classical wave consistency from stochastic distributional agreement.

\subsection{Experimental design}
\label{subsec:pairwise-protocol}

The tests answer two successive questions: do the wave rules recover familiar
LWR traffic dynamics, and does direct evolution of a random front population
give useful traffic distributions? Table \ref{tab:pairwise-comparison-map}
assigns the corresponding output and comparator. Analytical validation establishes recovery
of known LWR behavior. Numerical comparison asks whether direct law evolution
and an ensemble of road-grid realizations give similar stochastic outputs.

The exact Riemann and signal comparators use closed formulas and event logic
implemented independently of the proposed update. Godunov comparisons use
the same Greenshields fundamental diagram, continuous input-path law, observation geometry,
time horizon, and fixed random stream on both sides. The proposed solver
discretizes density state and physical time but not the physical road. The
Godunov solver discretizes road space and time and is repeated over the
stochastic ensemble. Accordingly, Mode-A quantities are proposed-versus-exact
differences, whereas Mode-B quantities are agreement, discrepancy, grid
sensitivity, or Monte Carlo precision. Appendix
\ref{app:numerical-checks} gives the definitions and acceptance checks.

\begin{table}[H]
\caption{One comparator assigned to each experiment.}
\label{tab:pairwise-comparison-map}
\centering
\small
\begin{tabular}{p{0.13\linewidth}p{0.28\linewidth}p{0.23\linewidth}p{0.24\linewidth}}
\hline
Experiment & Reconstructed output & Mode & Sole comparator \\
\hline
1 & Riemann shock path and density & A: analytical & Exact Riemann solution \\
2A & Greenshields signal waves & A: analytical & Exact wave-interaction solution \\
2B & Triangular signal waves & A: analytical & Exact triangular LWR \\
3 & Stochastic front statistics & B: numerical & Monte Carlo Godunov \\
4 & Travel-time reliability law & B: numerical & Monte Carlo Godunov \\
\hline
\end{tabular}
\end{table}

\subsection{Example 1: Riemann problem}
\label{subsec:rq1-riemann}

\noindent Comparison used in this experiment: proposed stochastic method versus
the exact LWR entropy solution.
The traffic question is whether the generator/front representation genuinely
contains the classical motion of one queue boundary. On a 4 km road, the
initial interface at \(x_0=2000\) m connects
\(\rho_L=30\) to \(\rho_R=90\) veh/km/lane. The Greenshields diagram uses
\(v_f=13.33\) m/s and \(\rhoj=150\) veh/km/lane.

This spatially localized datum is embedded through the atomic transition
measure
\[
 M_0(\dd x,\dd\rho,\dd\eta)
 =\delta_{x_0}(\dd x)\,
  \delta_{(\rho_L,\rho_R)}(\dd\rho,\dd\eta),
\]
rather than being forced into the stationary generator assumption used by the
many-front experiment. The distributional LWR front balance gives
\(\dot X=[q(\rho_R)-q(\rho_L)]/(\rho_R-\rho_L)\).
The proposed route integrates this atom velocity numerically and reconstructs
\(\rho_L+(\rho_R-\rho_L)H(x-X(t))\). The independent route evaluates the
closed entropy solution directly. This checks the characteristic/secant
conventions used by the framework; a deterministic atom is not a regular
Feller generator, so this experiment does not validate the stochastic Lax
closure or a commutator of singular operators.

The independent comparator evaluates
\[
 s_{\rm RH}=\frac{q(\rho_R)-q(\rho_L)}{\rho_R-\rho_L}
 =2.666\ \mathrm{m/s},
 \qquad x_s(t)=x_0+s_{\rm RH}t.
\]
The entropy ordering is strict:
\(q'(\rho_L)=7.998>s_{\rm RH}>q'(\rho_R)=-2.666\) m/s. The absolute and
relative speed differences are both zero at stored precision; the maximum
120 s path difference
is \(2.001\times10^{-11}\) m, the largest selected-profile
\(L^1\) difference is \(8.185\times10^{-13}\) veh/lane, and the maximum
vehicle-balance residual is \(1.215\times10^{-12}\) veh/lane. Changing the
front-transport step from 12 to 0.75 s changes only roundoff, as expected for a
constant atom velocity.

\begin{figure}[!htbp]
\centering
\includegraphics[width=\textwidth]{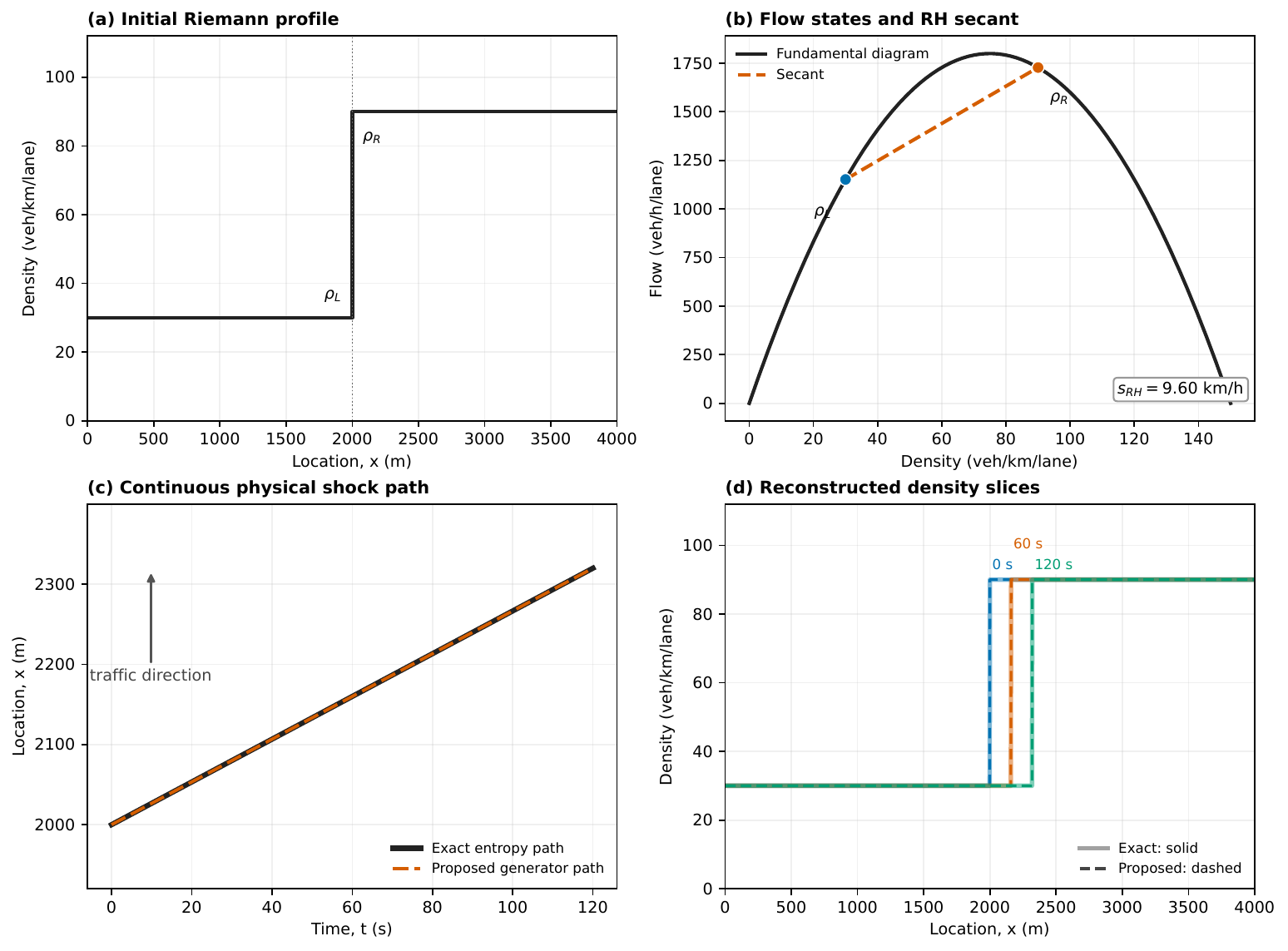}
\caption{Experiment 1 analytical validation: (a) initial Riemann density,
(b) Greenshields diagram with the state secant and Rankine--Hugoniot speed,
(c) continuous exact and proposed shock paths with time horizontal and
downstream location vertical, and (d) reconstructed density slices. The sole
comparator is the independent entropy solution.}
\label{fig:rq1-riemann-pairwise}
\end{figure}
\FloatBarrier

Figure \ref{fig:rq1-riemann-pairwise} connects the imposed density jump in
panel (a) to the flow secant in (b), the coincident shock paths in (c), and
the translated density profiles in (d). This provides a consistency and normalization check. Exact Riemann analysis
already solves this traffic problem more directly; the value of the
generator framework appears when the single atom is replaced by a random
population of state transitions.

\subsection{Signalized traffic}
\label{subsec:rq2-signal}

The second experiment retains a finite, analytically identifiable wave system
while changing fundamental-diagram regularity. A 600 m approach carries constant demand of
1200 veh/h/lane. The signal is red from 0 to 60 s and then green through the
300 s horizon. Red creates a stop-line queue and green launches the recovery
wave. Both subexperiments inject boundary waves, assign characteristic and secant
speeds, advance the finite wave measure through interaction events, and
reconstruct congestion extent, cumulative departures, and delay. This is a
formal deterministic boundary extension; the homogeneous stochastic closure
theorem does not by itself cover signal forcing.

\begin{figure}[!htbp]
\centering
\includegraphics[width=\textwidth]{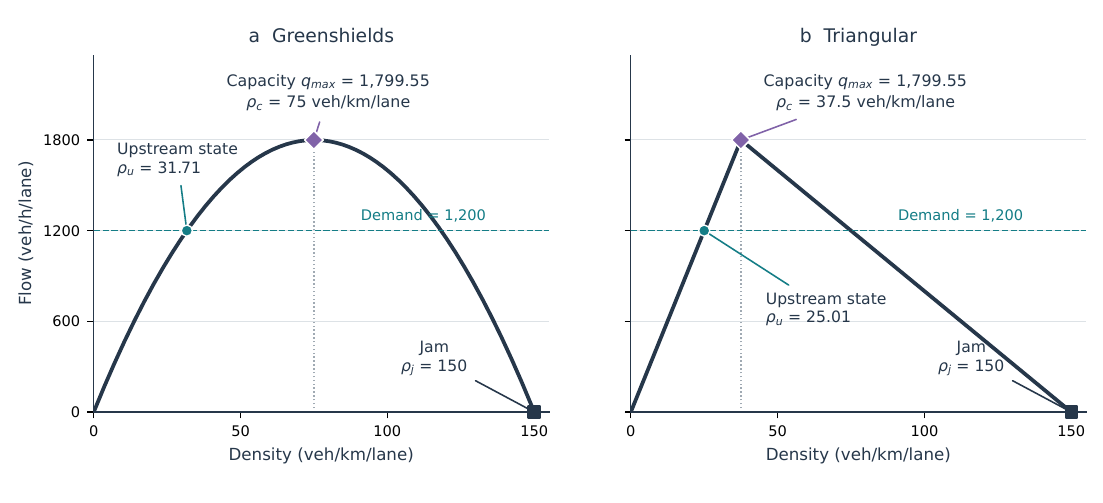}
\caption{Fundamental diagrams used in the signal experiments. Diamonds mark
the true maximum flow (capacity): Greenshields reaches it at 75 veh/km/lane,
and the triangular diagram at 37.5 veh/km/lane. Circles mark the distinct
free-branch densities carrying the same 1200 veh/h/lane demand. Jam density
has zero flow. These parameters differ from the faster road in Experiments
3--4 and the framework example.}
\label{fig:traffic-states}
\end{figure}

The diagrams in Figure \ref{fig:traffic-states} have the same capacity,
1799.55 veh/h/lane, but different critical densities. Demand is below capacity,
both maximum congested extents remain shorter than the 600 m approach, and
green lasts long enough for clearance. There is no upstream spillback in this
test.

\FloatBarrier

Four observables must be distinguished. If \(X(t)\) is the rear boundary,
the spatial queue length is \(\ell_c(t)=L-X(t)\): the road occupied by
stopped traffic and the moving discharge region behind the rear boundary.
This was termed congested extent, but it is a geometric length, not a count
of waiting vehicles or a claim that all vehicles in that region are stopped.
The stopped length \(\ell_0(t)\) extends from the rear boundary to the
upstream edge of the recovery wave, until they meet. For Greenshields this
edge belongs to a rarefaction fan; for the triangular diagram it is a contact. Cumulative departures
\(N_{\rm out}(t)\) count vehicles crossing the stop line; their derivative
\(q_{\rm out}(t)=\dot N_{\rm out}(t)\) is discharge flow. Finally,
\(E(t)=q_dt-N_{\rm out}(t)\) measures excess vehicles relative to the
initially uniform free-flow condition. Its integral is total delay.
Appendix \ref{app:signal-metrics} derives all four quantities.

\subsubsection{Greenshields diagram}
\label{subsubsec:rq2a-greenshields}

\noindent Comparison used in this experiment: proposed method versus
the independent exact Greenshields characteristic and wave-interaction
solution.
Here \(q(\rho)=v_f\rho(1-\rho/\rhoj)\), with the same
\(v_f=13.33\) m/s and \(\rhoj=150\) veh/km/lane as Experiment 1. Matching the
demand gives \(\rho_{\rm up}=31.710\) veh/km/lane and capacity
1799.55 veh/h/lane. During red, the queue atom joins the upstream state at jam
density and moves upstream at \(-2.818\) m/s. At green, the characteristic
emanating from jam density moves at \(-13.33\) m/s. After the two waves meet,
the queue front's right state is selected from the exact Greenshields
rarefaction, and its speed is updated by the corresponding secant of \(q\).

The independent construction solves the quadratic characteristic geometry in
closed form. It places the interaction at 76.084 s and 385.601 m, the maximum
congested extent of 230.998 m at 90.023 s, and clearance at 180.090 s. The proposed
reconstruction differs by at most \(5.491\times10^{-11}\) m along the queue
path and by zero at stored precision along the pre-interaction discharge wave.
The clearance-time difference is below \(10^{-8}\) s, the maximum
cumulative-departure and conservation residuals are below \(1.4\times10^{-9}\)
veh/lane, and independent integration of excess vehicles agrees with the
closed total delay of 1800.901 veh s/lane to \(10^{-9}\) veh s/lane.

\begin{figure}[!htbp]
\centering
\includegraphics[width=\textwidth]{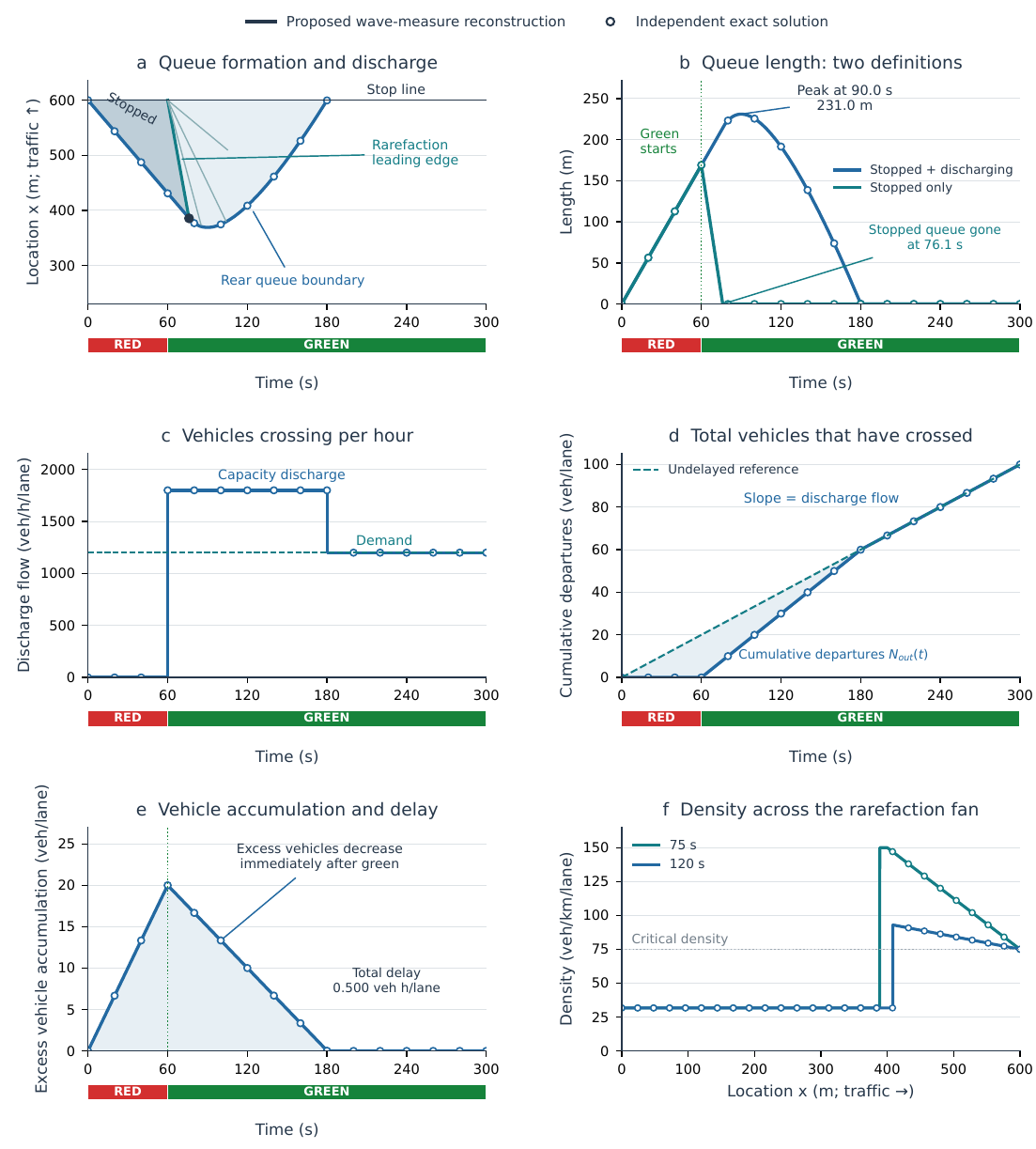}
\caption{Greenshields signal dynamics: (a) wave paths and the stopped/moving
regions, (b) spatial queue length (stopped plus discharging traffic) versus
stopped length, (c) discharge flow,
(d) cumulative departures versus an undelayed reference, (e) excess vehicles
and their delay integral, and (f) density profiles. Solid curves are the
proposed front-measure calculation; open markers are independently computed
exact values. Opaque red and green rectangles below each time axis show the
signal phase, not a traffic-state category.}
\label{fig:rq2a-greenshields}
\end{figure}
\FloatBarrier

Figure \ref{fig:rq2a-greenshields}(a,b) distinguishes the two queue lengths.
The stopped length decreases immediately when green begins at 60 s.
Before the rarefaction's leading edge reaches the rear at 76.084 s,
\[
 \ell_0(t)=\max\{L-v_f(t-60)-X(t),0\},\qquad
 \dot\ell_0=-13.33-(-2.818)=-10.512\ \mathrm{m/s}.
\]
Thus the stopped queue shrinks, rather than grows, after green and is gone
at 76.084 s. In contrast, the rear continues upstream until 90.023 s,
so the road length occupied by stopped \emph{and discharging} traffic can
still increase. Vehicles begin moving and spreading out within that region;
its spatial extent is not the number of excess vehicles.

Panels (c,d) show why capacity exceeding demand is fully consistent with
this geometry: discharge is 1799.55 veh/h/lane during queued green, above the
1200 veh/h/lane arrival rate, and cumulative departures catch up with the
undelayed reference. Accordingly, panel (e) shows
\(\dot E=q_d-q_{\max}=-599.55\) veh/h/lane: excess vehicles decrease
throughout queued green, while their accumulated delay continues increasing
until clearance. Panel (f) resolves the moving density states within the
rarefaction fan. The fan's leading edge, the rear boundary, stopped length,
and excess-vehicle count therefore describe different aspects of queue
dissipation and should not all be called queue length.

The agreement shows that the finite-wave specialization preserves
the characteristic, secant-speed, interaction, count, and conservation logic
of smooth LWR. It does not extend the spatially homogeneous stochastic closure
to signals; that extension remains a boundary-forcing problem.

\subsubsection{Triangular diagram}
\label{subsubsec:rq2b-triangular}

\noindent Comparison used in this experiment: deterministic wave
calculations for smooth, strictly concave approximations \(q_\varepsilon\)
versus the independent exact triangular-LWR solution.
The triangular diagram is
\(q_T(\rho)=\min\{v_f\rho,w(\rhoj-\rho)\}\), with
\(w=4.4433\) m/s and critical density 37.5 veh/km/lane. Its exact Newell/contact
construction gives \(\rho_{\rm up}=25.006\) veh/km/lane, an upstream queue
speed of \(-2.667\) m/s, a release contact speed of \(-4.443\) m/s, maximum
congested extent of 400.200 m, and clearance at 180.090 s.

The current closure theorem does not apply directly because \(q_T\) is
not differentiable at the capacity kink and is affine on each branch. We therefore
replace the minimum by the strictly concave smooth family
\(q_\varepsilon\) in Appendix \ref{app:signal-metrics}, build \(\B\) from
\(q_\varepsilon'\) and its secants, and compare the reconstructed wave measure
with the independently coded triangular solution. As
\(\varepsilon/q_{\max}\) decreases from 0.08 to 0.01, the flow supremum
difference falls from 70.18 to 8.97 veh/h/lane, maximum-queue difference from
1.126 to 0.018 m, clearance-time difference from 13.491 to 1.576 s, and
cumulative-departure path difference from 2.018 to 0.259 vehicle. The maximum
queue-path difference decreases from 144.47 to 20.43 m; this nonuniform metric
is dominated by the interval between the two nearby clearance times, when one
solution has already returned to zero queue.

\begin{figure}[pos=H]
\centering
\includegraphics[width=\textwidth]{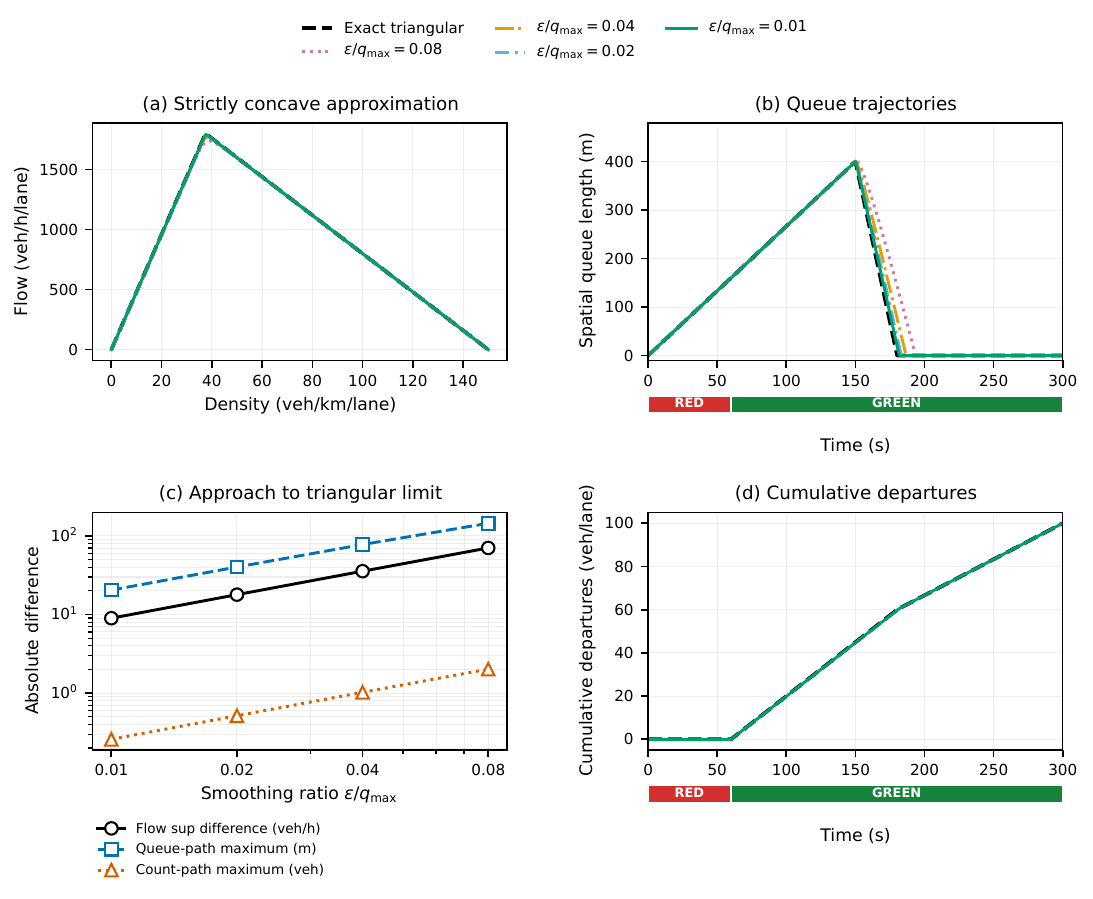}
\caption{Experiment 2B analytical validation: smooth strictly concave
approximations of the triangular diagram, queue paths, selected traffic
differences versus smoothing, and cumulative departures (vehicles/lane). Signal phases are marked
by colored rectangles on the time axes. No road-grid solution is included.}
\label{fig:rq2b-triangular}
\end{figure}

Figure \ref{fig:rq2b-triangular}(a) shows the rounded capacity kink narrowing
as smoothing decreases. Panel (b) shows the corresponding queue trajectories
approaching the triangular contact construction, with their largest separation
near clearance. Panel (c) quantifies decreasing flow, queue-path,
and departure discrepancies, while panel (d) shows the associated cumulative
departures approaching the exact curve. Small errors in maximum queue length
therefore do not imply equally small errors along the entire queue path.

\FloatBarrier

Table \ref{tab:rq2b-kink-cases} separates deterministic validity from
stochastic closure through three kink cases.
If all active states lie on one linear branch, \(\B\) is unique after choosing
that branch and the wave is a contact, but the current strict-concavity theorem
does not apply. For a transition across the critical density, the secant and
endpoint characteristic speeds are unique for the listed distinct states, and
the selected upward shock is entropic, although a direct nonsmooth generator
closure is still formal. If an active state is exactly critical,
\(q'(\rho_c)\) is not unique; \(\B\)'s drift part requires a one-sided choice,
subgradient, or split critical state, even though the secant to a distinct
state remains defined. Thus piecewise-linear concavity is rigorous for the
deterministic entropy calculation, but not yet for the present stochastic
closure at contacts or the kink.

These results quantify a deterministic smooth-flux approximation over the
tested finite horizons. In particular, release from jam density to the
capacity state is a contact on the triangular diagram's congested branch,
whereas a strictly curved approximation spreads the corresponding release
into a rarefaction fan. Smoothing can consequently change front positions
and interaction times even when the flux discrepancy is small. The test does
not establish uniform accuracy over longer times or distances, or
convergence of the stochastic generator and front-intensity law to an exact
triangular-model closure. Section~\ref{subsec:discussion-flux} explains the
pre-interaction kinematics and the limits of that comparison.

\begin{table}[H]
\centering
\caption{Triangular-kink cases. The deterministic entropy statements are rigorous; the direct nonsmooth stochastic-generator extension is not covered by the current closure theorem. Density pairs are in veh/km/lane.}
\label{tab:rq2b-kink-cases}
\footnotesize
\setlength{\tabcolsep}{2pt}
\begin{tabular}{>{\raggedright\arraybackslash}p{0.14\linewidth}>{\centering\arraybackslash}p{0.11\linewidth}>{\raggedright\arraybackslash}p{0.14\linewidth}>{\centering\arraybackslash}p{0.09\linewidth}>{\raggedright\arraybackslash}p{0.13\linewidth}>{\centering\arraybackslash}p{0.07\linewidth}>{\raggedright\arraybackslash}p{0.24\linewidth}}
\hline
Case & $(\rho_L,\rho_R)$ & Is $B$ unique? & Secant (m/s) & Contact? & Closure? & Status \\
\hline
I: one linear branch & (10.0, 25.0) & Yes, branch & 13.330 & Free contact & No & Deterministic rigorous; generator formal \\
II: cross-critical transition & (25.0, 100.0) & Yes, endpoints & -1.481 & No & No & Entropy shock rigorous; generator formal \\
III: state at critical density & (37.5, 100.0) & No at kink & -4.443 & Congested contact & No & Contact rigorous; unsplit $B$ undefined \\
\hline
\end{tabular}
\end{table}

\begin{table}[H]
\caption{Analytical validation of the proposed method. Each row uses only the
named exact traffic-wave comparator; no road-grid result enters the table.
Values are rounded for readability; differences are computed before rounding.
Maximum queue denotes spatial length, including moving discharge.}
\label{tab:analytical-validation-summary}
\centering\small
\begin{tabular}{>{\raggedright\arraybackslash}p{0.24\linewidth}rrr}
\toprule
Experiment and quantity & Exact & Proposed & Absolute difference \\
\midrule
Riemann shock speed (m/s) & 2.666 & 2.666 & 0 \\
Greenshields clearance (s) & 180.09 & 180.09 & \(<10^{-8}\) \\
Greenshields maximum queue (m) & 231.00 & 231.00 & \(<10^{-10}\) \\
Triangular-limit clearance, \(\varepsilon/q_{\max}=0.01\) (s)
 & 180.09 & 181.67 & 1.58 \\
Triangular-limit maximum queue, \(\varepsilon/q_{\max}=0.01\) (m)
 & 400.20 & 400.22 & 0.018 \\
\bottomrule
\end{tabular}
\end{table}

Table \ref{tab:analytical-validation-summary} summarizes the distinction:
the Riemann and Greenshields wave calculations agree to numerical precision,
whereas the triangular-limit rows retain a finite smoothing discrepancy,
especially in clearance time. Together, the two signal cases answer the
regularity question posed in RQ2. Smooth strict concavity is not
needed for deterministic triangular LWR, but it is used essentially by the
current closure proof and by the unambiguous drift construction at every
density state. Smooth approximation supplies a controlled numerical route;
it is evidence of deterministic limiting behavior, not a nonsmooth closure
theorem.

\subsection{Stochastic shock clustering}
\label{subsec:rq3-clustering}

\noindent Comparison used in this experiment: proposed method versus
the matched Monte Carlo Godunov ensemble.
The traffic question now changes from locating one wave to describing a
population of imposed stop-and-go-like density fronts. The 4 km observation
window is surrounded by 2 km buffers and evolved for 90 s under the
Greenshields diagram \(v_f=60\) km/h, \(\rhoj=150\) veh/km/lane. The initial
stationary spatial process is supported on 20--130 veh/km/lane. In normalized
state \(z\in(0,1)\), it recovers continuously according to
\(z_x=-\kappa z(1-z)\) and jumps upward with kernel density
\(\lambda(z,w)=\kappa\) for \(w>z\), where \(\kappa=4\) km\(^{-1}\).
This construction supplies a uniform stationary one-point law and admissible
one-sided traffic jumps without clipping.

The drift and kernel define \(\A_0\); \(\B_0\) is formed from
Greenshields characteristic and Rankine--Hugoniot speeds. The kinetic equation evolves \(F=p\lambda\), with \(p\) reconstructed
from its stationary adjoint on
31, 51, and 71 density states, with a 0.25 s base step.
The reported \(N=71\) law is the finest of these three primary comparison
resolutions, not a special physical state count. Its uniform density bins
have width \(110/71\simeq1.55\) veh/km/lane; \(N\) counts density states,
not road cells. The same law is reused in the reliability experiment to
avoid changing the traffic input between outputs. The finer \(N=141\)
audit below assesses residual state-resolution effects; \(N=71\) is a
finite-resolution reference, not a certified continuum-converged solution. Reconstruction gives
front intensity, size and severity laws, collision gain, count and spacing
distributions, and density moments. Representative continuous-coordinate wave
paths are generated sample by sample from the same PDMP input and deterministic
wave rules to visualize the physical meaning of the population balance. The
unlabeled law \(F\) does not uniquely determine those labeled paths, and the
paths are neither a second comparator nor an input to any reported
proposed-versus-Godunov discrepancy. The proposed statistical calculation does
not require road cells. The comparator analytically averages the same 800 continuous input
paths over 40, 20, and 10 m road cells and advances each realization with the
exact scalar Godunov flow. Fronts are then extracted by one fixed 0.05
veh/km/lane positive-increment rule.

\begin{figure}[!htbp]
\centering
\includegraphics[width=\textwidth]{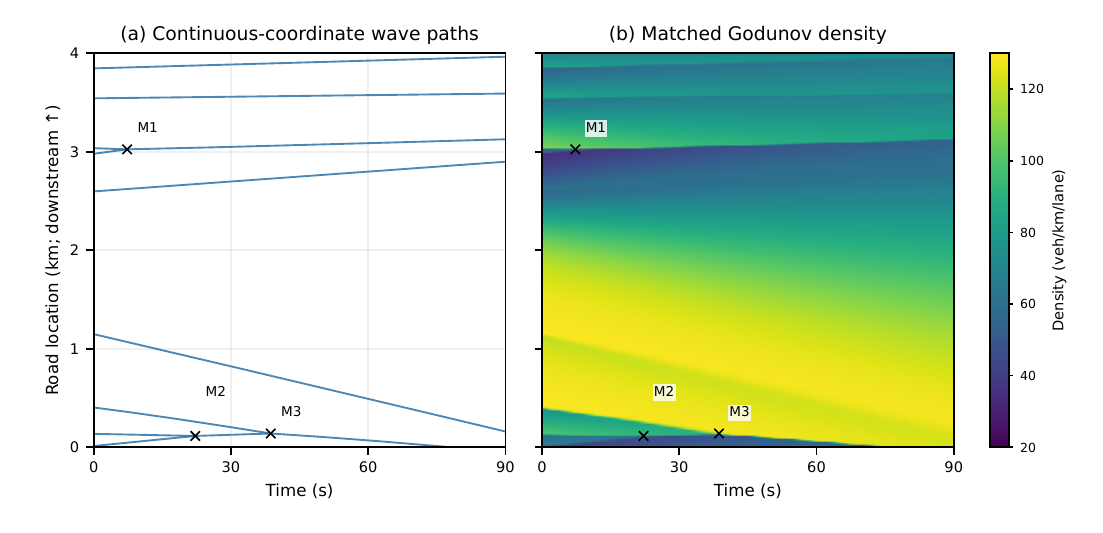}
\caption{One initial realization, two aligned views: (a) continuous-coordinate
wave paths and (b) the matched 10 m Godunov density field. Traffic travels
toward increasing road location. Crosses M1--M3 mark continuous-path merger
locations in both panels for orientation. The paths use the same initial
PDMP sample and deterministic wave rules; they are not uniquely reconstructed
from the unlabeled law \(F\). The density field is the numerical comparator.}
\label{fig:rq3-continuous-paths}
\end{figure}

Figure \ref{fig:rq3-continuous-paths} places the continuous-coordinate wave
paths beside the Godunov density field from the same initial realization,
with identical time and road-location axes. Converging paths on the left
correspond to density boundaries meeting on the right. The marked merger
locations are taken from the continuous-path calculation and provide visual
alignment, not independently detected Godunov events. This samplewise picture
explains what merging means before the ensemble statistics are considered.

\FloatBarrier

\begin{figure}[pos=H]
\centering
\includegraphics[width=\textwidth]{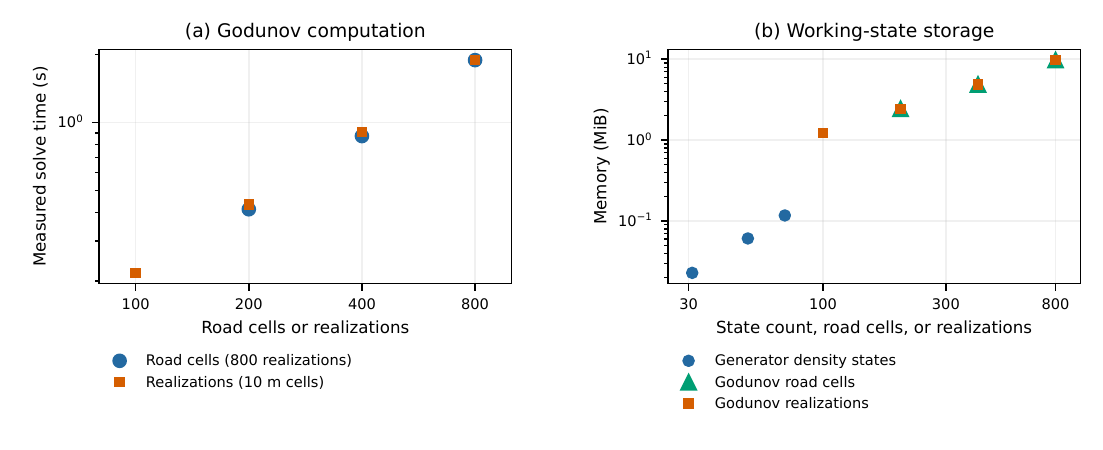}
\caption{Measured computational resources: (a) Godunov solve time with
road-cell count or ensemble size varied separately, and (b) working-state
storage for density-state, road-cell, and ensemble resolutions. Markers denote
measured configurations; no asymptotic scaling rate is fitted.}
\label{fig:rq3-godunov-density}
\end{figure}
\FloatBarrier

\begin{figure}[!htbp]
\centering
\includegraphics[width=\textwidth]{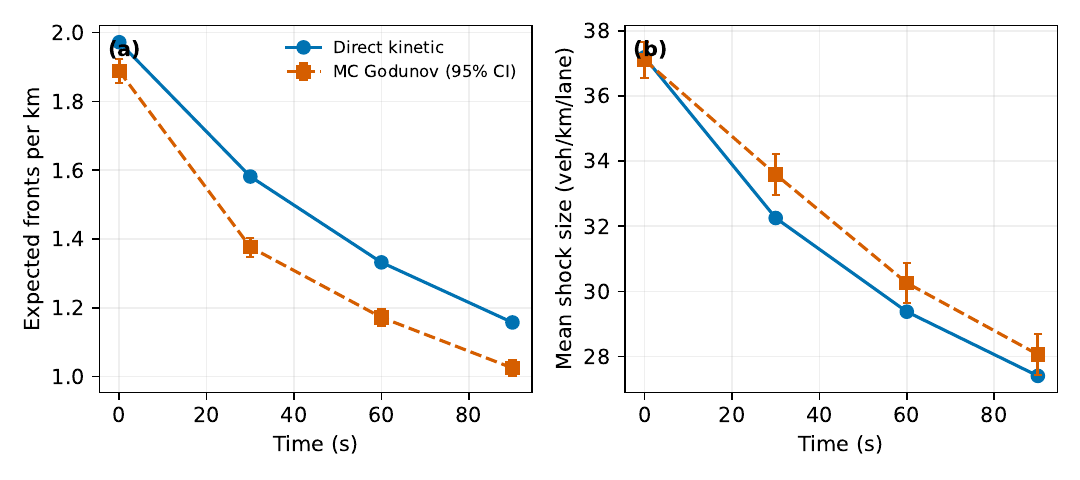}
\caption{Experiment 3 proposed-versus-Godunov evolution of front intensity and
jump amplitude. Godunov bands represent ensemble uncertainty; all curves use
the same stochastic input law.}
\label{fig:rq3-shock-statistics}
\end{figure}

\begin{figure}[!htbp]
\centering
\includegraphics[width=\textwidth]{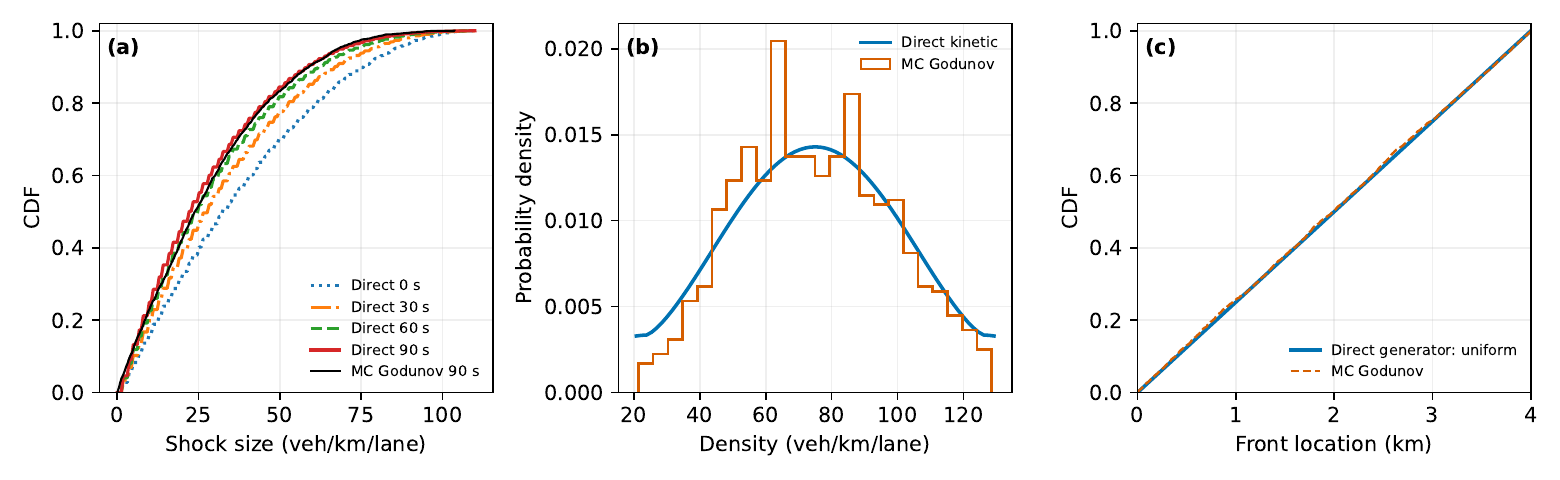}
\caption{Experiment 3 distributional comparison: evolving proposed shock-size
CDFs with the 90 s Godunov empirical CDF, proposed-versus-Godunov one-point
density laws, and the front-location CDF at 90 s. The direct location law is
uniform by spatial homogeneity.}
\label{fig:rq3-distributions}
\end{figure}

Figure \ref{fig:rq3-shock-statistics} shows a declining number of shocks and
smaller mean density jumps as the population evolves. Figure
\ref{fig:rq3-distributions} shows how that evolution changes the full
shock-size distribution, while the density and location panels compare the
laws at 90 s. The nearly uniform location CDF reflects spatial homogeneity,
not a prediction of a preferred queue location.

At 90 s, the 71-state calculation gives 1.157 fronts/km, mean jump size
27.405 veh/km/lane, severe-front probability
\(\Pr\{\Delta\rho\geq50\}=0.155\), and density standard deviation
25.179 veh/km/lane. The 10 m, 800-path Godunov ensemble gives 1.025
fronts/km, 28.063 veh/km/lane, 0.167, and 23.111 veh/km/lane, respectively.
The one-point density CDF maximum difference is 0.047, and its Wasserstein
distance is 2.034 veh/km/lane. For shock size, those values are 0.050 and
1.131 veh/km/lane; the shock-count and spacing CDF maximum differences are
0.144 and 0.090. Spatial homogeneity makes the direct front-location law
uniform on the 4 km observation window, with a mean of 2.000 km. The 3279 pooled
Godunov fronts at 90 s have mean location 1.984 km, a difference of
\(-0.016\) km, and a maximum CDF difference of 0.0138 from that uniform law.
This last comparison is a front-weighted descriptive statistic; no iid
\(p\)-value is assigned because fronts within one road realization are
dependent.

The integrated kinetic collision gain decreases as the population coarsens.
To obtain a like-for-like Godunov statistic, consecutive 0.5 s catalogs are
matched realization by realization with an order-preserving rule. A merger is
identified only when two adjacent parents and one child satisfy position,
support, outer-state, and amplitude checks; boundary entries and exits are
censored, and ambiguous interior changes remain unresolved. The direct versus
identified Godunov rates are 0.01226 versus 0.01214 mergers/(km s) over
0--30 s, 0.00778 versus 0.00557 over 30--60 s, and 0.00541 versus 0.00425 over
60--90 s. Across all paths, the rule identifies 2108 mergers, while 656 interior
losses and two gains remain unresolved. These are detection-dependent counts:
false matches and missed events are both possible, so the identified rate is
not a guaranteed lower bound. The direct curve counts all modeled binary
collisions; their difference combines model, discretization, and detection
effects and is not an isolated numerical error estimate.

\begin{figure}[!htbp]
\centering
\includegraphics[width=\textwidth]{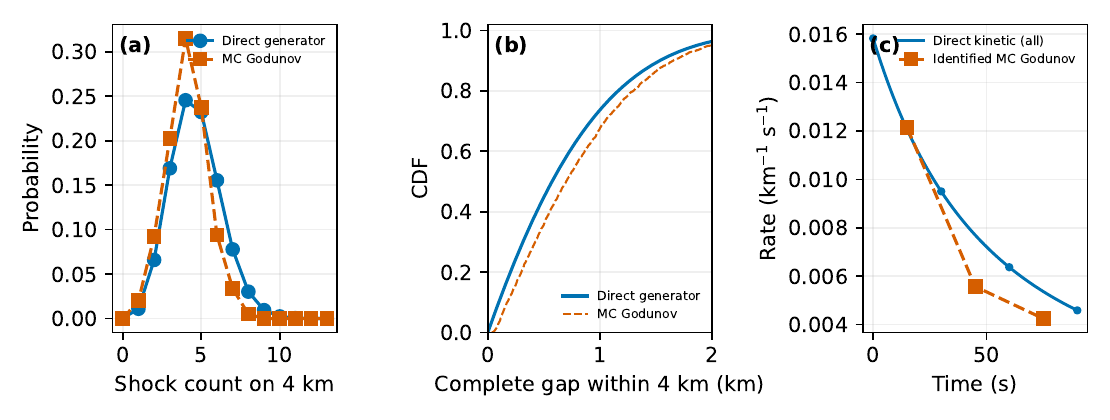}
\caption{Distributional comparison at 90 s and merger-rate
summary: shock count, complete within-window gaps, and kinetic all-collision versus
rule-identified two-parent/one-child Godunov rates. Unresolved and
boundary-censored catalog changes are not relabeled as mergers. Quantities are
proposed-versus-Godunov discrepancies, not road-grid errors relative to an
exact solution.}
\label{fig:rq3-count-spacing-mergers}
\end{figure}

Figure \ref{fig:rq3-count-spacing-mergers} complements the one-shock
statistics with the number of shocks on a road segment, distances between
neighboring shocks, and the decline in merger activity. The count discrepancy
is larger than the one-point density discrepancy, showing that agreement in
marginal density does not ensure agreement in the spatial arrangement of
congestion boundaries.

The spacing comparison also matches the observation rule. Only gaps whose
two endpoints fall within the 4 km window enter the empirical catalog.
The direct Palm spacing law is therefore weighted by \(L-s\), as derived in
Appendix \ref{app:front-law-diagnostics}. Comparing that catalog with an
unconditional next-front distance would favor long gaps that the window
cannot observe.

\FloatBarrier

\FloatBarrier

\FloatBarrier

Figure \ref{fig:rq3-godunov-density} separates solve time from stored state:
refining the Godunov road grid or increasing the ensemble increases both,
whereas the direct law stores density-state quantities rather than one road
field per realization. These measured configurations illustrate representation
cost, not a fitted asymptotic law. The road-grid sensitivity is material. At 90 s, detected front intensity is
0.860, 0.951, and 1.025 fronts/km at 40, 20, and 10 m, while working state
memory is 2.44, 4.88, and 9.77 MiB. At 10 m, increasing the ensemble from 100
to 800 reduces the front-intensity standard error from 0.0325 to 0.0117
fronts/km while increasing ensemble work eightfold. The
direct calculation changes from 1.100 to 1.157 fronts/km between 31 and 71
density states and uses approximately 0.023 to 0.118 MiB for the stored state;
with 71 states, halving the time step from 0.25 to 0.125 s changes final front
intensity by \(7.16\times10^{-8}\) fronts/km. These are representation and
precision sensitivities, not evidence of universal computational dominance.

Two direct-solver defects remain visible despite roundoff-level matrix
identities. Before normalization, the reconstructed marginal has mass
\(1.0540,\ 1.0413,\ 1.0337\) at 90 s for \(N=31,51,71\), respectively;
an additional internal \(N=141\) check gives \(1.0212\).
Table \ref{tab:state-audit} in Appendix \ref{app:front-law-diagnostics}
shows that refinement reduces but does not eliminate this mass defect.
The same table reports numerical small-front losses separately from mergers.
Figure \ref{fig:numerical-audit} contrasts the mass defect with
detector-threshold sensitivity: increasing the threshold reduces detected
shock counts and raises their mean size. These diagnostics
are reasons to retain numerical qualifications even when matrix identities
hold to roundoff. Transport of under-resolved small fronts into the state diagonal also removes
counts independently of binary collisions. Appendix
\ref{app:front-law-diagnostics} reports this separate loss budget and detector
sensitivity. These findings limit a continuum-accuracy claim: a conservative,
stationary matrix at a selected time does not establish convergence of the coupled
kinetic solver.

The experiment answers RQ3 within a controlled first-order setting. Direct
law evolution reproduces the scale and temporal coarsening of the matched
Godunov ensemble while exposing state-pair intensities and a collision term
without repeated road-grid solves. It does not show endogenous formation of
empirical stop-and-go instability: the admissible random disturbances are
imposed initially, and the closure remains restricted to a homogeneous road.

\subsection{Travel-time reliability}
\label{subsec:rq4-reliability}

\noindent Comparison used in this experiment: proposed method versus
instantaneous travel times reconstructed from the matched Monte Carlo
Godunov density profiles.
The traffic question is whether an evolved spatial generator can provide a
complete instantaneous travel-time distribution, rather than only a mean
density, without simulating another set of road-grid fields. As defined in
Section \ref{subsec:mapping-back}, instantaneous travel time uses the entire
road's speed profile at one specified time. Here that time is \(t^*=90\) s:
\begin{equation}
\label{eq:rq4-frozen-functional}
 T_L=\int_0^L\frac{1}{v(\rho(x,t^*))}\,\dd x,
 \qquad v(\rho)=\frac{q(\rho)}{\rho},\qquad L=4\ \mathrm{km}.
\end{equation}
It is not the first-passage time of a vehicle through a density field that
continues to evolve during the trip.

The stochastic traffic input, Greenshields diagram, \(\A_0\), \(\B\), and
90 s generator evolution are exactly those of Experiment 3. For the proposed
calculation, \(A_N(t^*)\) and its stationary law \(\pi\) are evaluated after the
Lax/kinetic evolution. Step 7 then treats \(1/v(\rho_i)\) as a spatial reward
and solves the conservative general-generator reward balance
\eqref{eq:reward-transport}; the Feynman--Kac transform
\eqref{eq:feynman-kac-travel} independently checks its moments and signs. The
comparator uses the matching 800 Godunov density profiles at the same time and evaluates the
same functional as \(\sum_i\Delta x/v(\rho_i(t^*))\). No deterministic or
same-process analytical travel-time surrogate is introduced.

All retained density states have positive speed. Using the same \(N=71\)
density-state reference selected in Section \ref{subsec:rq3-clustering}, the spatial-generator
matrix has nonnegative off-diagonal entries, maximum row-sum residual
\(1.42\times10^{-14}\), stationarity residual
\(2.01\times10^{-16}\), and unit reward-distribution mass to stored precision.
The proposed mean, standard deviation, and 95th percentile are 9.271, 2.253,
and 13.639 min. The 10 m Godunov ensemble gives 9.059, 2.034, and 13.007 min.
Thus, the proposed-minus-Godunov differences are 12.72, 13.14, and 37.89 s.
The proposed and Godunov on-time probabilities at a 10 min threshold are
0.697 and 0.730; their CDF maximum difference is 0.0501, and their Wasserstein
distance is 13.67 s.

\begin{figure}[!htbp]
\centering
\includegraphics[width=\textwidth]{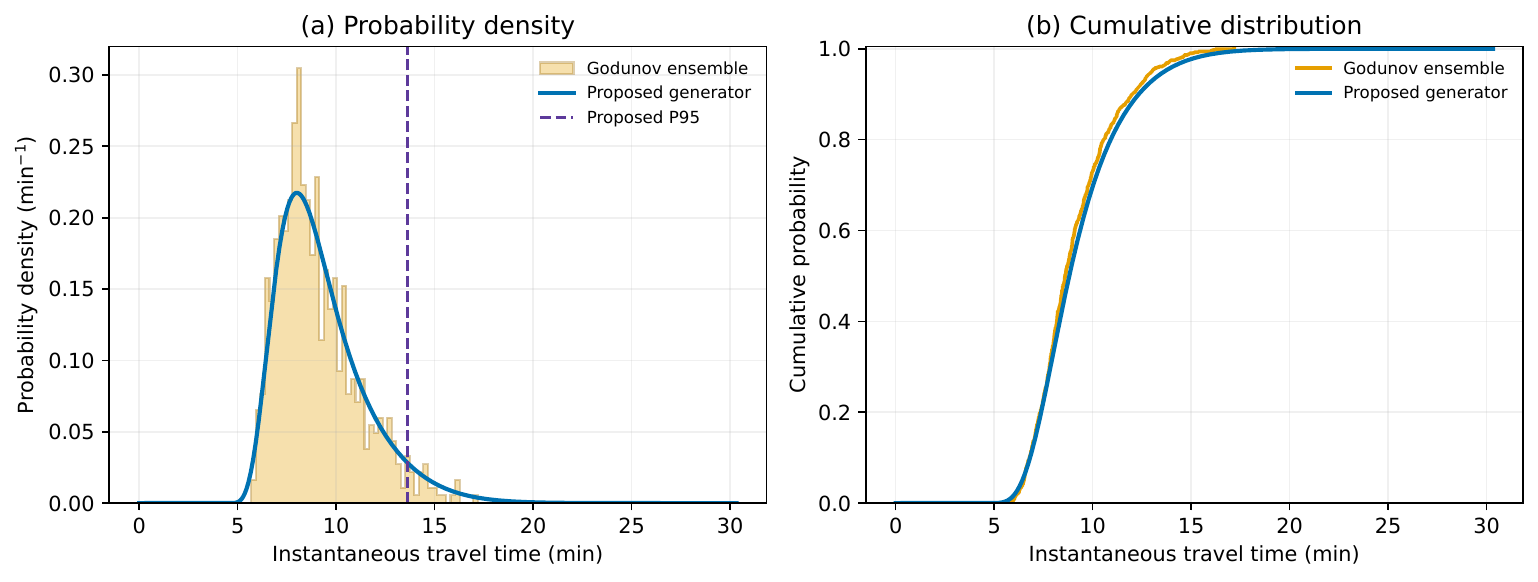}
\caption{Experiment 4 instantaneous travel-time density and CDF at 90 s. The
only comparator is the matched 10 m Monte Carlo Godunov ensemble at the same time;
there is no analytical travel-time reference.}
\label{fig:rq4-distribution}
\end{figure}

\begin{figure}[!htbp]
\centering
\includegraphics[width=\textwidth]{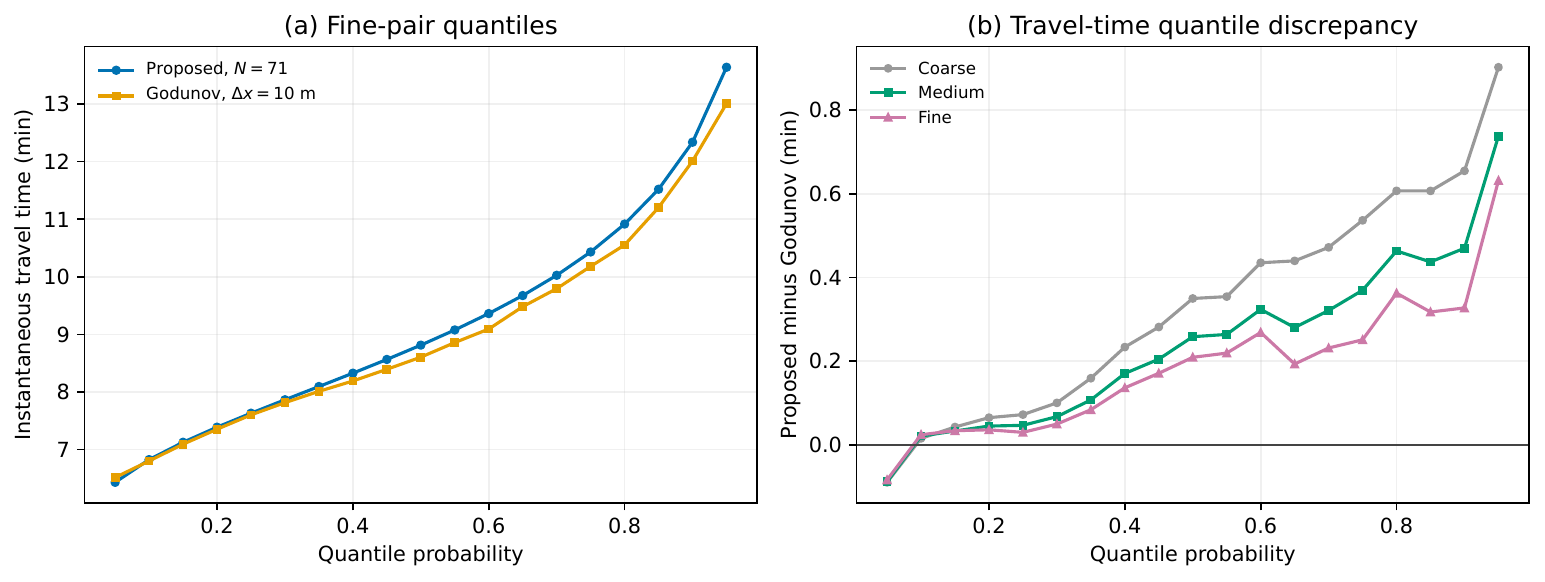}
\caption{Fine-pair travel-time quantiles and proposed-minus-Godunov quantile
differences over the tested state/road-resolution pairs.}
\label{fig:rq4-quantiles}
\end{figure}

Table \ref{tab:rq4-frozen-reliability} translates the distribution into
operational reliability measures: the proposed law has a longer upper tail,
a larger buffer allowance, and a lower probability of meeting the 10 min
threshold than the matched Godunov ensemble. Figure \ref{fig:rq4-distribution}
shows this shift in the density and CDF, and Figure \ref{fig:rq4-quantiles}
shows that the discrepancy is not a uniform time offset: the upper quantiles
differ more than the mean.

\begin{table}[H]
\caption{Instantaneous travel-time reliability at the fine resolution pair. The metrics
include coefficient of variation (CV), buffer time, buffer index (BI),
planning-time index (PTI), and the 10 min on-time probability.}
\label{tab:rq4-frozen-reliability}
\centering\small
\resizebox{\textwidth}{!}{
\begin{tabular}{lrrrrrrrrrr}
\toprule
Method & Mean & SD & CV & Median & P90 & P95 & Buffer & BI & PTI & $P(T\leq 10\,\mathrm{min})$ \\
 & \multicolumn{2}{c}{min} & & \multicolumn{4}{c}{min} & & & \\
\midrule
Proposed, $N=71$ & 9.271 & 2.253 & 0.243 & 8.813 & 12.338 & 13.639 & 4.368 & 0.471 & 3.410 & 0.697 \\
Godunov, $\Delta x=10$ m & 9.059 & 2.034 & 0.224 & 8.604 & 12.011 & 13.007 & 3.949 & 0.436 & 3.252 & 0.730 \\
\bottomrule
\end{tabular}
}
\end{table}

\FloatBarrier

\FloatBarrier

\begin{figure}[!htbp]
\centering
\includegraphics[width=\textwidth]{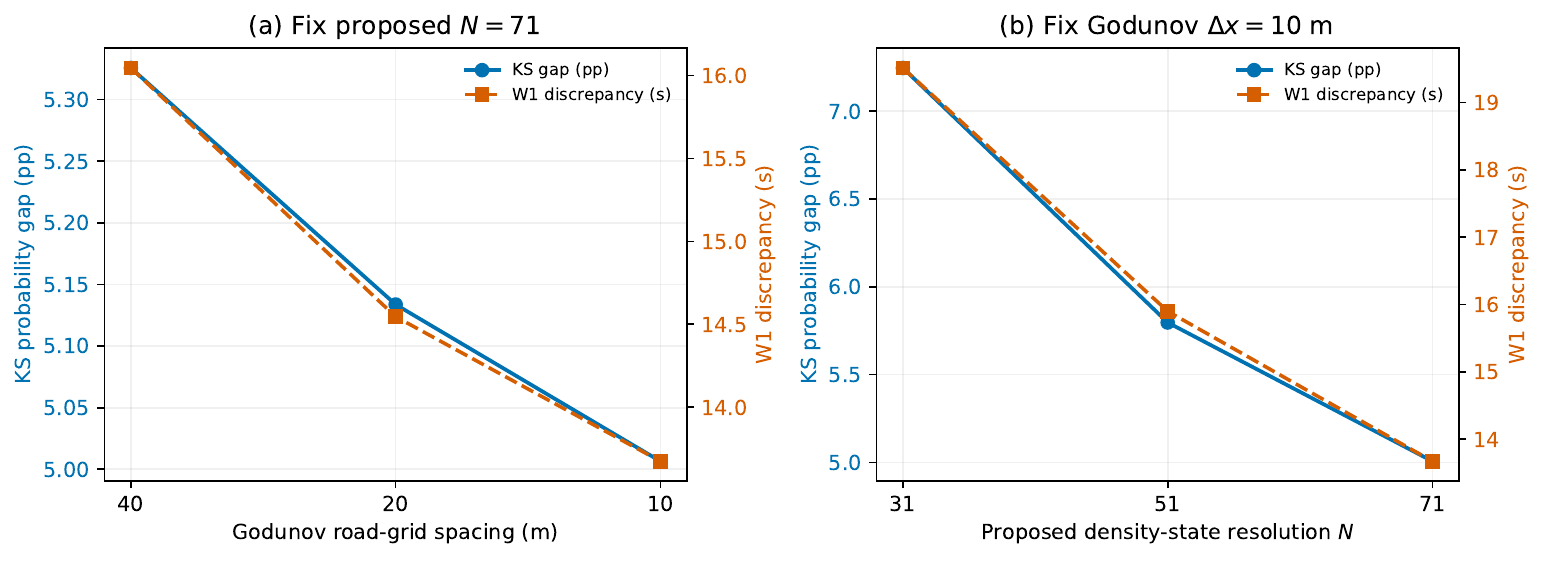}
\caption{One-at-a-time Experiment 4 sensitivity: Godunov road-cell size with
the proposed state resolution fixed at \(N=71\), and generator state
resolution with the Godunov grid fixed at 10 m.}
\label{fig:rq4-one-at-a-time}
\end{figure}

Figure \ref{fig:rq4-one-at-a-time} separates the two numerical controls.
Refining either representation reduces, but does not remove, the distributional
discrepancy over the tested range. Holding
the proposed calculation at \(N=71\) while refining the Godunov grid from 40
to 10 m changes the mean discrepancy from 15.17 to 12.72 s and the Wasserstein
distance from 16.05 to 13.67 s; the CDF gap is 0.0533, 0.0513, and 0.0501 over
40, 20, and 10 m. Holding Godunov at 10 m while increasing the generator state
resolution from 31 to 71 changes the mean discrepancy from 18.41 to 12.72 s,
the CDF gap from 0.0725 to 0.0501, and the Wasserstein distance from 19.51 to
13.67 s. These are finite-resolution sensitivities, not fitted convergence
orders. At the 10 m grid, using 100 rather than 800 stored fields changes the
empirical mean from 9.082 to 9.059 min and the P95 from 12.627 to 13.007 min,
illustrating the remaining tail-sampling variability.

\FloatBarrier

\begin{figure}[!htbp]
\centering
\includegraphics[width=\textwidth]{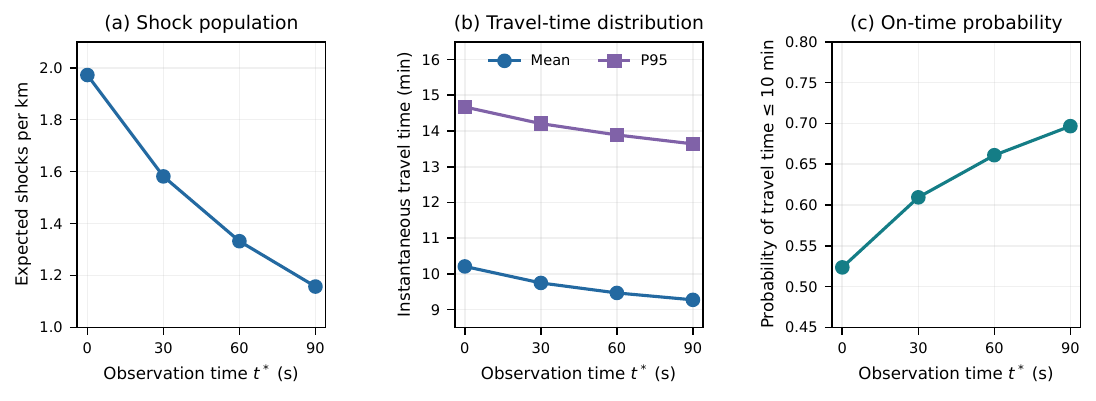}
\caption{Chronological evolution of the proposed outputs at four observation
times: (a) shock intensity, (b) mean and 95th-percentile instantaneous travel
time, and (c) probability of an instantaneous travel time at most 10 min.
Lines guide the eye through the same model evolution; they do not represent
a causal response to changing shock intensity.}
\label{fig:rq4-intensity-sensitivity}
\end{figure}

Figure \ref{fig:rq4-intensity-sensitivity} follows observation time from
0 to 90 s, connecting shock merging in panel (a) to instantaneous travel-time
mean and P95 in (b) and the 10 min on-time probability in (c). Each point
summarizes road conditions at that observation time, not a vehicle progressing
through its journey. As front intensity falls from 1.972 to
1.157 fronts/km between 0 and 90 s, the mean instantaneous travel time falls from 10.206 to
9.271 min, P95 from 14.673 to 13.639 min, and on-time probability rises from
0.524 to 0.697. This is an association along the prescribed model evolution,
not a causal demand experiment, and matching Godunov density profiles were retained
only at 90 s.

\FloatBarrier

The timing archive separates the upstream traffic solve from the instantaneous
travel-time calculation and its evaluation from Godunov density profiles. Reward-grid sensitivity
is assessed independently: the 401-node CDF differs from the 1601-node CDF
by 0.00776. These are finite-grid changes, not errors against a known travel-time
law. Timings describe the archived implementation and hardware, not a
universal performance ranking.

Table \ref{tab:numerical-comparison-summary} collects the main stochastic
outputs in traffic units. Relative to Godunov, the proposed law has more
shocks but smaller mean jumps, while its longer travel-time tail reduces
on-time probability. The two sets of differences concern distinct outputs
of the same evolved traffic law, not independent calibrations.

\begin{table}[H]
\caption{Proposed method versus the Godunov comparator in the stochastic
experiments. No entry is an exact-relative road-grid error. Differences are
proposed minus Godunov.}
\label{tab:numerical-comparison-summary}
\centering\small
\begin{tabular}{>{\raggedright\arraybackslash}p{0.26\linewidth}rrr
>{\raggedright\arraybackslash}p{0.12\linewidth}r}
\toprule
Quantity & Proposed & Godunov & Difference & Road grid & Ensemble \\
\midrule
Shocks/km at 90 s & 1.157 & 1.025 & 0.132 & 10 m & 800 \\
Mean shock size (veh/km/lane) & 27.405 & 28.063 & -0.658 & 10 m & 800 \\
Severe-shock probability & 0.155 & 0.167 & -0.012 & 10 m & 800 \\
Mean travel time (min) & 9.271 & 9.059 & 0.212 & 10 m & 800 \\
P95 travel time (min) & 13.639 & 13.007 & 0.631 & 10 m & 800 \\
On-time probability & 0.697 & 0.730 & -0.033 & 10 m & 800 \\
\bottomrule
\end{tabular}
\end{table}

Experiment 4 answers RQ4 for the functional the current framework actually
supports. The generator gives the entire instantaneous travel-time law and
standard reliability indices without another road-grid ensemble, with the
remaining distributional discrepancy reported explicitly. Dynamic vehicle
first passage through \(\rho(x,t)\), route choice, and signal or ramp control
require a space--time extension and are not inferred from this fixed-time comparison.

\subsection{Computational implications}
\label{subsec:computational-implications}

The experiments distinguish continuity in physical coordinates from an
absence of numerical approximation. The proposed method represents wave
positions and interactions without a finite-volume road mesh and evolves a
density-state law directly. Its implementation nevertheless uses a
density-state grid, collision quadrature, a physical-time integrator, and a
reward grid for the instantaneous travel-time calculation. Godunov instead stores and
advances road-cell values for each realization. Which representation is less
costly depends on the requested output: exact wave formulas remain best for
the deterministic tests, whereas the generator representation becomes
relevant when front or reliability distributions are the output themselves.

This interpretation closes the Results section without claiming that one
solver replaces the others. The following discussion relates the observed
agreements and discrepancies to the assumptions, applications, and unresolved
extensions of the framework.

\section{Discussion}
\label{sec:discussion}

The paired experiments evaluate stochastic shockwave dynamics under a
fixed deterministic fundamental diagram. Classical LWR already propagates
and merges shocks; the additional object here is the law of a random
ensemble. Under the stated assumptions, the generator/front representation
provides marginal density laws, state-pair intensities, front counts,
jump-size and wave-speed distributions, and merger gains and losses without
requiring a road-grid evolution for every realization.

The comparisons separate deterministic consistency from distributional
agreement. The Riemann and signal examples are deterministic wave tests;
the stochastic front and instantaneous travel-time calculations are compared
with matched Godunov ensembles. The latter travel-time calculation evaluates
a frozen-field functional, not experienced travel time through evolving
traffic. Spillback control, stochastic signal control, and field-validated
reliability prediction remain prospective applications rather than outcomes
established by these experiments.

\subsection{Meaning of the comparisons}
\label{subsec:comparative-scope}

The analytical results do not claim a new way to solve one deterministic
shock. The Riemann problem (RQ1) verifies the sign and normalization of the front-measure
specialization, and the Greenshields signal test (RQ2A) verifies that boundary-wave injection, characteristic
motion, front interaction, cumulative count, and conservation reduce to an
independent Greenshields construction. The triangular-limit test (RQ2B) answers a different question: the
deterministic triangular entropy problem remains well posed even though the
present stochastic closure theorem does not cover the capacity kink.

The stochastic results address the paper's central use case. In the stochastic shock-clustering experiment (RQ3), the
unknown is the law of a front population rather than one density realization.
The direct calculation gives state-pair intensities, collision gain and loss,
and count, size, spacing, severity, and density distributions. The matched
Godunov ensemble produces comparable empirical quantities only after 800
road-grid evolutions and a declared front detector. Their remaining
differences reflect both representations: density-state approximation in the
proposed method, and road-grid, extraction, and sampling sensitivity in the
Godunov ensemble. With no independent closed-form stochastic law, neither is
labeled the exact answer.

The instantaneous travel-time reliability experiment (RQ4) similarly compares two evaluations of one instantaneous travel-time
quantity. The evolved generator returns an entire reward distribution; the
Godunov side reconstructs the same functional from stored density fields. The
fine-pair CDF gap of 0.0501 and 13.67 s Wasserstein discrepancy show useful but
not exact agreement. The generator predicts a somewhat heavier travel-time
tail: its P95 is 37.89 s higher and its 10 min on-time probability 3.34
percentage points lower. This difference is operationally interpretable and
is not hidden by reporting only the mean.

The pairwise roles clarify the contribution. Against exact wave methods, the
generator representation supplies a deterministic front-measure specialization but no practical advantage for one tractable shock; the exact solution
is the natural method. Against a Godunov ensemble, the distinction is between
direct evolution of a density-state/front law and empirical aggregation after
road-grid evolution of every realization. The proposed framework is not
unique in representing continuous wave paths, and it does not replace exact
wave methods or Godunov/CTM. Its distinctive feature is the connected chain
from a random spatial traffic process to a closed generator/front law and then
to distributional traffic outputs without evolving a complete Eulerian road
grid for every realization.

\subsection{Triangular diagrams, contact waves, and closure scope}
\label{subsec:discussion-flux}

For the triangular fundamental diagram,
\begin{equation}
\label{eq:triangular-contact-discussion}
 q_{\mathrm T}(\rho)=\min\{v_f\rho,\,w(\rhoj-\rho)\},
 \qquad v_f,w>0,
\end{equation}
the free-flow and congested characteristic speeds are \(v_f\) and \(-w\),
respectively. Within either affine branch, characteristic and secant speeds
coincide. A discontinuity between states on that branch is a contact: it
propagates without the spreading associated with a strictly curved
rarefaction fan. Two contacts on the same branch have zero relative speed.
In particular, release from jam density to the capacity state is a contact
on the congested branch. Across the capacity kink, the secant between
distinct states remains defined, but the derivative at the critical state
is not unique.

This nonsmoothness is not a defect in deterministic LWR.
\citet{herbin2011entropy} distinguish nonuniqueness of weak solutions from
uniqueness of the entropy-selected solution for nondifferentiable
fundamental diagrams under the relevant assumptions. Newell and
Hamilton--Jacobi constructions can use the triangular geometry directly
\citep{newell1993,laval-leclercq2013}. The restriction concerns the present
\emph{smooth stochastic closure}: its proof and coefficient equations use
a smooth transformed convex flux, a unique derivative, and \(q''\).
Replacing \(q_{\mathrm T}\) by a rounded \(q_\varepsilon\) changes these
quantities and can alter contact transport, fan spreading, and collision
rates. It is an approximation to triangular dynamics, not an exact
nonsmooth closure.

A small flux discrepancy need not imply a small wave-position discrepancy
over an arbitrarily long propagation time. Consider the same fixed upward
state pair \(\rho_L<\rho_R\) supporting an isolated entropy front for both
fluxes before any interaction, and write
\(e_\varepsilon=q_\varepsilon-q_{\mathrm T}\). The Rankine--Hugoniot formula
gives
\begin{equation}
\label{eq:front-speed-approximation}
 s_\varepsilon-s_{\mathrm T}
 =\frac{e_\varepsilon(\rho_R)-e_\varepsilon(\rho_L)}
 {\rho_R-\rho_L}.
\end{equation}
With identical initial front locations and unchanged adjacent states,
\begin{equation}
\label{eq:front-position-approximation}
 \bigl|x_\varepsilon(t)-x_{\mathrm T}(t)\bigr|
 =t\bigl|s_\varepsilon-s_{\mathrm T}\bigr|
 \leq\frac{2t\|e_\varepsilon\|_\infty}{\rho_R-\rho_L},
\end{equation}
where the supremum is taken over the density domain. Thus a persistent
speed difference produces a positional offset that grows with propagation
time. If \(s_{\mathrm T}\neq0\), propagation over distance \(D\) requires
reference time \(D/|s_{\mathrm T}|\), showing the corresponding dependence
on distance. This calculation applies only to the fixed-state,
pre-interaction comparison. It is not a uniform error theorem for a
many-front stochastic field, and its bound is not uniform as the jump
amplitude tends to zero. For rarefactive state ordering, smoothing may
replace a contact by a fan, so a single-front comparison is insufficient.

In an interacting field, differences in front positions can change merger
times, boundary encounters, queue paths, and downstream observations.
Approximation differences may therefore become more consequential over
longer horizons or larger spatial domains, depending on the states visited,
especially near capacity. This does not imply that every error metric grows
monotonically. Nor does it establish that a long-horizon failure has been
measured in the present experiments. The small maximum-queue discrepancy in
Experiment~2B must also be distinguished from error along the entire queue
trajectory, as the reported queue-path, clearance-time, and cumulative-count
comparisons already illustrate.

The finite-horizon results support the examined deterministic approximation;
they do not establish convergence of \(\A_t\), \(F_t\), or their common
operator core as \(\varepsilon\to0\), or a uniform long-time or long-distance
accuracy guarantee. A direct triangular stochastic closure would require a
justified treatment of contact transport and the critical state, together
with a corresponding closure and interaction balance beyond the smooth
proof used here.

Spatial forcing is a second boundary of the theory. The zero-curvature form
allows an \(x\)-dependent term formally, but the homogeneous closure used for
the kinetic solver does not include signals, ramps, sources, junction supply
and demand, or control. The signal calculation is therefore a deterministic
finite-wave consistency test. A ramp-metering study in which spillback
probability is a decision criterion remains important, but it requires a
controlled junction/source extension before it can test the proposed law
rather than only a conventional traffic simulator.

\subsection{Traffic interpretation and limitations}
\label{subsec:discussion-limits}

The stochastic experiment should not be read as a first-order explanation of
spontaneous stop-and-go instability. Its stationary piecewise-deterministic
input deliberately contains decreasing recovery segments and upward fronts;
LWR transports and merges those disturbances. Reaction time, lane changes,
capacity drop, hysteresis, and instability formation remain outside the
model. The observed decline in front intensity is therefore a clustering
result conditional on imposed admissible disturbances.

The travel-time interpretation requires a distinction between instantaneous
and experienced travel time \citep{fhwa2012}. The former integrates inverse
speed along the road using conditions at \(t^*\); the latter follows a
vehicle through conditions that change during its journey. This paper
calculates the former. The latter would require a joint space--time
description along the vehicle trajectory, beyond the spatial transition law
at one specified time used here.
The within-model association between lower evolved front intensity and better
reliability is informative for the synthetic process, but it is not a causal
control result.

All stochastic inputs are synthetic. They were selected because they obey the
one-sided Markov assumptions without density clipping and expose drift,
transport, collision, and reward terms. The numerical comparisons show
agreement in scale, explicit distributional gaps, and sensitivity to both
density-state and road-grid resolution. They do not provide empirical
validation. A transportation application must estimate \(a\) and \(\lambda\)
jointly from detector or trajectory data, test spatial Markov adequacy, and
validate front counts and reliability out of sample.

Finally, the computational comparison is implementation- and output-specific.
The direct method avoids an Eulerian road mesh but still approximates density
state, collision integrals, physical time, and reward. The reported timings
favor direct law evaluation for these small stationary distributional tasks;
they do not establish asymptotic or universal superiority. The former
three-state pure-shock matrix is retained only in Appendix
\ref{app:spectral-integrability} as a sign, rate, and invariant regression
check. It is neither the continuum solver that evolves the population of congestion
shocks in the clustering experiment (RQ3), nor the evolved spatial generator
used to derive the instantaneous travel-time distribution in the reliability
experiment (RQ4).

These limits define a focused next agenda: nonsmooth and forced generator
closure, dynamic vehicle first passage, statistically estimated initial laws,
and field validation. The conclusion therefore states the contribution in
terms of a restricted but directly interpretable stochastic traffic law.

\section{Conclusion}
\label{sec:conclusion}

This paper develops a generator-based description of stochastic shockwave
dynamics in LWR with random initial density and a fixed deterministic
fundamental diagram. Every realization follows the same conservation law;
the evolved stochastic object is the distribution of density states and
congestion fronts. The traffic-variable adaptation of spatial Markov
closure yields a front balance and mappings to density laws, front counts,
sizes, speeds, merger statistics, and instantaneous travel-time
distributions. Numerical evaluation still requires density-state,
physical-time, collision, and reward approximations.

The Riemann and signal examples recover classical deterministic wave
motion, queue formation, discharge, and clearance. Smooth approximations
approach the triangular signal solution over the tested finite horizons,
but do not establish stochastic closure at contacts or the capacity kink,
or uniform accuracy over longer times and distances. The stochastic
experiments compare direct law evaluation with matched Godunov ensembles,
with remaining differences sensitive to resolution, extraction, and
sampling. The travel-time output describes a frozen traffic field at one
observation time, not an experienced trip through evolving conditions.

The contribution is direct computation and interpretation of traffic
distributions without a separate road-grid evolution for every realization,
not a new pathwise conservation law or universal superiority over existing
solvers. Extending closure to nonsmooth diagrams and controlled boundaries,
coupling the model to dynamic vehicle travel times, and estimating and
validating its initial spatial law from traffic observations are the next
steps toward operational use.

\clearpage
\appendix

The appendices collect proofs, operator calculations, numerical definitions,
and one deliberately limited finite-state algebraic check. They support the
traffic-facing argument in the main text; the finite three-state construction
is not counted as a fifth experiment and is not the solver used in the
stochastic comparisons.

\section{Hopf geometry}
\label{app:hopf-orientation}

This appendix establishes the deterministic geometry used by the stochastic
closure argument: concave Hopf duality, monotonicity of the rightmost
maximizer, and the resulting shock orientation. The representation is the
concave traffic-flow form of the classical Hopf--Lax construction and its
variational traffic interpretation
\citep{hopf1950,newell1993,daganzo2006,dafermos2016}. The monotone-selector
and Markov-splitting arguments are adapted to concave traffic variables from
the one-sided shock-clustering framework \citep{menon-srinivasan2010}. Let
\(U_x=\rho\).
With the downstream coordinate used in the main text, \(-U\) is the
cumulative vehicle-count (Moskowitz) function up to an additive constant, and
\(U\) solves
\begin{equation}
\label{eq:traffic-hj}
  U_t+q(U_x)=0.
\end{equation}

For \(v\in\mathbb R\), define the concave dual and its density selector by
\begin{equation}
\label{eq:concave-legendre}
  q_*(v)=\inf_{r\in[0,\rhoj]}\{vr-q(r)\},
  \qquad
  r_*(v)=\arg\min_{r\in[0,\rhoj]}\{vr-q(r)\}.
\end{equation}
Strict concavity makes \(r_*\) single valued. On the interior characteristic
range, \(q'(r_*(v))=v\); outside that range the selector takes the appropriate
boundary value. Under Assumptions \ref{ass:fd}--\ref{ass:initial-field}, the
concave Hopf--Lax formula \citep{hopf1950,lax1957,newell1993,daganzo2006} is
\begin{equation}
\label{eq:concave-hopf}
  U(x,t)=\sup_{s\in\mathbb R}
  \left\{U_0(s)+tq_*\!\left(\frac{x-s}{t}\right)\right\}.
\end{equation}
The maximizer is the backtracked road origin of the wave observed at
\((x,t)\); the remainder of this appendix supplies the conditions and
orientation proof used in Section \ref{sec:problem-description}.

Differentiating
\(U_t+q(U_x)=0\) at a smooth point yields
\[
  \rho_t+q'(\rho)\rho_x=0,
\]
which is the classical form of \eqref{eq:lwr}; the viscosity solution for \(U\)
has the entropy solution as its spatial derivative almost everywhere.

\subsection{Technical profile conditions}

The concise road-profile condition in Assumption
\ref{ass:initial-field} is used with the following right-process details. The
initial field has a canonical cadlag Borel-right strong-Markov realization.
After rebasing at any deterministic road origin \(x_0\), the restriction
\((\rho_0(x_0+s))_{s\geq0}\) is Borel right with respect to its completed,
right-continuous natural filtration. Almost surely, its potential satisfies
\begin{equation}
\label{eq:hopf-coercivity}
  U_0(s)+tq_*\!\left(\frac{x-s}{t}\right)\longrightarrow-\infty
  \quad\text{as }|s|\to\infty
\end{equation}
for every \(x\in\mathbb R\) and \(t>0\). This coercivity makes the Hopf
maximizer finite.

The direct splitting construction also uses the following local condition.
It is not needed for the global closure argument based on convex conjugacy.

\begin{assumption}[Interior selector for the splitting proof]
\label{ass:interior-selector}
Fix an active road--time region \(\mathcal R\). For every compact
\(K\Subset\mathcal R\), there is a compact interval
\(I_K\Subset(0,\rhoj)\) such that, almost surely,
\begin{equation}
\label{eq:interior-selector-hypothesis}
 r_*\!\left(\frac{x-a_{\rm H}(x,t)}{t}\right)\in I_K
 \qquad\text{for every }(x,t)\in K.
\end{equation}
The clock, integral functional, and running-maximum augmentation \(Z_x\)
defined in \eqref{eq:augmented-Z} is additionally assumed to have a cadlag
Borel-right strong-Markov realization. These conditions permit the splitting
argument of \citet{getoor1979}; they are not consequences of the traffic PDE.
\end{assumption}

\subsection{Concave duality}

The first step records the concave Fenchel relation and the characteristic
condition generated by the Hopf variational formula. These are standard
duality consequences of the Hopf--Lax framework
\citep{hopf1950,dafermos2016}, written here in density variables to make the
traffic sign convention explicit.

For \(q_*\) in \eqref{eq:concave-legendre},
\begin{equation}
\label{eq:concave-fenchel}
  q(r)=\inf_{v\in\mathbb R}\{vr-q_*(v)\},
  \qquad 0\leq r\leq\rhoj.
\end{equation}
Indeed, the definition of \(q_*\) gives
\(q_*(v)\leq vr-q(r)\), hence \(vr-q_*(v)\geq q(r)\). At an
interior state, choose \(v=q'(r)\). Strict concavity makes \(r\) the unique
minimizer in the definition of \(q_*(v)\), and equality follows. At an
endpoint the same statement follows with a supporting slope. Thus \(q_*\) is
concave and, wherever its minimizer is interior,
\begin{equation}
\label{eq:qstar-derivatives}
  q_*'(v)=r_*(v),
  \qquad q_*''(v)=\frac{1}{q''(r_*(v))}<0.
\end{equation}

The dynamic programming principle with the concave dual gives
\eqref{eq:concave-hopf}. Coercivity \eqref{eq:hopf-coercivity} makes the
maximizer set nonempty and compact, so its rightmost point
\begin{equation}
\label{eq:hopf-maximizer}
  a_{\rm H}(x,t)=\argmaxp_{s\in\mathbb R}
  \left\{U_0(s)+tq_*\!\left(\frac{x-s}{t}\right)\right\}
\end{equation}
is finite. Write
\begin{equation}
\label{eq:J-functional}
  J_{x,t}(s)=U_0(s)+tq_*\!\left(\frac{x-s}{t}\right).
\end{equation}
At a selected continuity point with an interior density,
\begin{equation}
\label{eq:J-first-order}
  0=\partial_sJ_{x,t}(s)
  =\rho_0(s)-r_*\!\left(\frac{x-s}{t}\right).
\end{equation}
Equations \eqref{eq:qstar-derivatives} and \eqref{eq:J-first-order} give
\begin{equation}
\label{eq:hopf-characteristic-relation}
  \rho(x,t)=\rho_0(a_{\rm H}(x,t)),
  \qquad
  q'(\rho(x,t))=\frac{x-a_{\rm H}(x,t)}{t}
\end{equation}
at a selected continuity point with interior density.

\subsection{Monotone maximizers}

This subsection proves monotonicity of the rightmost selector and exclusion
of upward-jump corners on the active interior region. The proof is included
to verify the traffic orientation and the precise endpoint scope.

\begin{proposition}[Monotone inverse map and shock orientation]
\label{prop:hopf-monotonicity}
For fixed \(t>0\), \(x\mapsto a_{\rm H}(x,t)\) is nondecreasing. A jump of
\(a_{\rm H}\) cannot produce a downward density jump; the density jump is
strictly upward when the two endpoint selector values differ. Moreover, an
upward jump of the initial density cannot itself maximize the Hopf functional
when the associated dual selector is in \((0,\rhoj)\). This final
corner-avoidance conclusion holds on the active region of Assumption
\ref{ass:interior-selector}.
\end{proposition}

Let \(K_t(x,s)=tq_*((x-s)/t)\). Where twice differentiable,
\begin{equation}
\label{eq:kernel-cross-partial}
  \partial_{xs}^2K_t(x,s)=-\frac{1}{t}
  q_*''\!\left(\frac{x-s}{t}\right)>0.
\end{equation}
Equivalently, for \(x_1<x_2\) and \(s_1<s_2\),
\begin{equation}
\label{eq:increasing-differences}
  K_t(x_2,s_2)+K_t(x_1,s_1)
  \geq K_t(x_2,s_1)+K_t(x_1,s_2).
\end{equation}
The inequality also holds across the endpoint regimes by concavity. Suppose
that rightmost maximizers \(a_1\) and \(a_2\) at \(x_1<x_2\) satisfied
\(a_1>a_2\). Apply \eqref{eq:increasing-differences} with the two maximizers
and add their optimality inequalities. Equality must hold throughout, which
would make \(a_1\) a maximizer at \(x_2\), contradicting the choice of \(a_2\)
as the rightmost maximizer. Therefore
\begin{equation}
\label{eq:aH-monotone}
  a_{\rm H}(x_1,t)\leq a_{\rm H}(x_2,t).
\end{equation}

Now suppose that \(\rho_0\) jumps upward at \(s_0\) and that the associated
dual selector is interior, so the dual term is differentiable in \(s\) at
\(s_0\). Then \(U_0\) has a convex corner there:
\[
  U'_{0,-}(s_0)<U'_{0,+}(s_0).
\]
The dual term in \eqref{eq:J-functional} is continuous in slope at \(s_0\), so
\(J'_{x,t,-}(s_0)<J'_{x,t,+}(s_0)\). A local maximum at a corner would require
\(J'_{x,t,-}(s_0)\geq0\geq J'_{x,t,+}(s_0)\), which is impossible. Thus, on
the active interior region, a rightmost Hopf maximizer is not an upward-jump
point of \(\rho_0\). No endpoint corner-avoidance claim is needed for the
global conjugacy proof.

Finally, if \(a_{\rm H}\) jumps upward as \(x\) increases, then
\((x-a_{\rm H})/t\) jumps downward. The global selector \(r_*(v)\) is
nonincreasing in \(v\), including the boundary-saturated regimes; hence the
selected density has a nonnegative jump. It is strictly positive if the two
selector values differ (in particular, when both are interior). Boundary
saturation can instead give equal density values. This proves Proposition
\ref{prop:hopf-monotonicity}.

As a consistency transformation, set
\begin{equation}
\label{eq:convex-conjugacy}
  u=-\rho,
  \qquad f(u)=-q(-u).
\end{equation}
Then \(f''(u)=-q''(-u)>0\), and upward traffic-density shocks become downward
shocks of the convex-flow variable \(u\). The direct proof above is retained
because it fixes all traffic signs without relying on a notational analogy.
For application of the closure theorem, extend \(f\) to a \(C^1\) strictly
convex flow on \(\mathbb R\). The scalar maximum principle confines the
solution to \([-\rhoj,0]\), so this extension is immaterial on the physical
density range.

\section{Markov closure}
\label{app:markov-closure}

This appendix gives the traffic-variable splitting argument. The one-sided
dependence and conditional splitting follow the framework cited above,
whereas the explicit clock coordinate, compact hit set, and localization from
a two-sided road to nonnegative coordinates are technical refinements
introduced here. Fix \(t>0\) and a compact road interval
\(X\) with \(X\times\{t\}\Subset\mathcal R\). Let \(I_0\Subset(0,\rhoj)\)
be the compact selector interval supplied by Assumption
\ref{ass:interior-selector}, fix \(x\in X\), suppress \(t\) from
the notation, and define
\begin{equation}
\label{eq:d-selector}
  d_x(s)=r_*\!\left(\frac{x-s}{t}\right).
\end{equation}
On the active interior range, \(d_x\) is strictly increasing and
\(s=x-tq'(d_x(s))\). From \eqref{eq:J-first-order}, in the
Lebesgue--Stieltjes sense,
\begin{equation}
\label{eq:J-integral-increment}
  J_x(s_2)-J_x(s_1)
  =\int_{s_1}^{s_2}\bigl(\rho_0(r)-d_x(r)\bigr)\dd r.
\end{equation}

\subsection{One-sided dependence}

We first isolate the deterministic fact behind Markov closure: increments of
the selected maximizer on either side of the observation point use only the
corresponding one-sided initial path. Probabilistic splitting is deferred to
the last subsection.

Let \(a=a_{\rm H}(x,t)\). Monotonicity implies
\(a_{\rm H}(x+h,t)\geq a\) for \(h>0\). Therefore
\begin{equation}
\label{eq:right-increment-functional}
  a_{\rm H}(x+h,t)-a
  =\argmaxp_{r\geq0}
  \{J_{x+h}(a+r)-J_{x+h}(a)\}.
\end{equation}
By \eqref{eq:J-integral-increment}, the right side depends on the initial
field only through \((\rho_0(a+r))_{r\geq0}\). The analogous maximization to
the left shows that
\((a_{\rm H}(x-h,t)-a)_{h>0}\) depends only on
\((\rho_0(a-r))_{r>0}\). This is a deterministic dependence statement; the
splitting theorem below turns it into conditional independence.

\subsection{Augmented Markov process}

To place the path functional in the Borel-right strong-Markov setting needed
for last-passage splitting \citep{getoor1979}, we augment the density by the
deterministic road coordinate, the dual selector, and the drawdown. The
explicit clock coordinate and the subsequent two-sided localization are
technical devices introduced in this paper.

Define the running maximum and drawdown
\begin{equation}
\label{eq:D-running-max}
  M_x(s)=\sup_{r\leq s}J_x(r),
  \qquad D_x(s)=M_x(s)-J_x(s)\geq0,
\end{equation}
and augment the initial process with the road coordinate as
\begin{equation}
\label{eq:augmented-Z}
  Z_x(s)=\bigl(s,\rho_0(s),d_x(s),D_x(s)\bigr).
\end{equation}
To see why this augmentation is Markov, let \(\xi\) be a stopping time for the
completed right-continuous natural filtration of \(\rho_0\), and put
\[
  K_\xi(r)=J_x(\xi+r)-J_x(\xi)
  =\int_0^r\bigl(\rho_0(\xi+v)-d_x(\xi+v)\bigr)\dd v.
\]
Then
\begin{equation}
\label{eq:D-update}
  D_x(\xi+r)
  =\max\left\{D_x(\xi),\sup_{0\leq v\leq r}K_\xi(v)\right\}
   -K_\xi(r).
\end{equation}
By the strong Markov property of \(\rho_0\), the conditional law of the future
\(\rho_0(\xi+\cdot)\) depends on the past only through \(\rho_0(\xi)\).
The coordinate \(\xi\) determines the future deterministic curve
\(d_x(\xi+\cdot)\), including any boundary-saturated portion. Equation
\eqref{eq:D-update} then shows that the conditional law of
\(Z_x(\xi+\cdot)\) depends only on \(Z_x(\xi)\), which establishes the Markov
update formula. The cadlag Borel-right strong-Markov realization needed for
Getoor's theorem is the explicit augmentation clause of Assumption
\ref{ass:interior-selector}; it is not inferred here merely from the
one-point Markov property of \(\rho_0\).
To apply the nonnegative-parameter right-process theorem to the two-sided
field, localize rather than choosing a single origin. For each \(n\geq1\), on
the event \(\{a_{\rm H}(x,t)>-n\}\), rebase the process at \(-n\) and apply
the splitting theorem to the nonnegative coordinate \(s+n\). Coercivity makes
\(a_{\rm H}(x,t)\) finite, so these events increase to probability one as
\(n\to\infty\). The localized conditional-independence identities therefore
extend to the full two-sided past by the usual bounded-test-function limit.

\subsection{Last-passage splitting}

We now identify the rightmost Hopf maximizer as the end of visits to a fixed
closed set and apply Getoor's Borel-right splitting theorem
\citep{getoor1979}. The compact formulation of the hit set is specific to the
interior-selector hypothesis adopted here.

Consider the fixed closed set
\begin{equation}
\label{eq:closure-hit-set}
  \mathcal H=\{(s,r,r,0):s\in\mathbb R,\ r\in I_0\}.
\end{equation}
The corner-avoidance result implies that the rightmost ultimate maximizer is a
continuity point and satisfies
\(\rho_0(a)=d_x(a)\). It also has \(D_x(a)=0\). Conversely, the last visit of
\(Z_x\) to \(\mathcal H\) is the last time at which \(J_x\) attains its ultimate
maximum. Consequently,
\begin{equation}
\label{eq:aH-last-hit}
  a_{\rm H}(x,t)=\sup\{s:Z_x(s)\in\mathcal H\}.
\end{equation}
The hit set is fixed and closed in the augmented state space because
\(I_0\) is compact. Its visit set is homogeneous under the canonical shift,
and its occurrence process is optional. Getoor's splitting theorem for the
end of visits of a cadlag Borel-right strong Markov process to this fixed set
therefore makes the pre-\(a\) and
post-\(a\) paths conditionally independent given \(Z_x(a)\)
\citep{getoor1979}.

At the hit,
\[
  Z_x(a)=\bigl(a,d_x(a),d_x(a),0\bigr).
\]
This state is determined by \(a\). Because the selected state lies in the
active interval, Equation \eqref{eq:hopf-characteristic-relation} gives the
equivalent relation
\begin{equation}
\label{eq:a-density-equivalence}
  a=x-tq'(\rho(x,t)).
\end{equation}
Thus the conditioning state is equivalently the present density
\(\rho(x,t)\), not an unobserved accumulated potential.

Combine this splitting with \eqref{eq:right-increment-functional}. Given
\(\rho(x,t)\), every right increment of \(a_{\rm H}\) is conditionally
independent of all left increments. Therefore \(x\mapsto a_{\rm H}(x,t)\) is
Markov, possibly with \(x\)-dependent transition laws. At a shock, the
rightmost maximizer selects the cadlag (right-continuous) entropy
representative; the identity is first obtained at continuity points and
extended to shock locations by right continuity. The resulting one-to-one
relation \eqref{eq:a-density-equivalence} on the active interval transfers the
Markov property to \(\rho(\cdot,t)\). Proposition
\ref{prop:hopf-monotonicity}
supplies the upward-jump orientation and completes the proof of Theorem
\ref{thm:traffic-closure} on the active region. The global theorem, including
boundary-selected states, follows independently from the convex-conjugacy
argument below, which invokes the one-sided Markov-closure result cited at the
start of this appendix.

The conjugacy \eqref{eq:convex-conjugacy} gives that shorter verification: it
places the problem directly under the corresponding spectrally negative
Markov-closure result. Neither route proves Feller preservation, which is why
Assumption \ref{ass:generator-regularity} is stated separately.

\section{Generator adjoints}
\label{app:generator-adjoints}

This appendix supplies the operator calculations used in the continuum
framework.  The spatial drift--jump generator, signed temporal operator, and
their forward equations are the traffic-variable counterparts of the
shock-clustering generator framework \citep{menon-srinivasan2010}; the compact
state space, boundary behavior, and integrability conditions recorded below
are the technical regime adopted in this paper.  We first compute the spatial
adjoint, then derive the temporal operator and its adjoint.

The abbreviated regularity statement in Assumption
\ref{ass:generator-regularity} is used with a common core containing
\(C_c^1((0,\rhoj))\), local continuity of the jump kernel in its lower state,
and the bounded-variation condition
\begin{equation}
\label{eq:lambda-integrability}
  \int_\rho^{\rhoj}
  \bigl(1\wedge|\eta-\rho|\bigr)
  \lambda(\rho,\dd\eta,t)<\infty.
\end{equation}
When a front count and a jump density are used, the stronger finite-activity
condition is
\begin{equation}
\label{eq:lambda-finite-activity}
  \Lambda(\rho,t):=\int_\rho^{\rhoj}
  \lambda(\rho,\dd\eta,t)<\infty,
\end{equation}
locally uniformly in \((\rho,t)\). Boundary coefficients keep the process in
\([0,\rhoj]\); in particular, \(a(0,t)=0\), and the upward kernel has no mass
outside the physical density interval.

For bounded-variation paths the jump integral is uncompensated. If
\(\varphi\in C_c^1((0,\rhoj))\), then
\[
  |\varphi(\eta)-\varphi(\rho)|
  \leq\|\varphi'\|_\infty|\eta-\rho|.
\]
The first-moment condition \eqref{eq:lambda-integrability}, rather than only a
second-moment condition, therefore guarantees local absolute convergence of
the small-jump integral in \eqref{eq:spatial-generator}. Finite activity
\eqref{eq:lambda-finite-activity} additionally makes the number of fronts per
unit road length finite. To remain in the physical state space, the negative
drift must satisfy \(a(0,t)=0\); negative drift at \(\rhoj\) points inward and
need not vanish. The upward kernel has no mass beyond \(\rhoj\) and vanishes
when no higher state is available.

\subsection{Spatial adjoint}

This subsection converts the spatial generator into the stationary forward
balance used in the main text.  The integration by parts, compact-state
limits, and notation below specialize the drift--jump form to the present
traffic model.

For a smooth compactly supported density \(p\), integration by parts gives
\[
  \int_0^{\rhoj}p(\rho)[-a(\rho)\varphi'(\rho)]\dd\rho
  =\int_0^{\rhoj}\varphi(\rho)\partial_\rho(ap)(\rho)\dd\rho.
\]
Fubini's theorem gives
\begin{align*}
 &\int_0^{\rhoj}p(\rho)
  \int_\rho^{\rhoj}
  (\varphi(\eta)-\varphi(\rho))\lambda(\rho,\eta)
  \dd\eta\dd\rho\\
 &\quad=\int_0^{\rhoj}\varphi(\rho)
  \left\{
    \int_0^\rho p(m)\lambda(m,\rho)\dd m
    -p(\rho)\int_\rho^{\rhoj}\lambda(\rho,\eta)\dd\eta
  \right\}\dd\rho.
\end{align*}
Hence
\begin{equation}
\label{eq:A-adjoint-full}
  (\A^\dagger p)(\rho)
  =\partial_\rho(ap)(\rho)
  +\int_0^\rho p(m)\lambda(m,\rho)\dd m
  -p(\rho)\int_\rho^{\rhoj}\lambda(\rho,\eta)\dd\eta.
\end{equation}
Substitution of \(F=p\lambda\) yields
\begin{equation}
\label{eq:stationary-adjoint-expanded}
  0=\A_t^\dagger p_t
  =\partial_\rho(ap)(\rho,t)
    +\int_0^\rho F(m,\rho,t)\dd m
    -\int_\rho^{\rhoj}F(\rho,\eta,t)\dd\eta.
\end{equation}

\subsection{Temporal operator and adjoint}

This subsection translates physical front motion into an operator acting in
time and then computes its forward adjoint.  The construction specializes the
signed temporal balance to a concave traffic flow; the sign conventions are
spelled out here because spatial jumps are upward in density.

On a smooth branch, \(\rho_x=-a(\rho,t)\), and the LWR equation gives
\begin{equation}
\label{eq:smooth-time-density}
  \rho_t=-q'(\rho)\rho_x=q'(\rho)a(\rho,t).
\end{equation}

\begin{definition}[Temporal compatibility operator]
\label{def:temporal-operator}
For \(\varphi\in\mathcal D\), define
\begin{equation}
\label{eq:temporal-generator}
  (\B_t\varphi)(\rho)
  =q'(\rho)a(\rho,t)\varphi'(\rho)
  -\int_\rho^{\rhoj}
  \RH{\rho}{\eta}
  \bigl(\varphi(\eta)-\varphi(\rho)\bigr)
  \lambda(\rho,\dd\eta,t).
\end{equation}
The operator \(\B_t\) is signed and need not be a Markov generator in physical
time.
\end{definition}

On a continuous branch, LWR gives the drift term
\(q'(\rho)a(\rho,t)\partial_\rho\). A front from \(\rho\) to \(\eta\) moves a
distance \(\RH{\rho}{\eta}\dd t\). Relative to a fixed observation point this
shifts the spatial discontinuity in the opposite direction, giving the signed
jump coefficient \(-\RH{\rho}{\eta}\lambda(\rho,\eta)\). This proves
\eqref{eq:temporal-generator}.

Taking its formal adjoint as above gives
\begin{align}
\label{eq:B-adjoint-full}
  (\B^\dagger p)(\rho)
  &=-\partial_\rho(q'ap)(\rho)
  -\int_0^\rho
    \RH{m}{\rho}p(m)\lambda(m,\rho)\dd m\notag\\
  &\quad+p(\rho)\int_\rho^{\rhoj}
    \RH{\rho}{\eta}\lambda(\rho,\eta)\dd\eta.
\end{align}
Equivalently, in terms of \(F\),
\begin{align}
\label{eq:temporal-adjoint-expanded}
  \partial_t p(\rho,t)=\B_t^\dagger p_t
  &=-\partial_\rho(q'ap)(\rho,t)
    -\int_0^\rho\RH{m}{\rho}F(m,\rho,t)\dd m \notag\\
  &\quad+\int_\rho^{\rhoj}\RH{\rho}{\eta}F(\rho,\eta,t)\dd\eta.
\end{align}
The two balances have distinct roles:
\[
  \A_t^\dagger p_t=0
  \quad\text{and}\quad
  \partial_tp_t=\B_t^\dagger p_t.
\]
The first expresses translation invariance along the road at a fixed time;
the second evolves that invariant marginal in physical time.

For completeness, Assumption \ref{ass:kinetic-regularity} is used here with
the following technical clauses. The coefficients are independent of \(x\),
\(p(\rho,t)>0\), and all derivatives have locally uniform integrable bounds
that permit differentiation and rearrangement of the displayed integrals.
The operators \(\A_t\), \(\B_t\), and the multiplication operators in
\eqref{eq:MN-continuum} share an invariant core
\(\mathcal D\subset C_c^2((0,\rhoj))\). The weak bounded-variation evolution
identity and mixed generator limits are valid on this core, and the backward
system \eqref{eq:backward-equations} has an operator-determining family of
common-core solutions. Finally, phase-space boundary flows vanish at
\(\rho=0\), \(\eta=\rhoj\), and as \(\eta\downarrow\rho\).

\subsection{Spatial-law reconstruction}

The one-point marginal and downstream transition law also reconstruct
multi-location traffic probabilities. For locations
\(x_1<\cdots<x_k\), the standard Markov product formula
\citep{ethier-kurtz1986} is
\begin{align}
\label{eq:fidi}
  &\mathbb P\{\rho(x_1,t)\in\dd r_1,\ldots,
  \rho(x_k,t)\in\dd r_k\}\notag\\
  &\quad=p_t(\dd r_1)
  P_{x_2-x_1}(r_1,\dd r_2)\cdots
  P_{x_k-x_{k-1}}(r_{k-1},\dd r_k).
\end{align}
Thus the framework supplies joint congestion probabilities across detector
locations, rather than only independent one-location summaries.

\section{Common-core compatibility}
\label{app:zero-curvature}

The zero-curvature identity is the inherited space--time compatibility
relation introduced above; the
common-core and domain-preservation qualifications stated here are imposed for
this manuscript.

\begin{theorem}[Consistent space--time evolution]
\label{thm:generator-lax}
The road-wise and detector-wise transition laws are compatible only if
\begin{equation}
\label{eq:zero-curvature}
  \partial_t\A-\partial_x\B=[\A,\B]
  \quad\text{on }\mathcal D.
\end{equation}
This is a conditional continuum theorem: it assumes the common domain in
Assumption \ref{ass:kinetic-regularity} and does not assert that this domain is
automatically preserved.
\end{theorem}

For the homogeneous road considered in the main analysis,
\(\partial_x\B=0\), and \eqref{eq:zero-curvature} reduces to
\begin{equation}
\label{eq:stationary-lax}
  \partial_t\A=[\A,\B].
\end{equation}

The formal zero-curvature calculation follows from backward-equation
compatibility.  The common-core hypothesis used in Theorem
\ref{thm:generator-lax} is, however, a
manuscript-specific assumption introduced to distinguish that formal identity
from a justified operator statement.  The cross-differentiation in the
theorem is therefore a necessary compatibility calculation. It is important
that the one-point identity
\(\partial_tP_t=P_t\B_t\) alone does \emph{not} imply a semigroup Lax
equation: such a conclusion would require an additional exact two-point BV
evolution identity. Rather than hide that substantive step in a formal
product-rule argument, Assumption \ref{ass:kinetic-regularity} states the
operator-determining common-core compatibility explicitly. The theorem then
establishes, under those compatibility hypotheses, that the corresponding
curvature vanishes on \(\mathcal D\). Establishing that compatibility and
preservation of the core from rough initial data remains outside the present
continuum theorem.

For a backward observable in the common core, write
\begin{equation}
\label{eq:backward-equations}
  \partial_x\varphi+\A\varphi=0,
  \qquad
  \partial_t\varphi+\B\varphi=0.
\end{equation}
Differentiating the first equation with respect to \(t\) and the second with
respect to \(x\) gives
\[
  \varphi_{xt}=-(\partial_t\A)\varphi+\A\B\varphi,
  \qquad
  \varphi_{tx}=-(\partial_x\B)\varphi+\B\A\varphi.
\]
Equality of mixed derivatives shows that the operator in
\eqref{eq:zero-curvature} annihilates every common solution. The
operator-determining clause promotes the identity to the full common domain
\(\mathcal D\), which proves Theorem \ref{thm:generator-lax} conditionally.

\section{Commutator expansion}
\label{app:commutator-expansion}

This appendix expands the generator commutator into drift, transport, and
binary-jump contributions. The quadratic binary-jump commutator structure is
inherited from the shock-clustering generator calculation cited at the start
of this appendix; the upper-density-jump ordering, concave traffic
flow signs, and front-number normalization in the expansion below are supplied
for the present traffic formulation. The displayed calculation retains the
present notation so that every sign in the main kinetic equations can be
checked directly.

Under Assumption \ref{ass:kinetic-regularity}, the conditional jump density
satisfies
\begin{align}
\label{eq:lambda-kinetic}
  \partial_t\lambda
  &+\partial_\rho(\lambda W_\rho)
   +\partial_\eta(\lambda W_\eta) \notag\\
  &=\Qtr(\lambda,\lambda)
  +\lambda\left[
    \bigl(q'(\rho)-\RH{\rho}{\eta}\bigr)\partial_\rho a(\rho,t)
    +q''(\rho)a(\rho,t)
  \right],
\end{align}
where
\begin{align}
\label{eq:Qtr}
  \Qtr(\lambda,\lambda)(\rho,\eta)
  &=\int_\rho^\eta
    \bigl(\RH{\rho}{m}-\RH{m}{\eta}\bigr)
    \lambda(\rho,m)\lambda(m,\eta)\dd m \notag\\
  &\quad-\int_\eta^{\rhoj}
    \bigl(\RH{\rho}{\eta}-\RH{\eta}{m}\bigr)
    \lambda(\rho,\eta)\lambda(\eta,m)\dd m \notag\\
  &\quad-\int_\rho^{\rhoj}
    \bigl(\RH{\rho}{m}-\RH{\rho}{\eta}\bigr)
    \lambda(\rho,\eta)\lambda(\rho,m)\dd m .
\end{align}
The last term is a generator-rate correction, not the upstream loss term in
the physical front-number balance.

Suppress \(t\) and write
\begin{equation}
\label{eq:generic-coefficients}
  b(\rho)=-a(\rho),\quad
  c(\rho)=q'(\rho)a(\rho),\quad
  n(\rho,\eta)=\lambda(\rho,\eta),\quad
  k(\rho,\eta)=-\RH{\rho}{\eta}\lambda(\rho,\eta).
\end{equation}
Let \(D_b=b\partial_\rho\) and
\(J_n\varphi=\int_{\eta>\rho}(\varphi(\eta)-\varphi(\rho))
n(\rho,\eta)\dd\eta\), with analogous notation for \(D_c,J_k\). Then
\begin{equation}
\label{eq:commutator-four-parts}
  [\A,\B]=[D_b,D_c]+[D_b,J_k]+[J_n,D_c]+[J_n,J_k].
\end{equation}

The drift--drift term is
\begin{align*}
  [D_b,D_c]
  &=(bc'-cb')\partial_\rho\\
  &=\{-a(q''a+q'a')-q'a(-a')\}\partial_\rho
  =-q''a^2\partial_\rho.
\end{align*}
Since the differential coefficient of \(\partial_t\A\) is \(-\partial_ta\),
this proves \eqref{eq:a-drift-equation}. Separating variables in that scalar
ODE gives, for \(a(\rho,t_0)=a_0(\rho)\geq0\),
\begin{equation}
\label{eq:a-explicit-solution}
  a(\rho,t)=
  \frac{a_0(\rho)}
  {1-q''(\rho)a_0(\rho)(t-t_0)}.
\end{equation}
Because \(q''<0\), the denominator is at least one for \(t\geq t_0\).

For the mixed terms, differentiation under the integral gives
\begin{align}
\label{eq:drift-jump-detail}
 D_bJ_k\varphi
 &=b(\rho)\int_\rho^{\rhoj}
 \{(\varphi(\eta)-\varphi(\rho))k_\rho
    -\varphi'(\rho)k\}\dd\eta,\\
\label{eq:jump-drift-detail}
 J_kD_b\varphi
 &=\int_\rho^{\rhoj}
 \{b(\eta)\varphi'(\eta)-b(\rho)\varphi'(\rho)\}
 k(\rho,\eta)\dd\eta,\\
\label{eq:jump-drift-c-detail}
 J_nD_c\varphi
 &=\int_\rho^{\rhoj}
 \{c(\eta)\varphi'(\eta)-c(\rho)\varphi'(\rho)\}
 n(\rho,\eta)\dd\eta,\\
\label{eq:drift-jump-n-detail}
 D_cJ_n\varphi
 &=c(\rho)\int_\rho^{\rhoj}
 \{(\varphi(\eta)-\varphi(\rho))n_\rho
    -\varphi'(\rho)n\}\dd\eta.
\end{align}
Equations \eqref{eq:drift-jump-detail}--\eqref{eq:drift-jump-n-detail}
display both mixed commutators with their correct order:
\([D_b,J_k]=D_bJ_k-J_kD_b\) and
\([J_n,D_c]=J_nD_c-D_cJ_n\).
The lower-limit contribution in \eqref{eq:drift-jump-detail} is zero because
\(\varphi(\eta)-\varphi(\rho)=0\) at \(\eta=\rho\). Integrating the
\(\varphi'(\eta)\) terms by parts in \(\eta\), and applying the no-flow
boundary conditions, collects the mixed contribution into transport in the
two phase variables plus the linear source in
\eqref{eq:lambda-kinetic}. The resulting coefficient of
\(\varphi(\eta)-\varphi(\rho)\) is
\begin{align}
\label{eq:mixed-R}
  R_{\rm mix}(\rho,\eta)
  &=-\partial_\rho(\lambda W_\rho)
    -\partial_\eta(\lambda W_\eta)\notag\\
  &\quad+\lambda\left[
    (q'(\rho)-\RH{\rho}{\eta})\partial_\rho a(\rho)
    +q''(\rho)a(\rho)
  \right].
\end{align}

It remains to expand the jump--jump term. Successive transitions divide into
three ordered configurations:
\[
  \rho<m<\eta,\qquad
  \rho<\eta<m,\qquad
  \rho<m\ \text{with a common lower state}.
\]
Inserting \(k=-\RH{\rho}{\eta}n\) from
\eqref{eq:generic-coefficients}, applying Fubini's theorem, and collecting the
coefficient of \(\varphi(\eta)-\varphi(\rho)\) gives exactly
\(\Qtr(n,n)\) in \eqref{eq:Qtr}. Therefore
\[
  \partial_t\lambda=R_{\rm mix}+\Qtr(\lambda,\lambda),
\]
which is \eqref{eq:lambda-kinetic}. The common-lower-state term in
\(\Qtr\) appears because \(\lambda\) is a conditional transition rate. The
direct count of physical fronts has a different third term, derived next.

\clearpage
\section{Shock-number balance}
\label{app:number-density}

This appendix derives the physical shock-number balance from the spatial
Markov description.  The identity \(F=p\lambda\), the three collision
configurations, and the adjoint cancellation are the traffic-sign
transcriptions of the shock-clustering number-density framework
\citep{menon-srinivasan2010}.
The explicit conversion from the raw \(p\lambda\lambda\) rates to an operator
written only in \(F\), together with its dimensional audit, is supplied here
to clarify the normalization used in this paper.

Fix a physical time and suppress it in the notation. In a stationary road
field,
\begin{equation}
\label{eq:campbell-intensity}
  \mathbb E\,N_L(\dd\rho,\dd\eta)
  =L p(\rho)\lambda(\rho,\eta)
  \dd\rho\dd\eta
  =L F(\rho,\eta)\dd\rho\dd\eta,
\end{equation}
where \(N_L\) counts fronts in a road segment of length \(L\). This is the
Campbell/Palm meaning of \eqref{eq:F-definition} in the present traffic
ordering.

\subsection{Merger gain}

The first collision configuration creates a \(\rho\to\eta\) front when two
adjacent fronts merge.  The following calculation gives its raw growth rate
after reversing the state ordering.

Take \(\rho<m<\eta\) and two small spacings
\(\Delta x_1,\Delta x_2>0\). By the spatial Markov property,
\begin{align}
\label{eq:triple-probability}
 &\mathbb P\{\rho(0)\in\dd\rho,
 \rho(\Delta x_1)\in\dd m,
 \rho(\Delta x_1+\Delta x_2)\in\dd\eta\}\notag\\
 &\quad=p(\rho)\lambda(\rho,m)\lambda(m,\eta)
 \dd\rho\dd m\dd\eta\,
 \Delta x_1\Delta x_2+o(\Delta x_1\Delta x_2).
\end{align}
The upstream \(\rho\to m\) front moves faster than the downstream
\(m\to\eta\) front by
\begin{equation}
\label{eq:gain-relative-speed}
  K_1=\RH{\rho}{m}-\RH{m}{\eta}>0.
\end{equation}
Taking \(\Delta x_2=K_1\Delta t\), dividing the triple-event probability by
\(\Delta x_1\Delta t\) to obtain a rate per unit road length and physical
time, and then integrating over \(m\) gives
\begin{equation}
\label{eq:C1-raw}
  C_1(\rho,\eta)=\int_\rho^\eta K_1
  p(\rho)\lambda(\rho,m)\lambda(m,\eta)\dd m.
\end{equation}

\subsection{Merger losses}

The remaining configurations remove a specified \(\rho\to\eta\) front by a
merger on either side.  The following calculations give the two decay rates
in the traffic state ordering.

If \(\rho<\eta<m\), the \(\rho\to\eta\) front catches the
\(\eta\to m\) front at relative speed
\(K_2=\RH{\rho}{\eta}-\RH{\eta}{m}>0\). Its loss rate is
\begin{equation}
\label{eq:C2-raw}
  C_2(\rho,\eta)=-p(\rho)\lambda(\rho,\eta)
  \int_\eta^{\rhoj}K_2\lambda(\eta,m)\dd m.
\end{equation}
If \(m<\rho<\eta\), an upstream \(m\to\rho\) front catches the
\(\rho\to\eta\) front at speed
\(K_3=\RH{m}{\rho}-\RH{\rho}{\eta}>0\). The adjacent pair starts in state
\(m\), so its loss rate is
\begin{equation}
\label{eq:C3-raw}
  C_3(\rho,\eta)=-\lambda(\rho,\eta)
  \int_0^\rho K_3 p(m)\lambda(m,\rho)\dd m.
\end{equation}
Without the relative-speed factors and integrations, the three raw
probability-rate integrands are
\begin{align}
\label{eq:C-p-lambda}
  C_1&=p(\rho)\lambda(\rho,m)\lambda(m,\eta),\notag\\
  C_2&=-p(\rho)\lambda(\rho,\eta)\lambda(\eta,m),\\
  C_3&=-p(m)\lambda(m,\rho)\lambda(\rho,\eta).\notag
\end{align}
Equations \eqref{eq:C1-raw}--\eqref{eq:C3-raw} are the three expressions
summarized in \eqref{eq:C-p-lambda}.

The cited collision rates are naturally expressed as
\(p\lambda\lambda\).  To write the present equation entirely in terms of the
physical number density, we make the normalization explicit. If \(p>0\),
substitute \(\lambda(r,s)=F(r,s)/p(r)\). The gain term becomes
\[
  p(\rho)\lambda(\rho,m)\lambda(m,\eta)
  =\frac{F(\rho,m)F(m,\eta)}{p(m)}.
\]
Similarly,
\[
  p(\rho)\lambda(\rho,\eta)\lambda(\eta,m)
  =\frac{F(\rho,\eta)F(\eta,m)}{p(\eta)},
\]
and
\[
  p(m)\lambda(m,\rho)\lambda(\rho,\eta)
  =\frac{F(m,\rho)F(\rho,\eta)}{p(\rho)}.
\]
Substitution in the three raw rates gives the collision operator used in the
main front-frequency balance:
\begin{align}
\label{eq:Ctr}
  \Cp(F,F)(\rho,\eta)
  &=\int_\rho^\eta
    \bigl(\RH{\rho}{m}-\RH{m}{\eta}\bigr)
    \frac{F(\rho,m)F(m,\eta)}{p(m)}\dd m \notag\\
  &\quad-F(\rho,\eta)\int_\eta^{\rhoj}
    \bigl(\RH{\rho}{\eta}-\RH{\eta}{m}\bigr)
    \frac{F(\eta,m)}{p(\eta)}\dd m \notag\\
  &\quad-F(\rho,\eta)\int_0^\rho
    \bigl(\RH{m}{\rho}-\RH{\rho}{\eta}\bigr)
    \frac{F(m,\rho)}{p(\rho)}\dd m .
\end{align}
This proves the collision operator \eqref{eq:Ctr}.

Between collisions, the two end states follow the phase velocities
\(W_\rho,W_\eta\), so conservation in \((\rho,\eta)\)-space adds the divergence
terms in \eqref{eq:F-transport}.  The same result follows from the
forward-equation cancellation displayed next. The cancellation is worth
showing because \(F\) is not the conditional rate \(\lambda\).
Differentiating \(F=p\lambda\), using
\eqref{eq:lambda-kinetic}, and suppressing the common arguments gives
\begin{align}
\label{eq:F-product-expansion}
 E&:=\partial_tF+\partial_\rho(FW_\rho)+\partial_\eta(FW_\eta)\notag\\
 &=p\Qtr(\lambda,\lambda)+\lambda\,\partial_tp\notag\\
 &\quad+\lambda\left[
   (q'(\rho)-\RH{\rho}{\eta})\partial_\rho(ap)
   +p q''(\rho)a
 \right].
\end{align}
Substitute the temporal adjoint balance and expand
\(\partial_\rho(q'ap)=q''ap+q'\partial_\rho(ap)\). The terms containing
\(q''ap\) cancel, leaving
\begin{align}
\label{eq:marginal-cancellation}
 &\partial_tp+(q'(\rho)-\RH{\rho}{\eta})\partial_\rho(ap)
   +p q''(\rho)a\notag\\
 &\quad=-\RH{\rho}{\eta}\partial_\rho(ap)
 -\int_0^\rho\RH{m}{\rho}F(m,\rho)\dd m
 +\int_\rho^{\rhoj}\RH{\rho}{m}F(\rho,m)\dd m\notag\\
 &\quad=-\int_0^\rho
   \bigl(\RH{m}{\rho}-\RH{\rho}{\eta}\bigr)F(m,\rho)\dd m\notag\\
 &\qquad+\int_\rho^{\rhoj}
   \bigl(\RH{\rho}{m}-\RH{\rho}{\eta}\bigr)F(\rho,m)\dd m,
\end{align}
where the last equality uses the stationary adjoint balance
\(\partial_\rho(ap)=-\int_0^\rho F(m,\rho)\dd m
+\int_\rho^{\rhoj}F(\rho,m)\dd m\).

Finally, the common-lower-state term in the conditional-rate operator is
\begin{equation}
\label{eq:Q-common-lower-cancellation}
 -p(\rho)\lambda(\rho,\eta)
  \int_\rho^{\rhoj}
  \bigl(\RH{\rho}{m}-\RH{\rho}{\eta}\bigr)
  \lambda(\rho,m)\dd m
 =-\lambda(\rho,\eta)
  \int_\rho^{\rhoj}
  \bigl(\RH{\rho}{m}-\RH{\rho}{\eta}\bigr)
  F(\rho,m)\dd m.
\end{equation}
It cancels the last integral in \eqref{eq:marginal-cancellation}. The first
integral that remains is exactly the upstream physical loss
\eqref{eq:C3-raw}; the other two terms of \(p\Qtr\) are
\eqref{eq:C1-raw} and \eqref{eq:C2-raw}. This proves
\eqref{eq:F-transport} term by term. Both adjoint equations are essential;
setting \(\partial_tp=0\) would be incorrect.

As a manuscript-specific check on the explicit \(F\)-normalization, consider
dimensions. If density has unit \(D\), then
\([p]=D^{-1}\), \([\lambda]=L^{-1}D^{-1}\), and
\([F]=L^{-1}D^{-2}\). A term \(FF\dd m/p\), multiplied by a relative speed,
has the same units as \(\partial_tF\). A bare product \(FF\dd m\) does not.

Normalizing by \(\Gamma(t)\) and pushing the selected state pair through the
amplitude map \(\eta-\rho\) and the Rankine--Hugoniot speed map gives
\begin{align}
\label{eq:amplitude-pushforward}
  \mathbb P_{\rm shock}(\Delta\rho\in\dd\xi)
  &=\frac{1}{\Gamma(t)}
  \int_0^{\rhoj-\xi}F(\rho,\rho+\xi,t)\dd\rho\,\dd\xi,\\
\label{eq:speed-pushforward}
  \mathbb P_{\rm shock}(s\in\dd z)
  &=\frac{1}{\Gamma(t)}
  \int_{\rho<\eta}
  \delta\!\left(z-\RH{\rho}{\eta}\right)
  F(\rho,\eta,t)\dd\rho\dd\eta\,\dd z.
\end{align}

Finally, integrate \eqref{eq:campbell-intensity} over all state pairs. This
last normalization is a direct consequence derived here from the stated
Campbell intensity, rather than an additional kinetic assumption. The
expected number of fronts in length \(L\) is \(L\Gamma(t)\), and division by
that quantity yields the normalized Palm law \eqref{eq:shock-palm} and its
push-forwards \eqref{eq:amplitude-pushforward}--\eqref{eq:speed-pushforward}.

\section{Collision-rate positivity}
\label{app:collision-positivity}

This appendix verifies that every merger rate used above is nonnegative and
that the phase velocities have the signs asserted in the main text.  These
claims are elementary consequences of the strict-concavity form of the Lax
entropy ordering \citep{lax1957,dafermos2016}; the compact set of inequalities
below is recorded here for the present traffic convention.

\begin{lemma}[Traffic Lax and secant inequalities]
\label{lem:lax-concave}
If an entropy shock joins \(\rho<\eta\), then
\begin{equation}
\label{eq:lax-traffic}
  q'(\rho)>\RH{\rho}{\eta}>q'(\eta).
\end{equation}
More generally, if \(r<s<z\), then
\begin{equation}
\label{eq:ordered-secant-speeds}
  \RH{r}{s}>\RH{s}{z}.
\end{equation}
\end{lemma}

\begin{proof}
The mean value theorem gives
\(\RH{\rho}{\eta}=q'(\xi)\) for some \(\xi\in(\rho,\eta)\). Since \(q'\) is
strictly decreasing, the two inequalities in \eqref{eq:lax-traffic} follow.
Writing the two secants as averages of the decreasing function \(q'\) over
adjacent intervals gives \eqref{eq:ordered-secant-speeds}.
\end{proof}

For \(r<s\),
\begin{equation}
\label{eq:secant-average}
  \RH{r}{s}=\frac{1}{s-r}\int_r^s q'(z)\dd z.
\end{equation}
Since \(q'\) is strictly decreasing, averages over intervals farther to the
left are larger. Consequently,
\begin{align}
\label{eq:three-relative-speed-inequalities}
  \RH{\rho}{m}&>\RH{m}{\eta}
    &&(\rho<m<\eta),\notag\\
  \RH{\rho}{\eta}&>\RH{\eta}{m}
    &&(\rho<\eta<m),\\
  \RH{m}{\rho}&>\RH{\rho}{\eta}
    &&(m<\rho<\eta).\notag
\end{align}
These are precisely the three positive relative speeds in the number-density
operator. They also imply the boundary quasi-positivity of the finite matrix
flow through \eqref{eq:boundary-vector-field}.

\begin{lemma}[Isolated-front state transport]
\label{lem:phase-velocity-signs}
For the phase velocities in \eqref{eq:phase-velocities},
\(W_\rho\geq0\) and \(W_\eta\leq0\). Along a collision-free phase-space
characteristic,
\begin{equation}
\label{eq:shock-amplitude-decay}
  \frac{\dd}{\dd t}(\eta-\rho)=W_\eta-W_\rho\leq0.
\end{equation}
\end{lemma}

\begin{proof}
The phase-velocity inequalities use the same ordering with a tangent in place
of one secant. Equation \eqref{eq:lax-traffic} and \(a\geq0\) give
\(W_\rho\geq0\) and \(W_\eta\leq0\); subtracting the two state-transport
equations gives \eqref{eq:shock-amplitude-decay}. These signs do not depend on
whether the front itself moves upstream or downstream.
\end{proof}

\section{Numerical definitions and checks}
\label{app:numerical-checks}

This appendix records the pairwise metrics and identities used to accept or
reject each experiment. It complements the traffic interpretation in the main
text with enough numerical detail to reproduce every table. Configurations,
raw arrays, and scripts are archived with the paper; no value in the Results
section was entered independently of those files. The analytical-validation
definitions below contain no road-grid result, and the stochastic-comparison
definitions contain no exact-solution reference.

\subsection{Riemann metrics}
\label{app:riemann-metrics}

The Riemann check compares the deterministic front-measure reconstruction
only with the independent exact entropy solution. If
\(x_s^{\rm P}(t)\) and \(x_s^{\rm E}(t)\) denote the proposed and exact
front positions, respectively, the reported path discrepancy is
\begin{equation}
\label{eq:riemann-path-metric}
 E_x=\max_{0\leq t\leq T}
 |x_s^{\rm P}(t)-x_s^{\rm E}(t)|.
\end{equation}
At each stored time, both profiles are evaluated analytically on a common
reporting grid used only for quadrature and plotting. Their density-profile
discrepancy is
\begin{equation}
\label{eq:riemann-l1-metric}
 E_\rho(t)=\int_{x_a}^{x_b}
 |\rho^{\rm P}(x,t)-\rho^{\rm E}(x,t)|\,\dd x.
\end{equation}
Vehicle balance includes the exact accumulated flow through both domain
boundaries. The speed discrepancies are
\(|s^{\rm P}-s_{\rm RH}|\) and
\(|s^{\rm P}-s_{\rm RH}|/|s_{\rm RH}|\). Time-step sensitivity repeats the
weak atom-moment update over the configured integration steps; because the
right-hand side is constant, all changes remain at floating-point roundoff
and no empirical convergence order is claimed.

\subsection{Signal metrics}
\label{app:signal-metrics}

The signal experiment uses one common demand flow and separate free-branch
densities obtained by solving \(q(\rho_{\rm up})=1200\)
veh/h/lane. This avoids attributing a demand difference to fundamental-diagram curvature.
The proposed calculation represents the red-phase queue front as an atom of
the wave measure. Its temporal operator assigns the Rankine--Hugoniot secant
speed to that atom and characteristic speeds to the green-release wave. The
independent comparators are closed characteristic and event constructions;
no finite-volume road grid enters either subexperiment.

The following formulas specialize classical characteristic and cumulative-count
analysis \citep{newell1993,daganzo2006,laval-leclercq2013} to the declared
experiment. They are benchmark derivations, not new traffic laws. Write
\(g\) for red duration, \(d=q_d\) for demand, \(C=q_{\max}\) for capacity,
\(r=\rho_{\rm up}\), \(j=\rho_j\), and \(L\) for approach length, using
metres, seconds and vehicles per lane. Assume \(0<d<C\), an initially uniform
free state, no startup lost time, no upstream spillback, and green through
clearance. The initially present vehicles must be included in vehicle balance.

For Greenshields, let \(\alpha=r/j<1/2\) and
\(\tau=t-g\). The rear boundary and release wave meet at
\(t_i=g/(1-\alpha)\). The independently derived rear path is
\begin{equation}
\label{eq:signal-rear-closed}
 X(t)=
 \begin{cases}
 L-v_f\alpha t,&0\leq t\leq t_i,\\
 L+v_f(1-2\alpha)\tau
 -2v_f\sqrt{\alpha(1-\alpha)g\tau},&t_i<t<t_c,\\
 L,&t\geq t_c .
 \end{cases}
\end{equation}
Indeed the rarefaction state on its right is
\(\rho^+=\tfrac j2[1-(X-L)/(v_f\tau)]\), so Rankine--Hugoniot gives
\(X'=v_f(1/2-\alpha)+(X-L)/(2\tau)\).
Matching at \(t_i\) yields \eqref{eq:signal-rear-closed}. Its turning point and
clearance are
\begin{equation}
\label{eq:signal-events-closed}
 t_p=g+\frac{\alpha(1-\alpha)g}{(1-2\alpha)^2},
 \qquad
 t_c=\frac{g}{(1-2\alpha)^2}=\frac{Cg}{C-d}.
\end{equation}
At the turning point, \(\rho^+=j-r\), hence
\(q(\rho^+)=q(r)=d\) and the front speed is zero. During green before
\(t_i\), the stopped length is
\(\ell_0(t)=L-v_f(t-g)-X(t)\); it is zero afterward.
During red, \(\ell_0=\ell_c=L-X\). These definitions explain why the
congested extent can grow even while stopped traffic disappears.

For the triangular diagram, put \(k=d/(j-r)>0\). Before release meets the
rear, \(X=L-kt\); the release travels at \(-w\), giving
\(t_i=wg/(w-k)\). Thereafter \(X=L-kt_i+v_f(t-t_i)\) until clearance.
The maximum extent is \(kt_i\), and
\(t_c=t_i(1+k/v_f)=Cg/(C-d)\). Equal demand and capacity therefore imply
equal clearance times for the two diagrams, not equal congestion lengths.

For both exact diagrams, the departure count measured from \(t=0\) is
\begin{equation}
\label{eq:signal-departures}
 N_{\rm out}(t)=
 \begin{cases}
 0,&0\leq t\leq g,\\
 C(t-g),&g<t<t_c,\\
 dt,&t\geq t_c.
 \end{cases}
\end{equation}
Its derivative is respectively \(0,C,d\), apart from the switching instants.
The excess inventory is \(E(t)=dt-N_{\rm out}(t)\), and the physical
inventory satisfies
\(\int_0^L\rho(x,t)\dd x=rL+E(t)\). Thus, once the horizon exceeds clearance,
\begin{equation}
\label{eq:total-delay-numerical}
 D_{\rm tot}=\int_0^T E(t)\dd t
 =\tfrac12 dg\,t_c .
\end{equation}
The proposed delay is independently integrated from its reconstructed
inventory, whereas the exact comparator uses this closed formula.
Off-reporting-grid quadrature checks the physical inventory, including
splits at all current fronts. This avoids accepting a reporting-grid
interpolant as an exact continuous solution.

For the triangular branches \(a=v_f\rho\) and
\(b=w(\rhoj-\rho)\), the smooth approximation is
\begin{align}
\label{eq:smooth-triangular}
 m_\varepsilon(\rho)
 &=\frac12\{a+b-\sqrt{(a-b)^2+\varepsilon^2}\},\\
 q_\varepsilon(\rho)
 &=m_\varepsilon(\rho)
 -\left[m_\varepsilon(0)
 +\frac{\rho}{\rhoj}\{m_\varepsilon(\rhoj)-m_\varepsilon(0)\}\right].
\notag
\end{align}
The affine correction makes \(q_\varepsilon(0)=q_\varepsilon(\rhoj)=0\)
without altering strict concavity. Direct differentiation gives
\[
 q_\varepsilon''(\rho)
 =-\frac{(v_f+w)^2\varepsilon^2}
 {2\{[(v_f+w)\rho-w\rhoj]^2+\varepsilon^2\}^{3/2}}<0.
\]
The smooth family is a numerical choice made here. If
\(\delta(\rho)=q_T(\rho)-m_\varepsilon(\rho)\), then
\(0\leq\delta\leq\varepsilon/2\). The affine endpoint correction is a convex
combination of \(\delta(0)\) and \(\delta(j)\), so
\(\|q_\varepsilon-q_T\|_\infty\leq\varepsilon/2\), and the maximum-flow
difference obeys the same bound.
For the triangular case, each \(q_\varepsilon\) reconstruction is compared
only with the independently coded exact triangular Newell/contact solution.
The comparison reports the flow supremum difference on a dense density grid,
maximum queue-front and cumulative-count differences on the common 0.25 s
reporting grid, and differences in peak and clearance event times. These
sampled maxima are not asserted to be continuous-time suprema. The decreasing
differences establish a deterministic smooth-flow approximation trend over
the tested values only; they do not prove convergence of the stochastic
generator closure at the triangular kink.

\subsection{Front-law diagnostics}
\label{app:front-law-diagnostics}

For the state-triangle midpoint cells, the direct shock intensity and
shock-size moments are
\begin{align}
\label{eq:discrete-front-moments}
 \Gamma_N(t)&=\sum_{i<j}F_{ij}(t)(\Delta z)^2,\\
 \overline{\Delta\rho}_N(t)
 &=\Gamma_N(t)^{-1}\sum_{i<j}
 D(z_j-z_i)F_{ij}(t)(\Delta z)^2 .
\notag
\end{align}
The severity probability uses the same normalized weights over cells with
\(D(z_j-z_i)\geq50\) veh/km/lane. The integrated collision-gain term is the
binary merger rate. Because one binary merger creates one composite front and
removes two parents, the discrete collision check is
\begin{equation}
\label{eq:collision-count-check}
 \sum_{i<j}\mathcal C_{ij}(\Delta z)^2
 +\sum_{i<j}\mathcal C^{\rm gain}_{ij}(\Delta z)^2=0
\end{equation}
up to quadrature and roundoff.

Two additional distributions are evaluated from the row generator at the observation time.
If its upward entries are multiplied by \(s\) while its diagonal is left
unchanged, then
\[
 G_L(s)=\pi\exp(LA_N(s))\mathbf1
\]
is the probability-generating function of the upward-front count. A discrete
Fourier inversion with \(M\) modes returns counts modulo \(M\); aliasing
must therefore be checked, not ignored. For the 71-state generator at 90 s,
increasing \(M\) from 48 to 96 changes the CDF by less than \(10^{-12}\),
and the PMF mean agrees with \(L\Gamma_N=4.628676\) to \(10^{-10}\). For inter-front
spacing, set the upward off-diagonal entries to zero while retaining their
contribution to the diagonal; the resulting killed matrix \(Q_0\) gives
\[
 \mathbb P\{S>x\}=\nu\exp(xQ_0)\mathbf1,
\]
where \(\nu\) is the post-jump Palm distribution. These are standard
finite-chain marking and killing constructions \citep{davis1993,ethier-kurtz1986};
the traffic use and state-space discretization are implemented here.

The complete gaps in a finite observation window have a different law from
\(S\). By stationary Palm counting \citep{daley-verejones2003}, a gap of
length \(s<L\) can start within only \(L-s\) units of road. Its pooled density
is therefore proportional to \((L-s)f_S(s)\), without an independence
assumption on consecutive gaps. Writing \(H(s)=\Pr\{S>s\}\), integration by
parts yields
\begin{equation}
\label{eq:complete-gap-law}
 \Pr_{\rm complete}\{S\leq r\}
 =\frac{L-(L-r)H(r)-\int_0^r H(u)\dd u}
 {L-\int_0^L H(u)\dd u},\qquad 0\leq r\leq L.
\end{equation}
The denominator is \(\mathbb E[(L-S)_+]\), and the expected number of
complete gaps is \(\Gamma_N\mathbb E[(L-S)_+]\).
The implementation evaluates the survival integral by the killed-matrix
resolvent and compares both CDFs on a common 201-point spacing grid.
The reported maximum is over that grid, not a certified continuum supremum.

The Monte Carlo Godunov comparator clusters contiguous positive
density increments above the predeclared 0.05 veh/km/lane threshold. A 2 km
buffer on each side of the 4 km observation window prevents boundary
characteristics from reaching reported fronts over 90 s. The same continuous
PDMP path is sampled once per realization and analytically averaged over each
cell of every tested road grid, so changing \(\Delta x\) does not change the underlying stochastic
input. Confidence intervals for expected counts use realization-level
standard errors; intervals for the pooled mean shock size use a
realization-clustered ratio-estimator standard error.

For the merger statistic, consecutive 0.5 s catalogs are connected by an
order-preserving dynamic program. An identified merger requires two adjacent
parents and one child to satisfy a characteristic-travel-plus-two-cell
position bound, intersecting swept support, consistent outer states, and a
consistent composite amplitude. Unmatched fronts within the same tolerance of
an observation boundary are censored as entries or exits. All other losses or
gains remain explicitly unresolved. The identified rate is not ground truth or a guaranteed lower bound:
finite-volume smoothing can hide real mergers or create false associations.
Because the direct kinetic rate counts all modeled binary collisions, its
difference from this catalog statistic is not an isolated error estimate. The complete tolerances and transition ledger are archived with the
processed data.

Spatial homogeneity makes the direct front-location distribution uniform on
the observation window. The Godunov diagnostic pools detected 90 s front
locations and reports its mean and maximum CDF difference from that uniform
law. It is descriptive and front weighted; no iid significance test is used
because multiple fronts from one realization are dependent. The experiment
evolves prescribed random disturbances and is not a linear-instability test.

The labeled continuous-coordinate paths shown for traffic interpretation are
samplewise deterministic wave visualizations initialized from the same PDMP
law. They are not uniquely recoverable from the unlabeled population density
\(F\), are not an additional validation comparator, and are not used when
forming any proposed-versus-Godunov statistic.

Two discretization effects are audited separately. Let
\(Z_N=\sum_i p_i^{\rm raw}\Delta z\) denote the mass reconstructed by the
discrete stationary adjoint before normalization. Using \(p_i=p_i^{\rm raw}/Z_N\)
enforces unit mass, but does not also preserve the original continuum drift
and adjoint exactly. Independently, the finite-volume state transport sends
small fronts into the diagonal. With zero outer-boundary input its
instantaneous count budget is
\begin{equation}
\label{eq:front-loss-audit}
 -\dot\Gamma_N=R_N^{\rm collision}+R_N^{\rm diagonal}.
\end{equation}
The collision contribution obeys \eqref{eq:collision-count-check};
the second is numerical small-front loss, not an additional merger mechanism.
Table \ref{tab:state-audit} reports both effects across density-state resolutions.
At \(N=71,t=90\) s, the raw mass excess is 3.37\%, and diagonal loss contributes
7.45\% of the instantaneous total count decay. Both decrease under refinement,
but remain nonzero at \(N=141\), so no continuum convergence theorem or
mesh-independent accuracy is inferred.

\begin{table}[tbp]
\centering\small
\caption{Direct-solver audit at 90 s. Both rates are in fronts/(km s).
The 141-state run is an internal resolution check, not an additional
cross-method comparator. Raw mass is measured before normalization.}
\label{tab:state-audit}
\begin{tabular}{rrrrr}
\toprule
$N$ & Raw mass & Fronts/km & Collision rate & Diagonal loss \\
\midrule
31 & 1.0540 & 1.1004 & 0.00424 & 0.00078 \\
51 & 1.0413 & 1.1393 & 0.00448 & 0.00050 \\
71 & 1.0337 & 1.1572 & 0.00459 & 0.00037 \\
141 & 1.0212 & 1.1809 & 0.00472 & 0.00019 \\
\bottomrule
\end{tabular}

\end{table}

\begin{figure}[tbp]
\centering
\includegraphics[width=\textwidth]{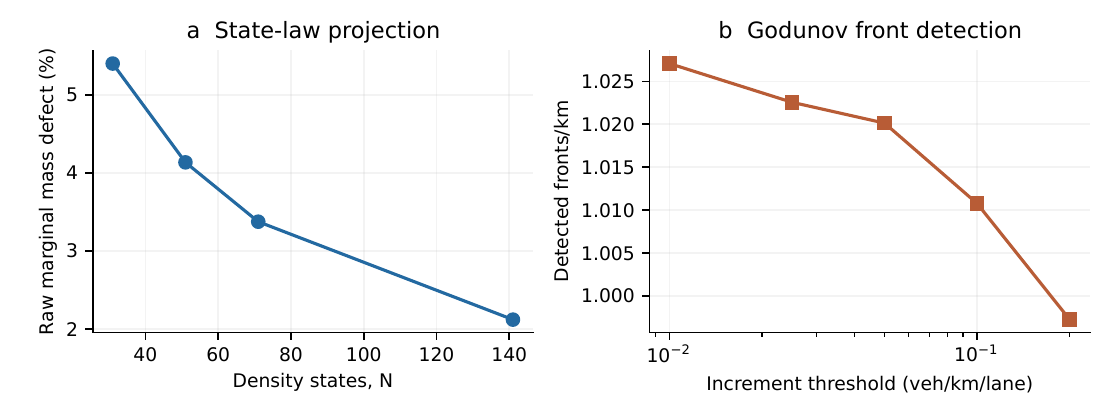}
\caption{Numerical audit: (a) marginal mass defect before direct-solver
normalization and (b) Godunov detector sensitivity at the same observation time on the
interior 3.6 km. These are internal diagnostics of the two representations.}
\label{fig:numerical-audit}
\end{figure}

A separate detector sensitivity reprocesses the same 800 stored 10 m
density profiles, accepting front centroids in the interior interval
\([0.2,3.8]\) km to avoid truncated edge supports. Increasing the positive
increment threshold from 0.01 to 0.20 veh/km/lane changes front intensity
from 1.0271 to 0.9972/km and mean amplitude from 27.753 to 28.476 veh/km/lane.
At the declared 0.05 threshold these interior values are 1.0201/km and
27.933 veh/km/lane; they need not equal the full-window catalog.
Figure \ref{fig:numerical-audit}(a) shows the decreasing mass defect;
panel (b) shows how the extraction threshold changes the detected population.
This is extraction sensitivity, not a correction to true front counts.

\subsection{Travel-time validation}
\label{app:travel-time-validation}

The travel-time quantity is the instantaneous travel-time functional in
\eqref{eq:rq4-frozen-functional}. The proposed calculation is accepted only if
\(A_N\mathbf1=0\), off-diagonal rates are nonnegative,
\(\pi A_N=0\), and \(\Phi(0,L)=1\). Nonnegative reward implies that
\(\Phi\) is nonincreasing for \(\theta\geq0\); the implementation checks this
property on the eight configured nonnegative \(\theta\) values. Its moment
check uses
\begin{equation}
\label{eq:fk-moment-check}
 \mathbb E[T_L]=-\partial_\theta\Phi(0,L),\qquad
 \mathbb E[T_L^2]=\partial_{\theta\theta}\Phi(0,L),
\end{equation}
and compares those derivatives with the augmented matrix equations stated in
Section \ref{subsec:mapping-back}. An independent two-state check also verifies
\(\operatorname{Var}(T_L)=2\int_0^L(L-h)\operatorname{Cov}_\pi(c(\rho(0)),c(\rho(h)))\dd h\).
The conservative reward scheme
must retain nonnegative mass summing to one, keep a nondecreasing CDF in
\([0,1]\), and have negligible upper-grid overflow. Every retained state has
positive speed; otherwise the finite reward would require an explicitly
represented probability mass at infinite travel time.

The reliability metrics are defined from the instantaneous travel-time distribution:
buffer time \(T_{95}-\mathbb E[T]\), buffer index
\((T_{95}-\mathbb E[T])/\mathbb E[T]\), and planning-time index
\(T_{95}/T_{\rm free}\). The on-time probability uses the declared 10 min
threshold. The sole comparator is the matched Monte Carlo Godunov density-profile
ensemble from Experiment 3. For a stored field with cells of width
\(\Delta x\), its functional is evaluated as
\begin{equation}
\label{eq:godunov-frozen-travel}
 T_L^{\rm G}=\sum_{i=1}^{L/\Delta x}
 \frac{\Delta x}{v(\rho_i(t^*))}.
\end{equation}
The comparison reports proposed-minus-Godunov differences in moments,
quantiles, and reliability indices, together with the CDF maximum difference
and one-dimensional Wasserstein distance. No deterministic or analytical
travel-time reference is used. One-at-a-time comparisons fix \(N=71\) while
varying \(\Delta x\), and fix \(\Delta x=10\) m while varying \(N\); ensemble
prefixes describe sampling sensitivity; and reward-grid refinement describes
the CDF calculation's internal precision. These calculations never substitute
\(1/\mathbb E[v]\) for \(\mathbb E[1/v]\).

\subsection{Timing and reproducibility}
\label{app:timing-protocol}

Timing comparisons use one machine and software environment, fixed
configurations, and repeated wall-clock measurements. Setup, path generation,
solve, plotting, and compilation are recorded separately when the
implementation exposes them. Tables report medians and either interquartile
ranges or full repeated ranges. These small synthetic experiments establish
measured cost for the stated implementation; they do not establish
asymptotic or universal speed superiority.

\section{Finite pure-shock algebraic check}
\label{app:spectral-integrability}

This appendix keeps the former finite-generator material in its limited
role. The finite Markov \(N\)-wave construction is inherited from
\citet{menon2011}; the ascending-density, upper-triangular form below is its
traffic-variable reflection under \(u=-\rho\). It verifies signs,
nonnegative pure-shock rates, and a matrix invariant. It is not a traffic
experiment, a CTM road grid, or the state-resolution approximation used in
Section \ref{subsec:state-space-computation}.
In plain language, it checks that a deliberately tiny table of transition
rates obeys the inherited algebra; it does not predict a road density field
or identify shock statistics from eigenvalues.

\subsection{Continuum identity}

We first verify the continuum multiplication--generator commutator on the
common core. This calculation transcribes the spectral identity into the
density ordering and concave-flow sign convention used here, while retaining
the operator-domain qualification needed for an unbounded generator.

Let \(M\) and \(N\) be the multiplication operators
\begin{equation}
\label{eq:MN-continuum}
  (M\varphi)(\rho)=\rho\varphi(\rho),
  \qquad
  (N\varphi)(\rho)=q(\rho)\varphi(\rho).
\end{equation}
On the common core,
\begin{equation}
\label{eq:commutator-identity}
  [\A,N]-[M,\B]=0.
\end{equation}
Consequently, whenever the homogeneous compatibility equation
\eqref{eq:stationary-lax} applies,
\begin{equation}
\label{eq:spectral-lax}
  \partial_t(\A-\mu M)
  =[\A-\mu M,\B+\mu N],
  \qquad \mu\in\mathbb C.
\end{equation}
This is a conditional continuum identity under the common-domain assumptions;
no continuum characteristic determinant is asserted.

For a test function in the common core, direct calculation gives
\begin{align}
\label{eq:AN-continuum}
  [\A,N]\varphi(\rho)
  &=-a(\rho)q'(\rho)\varphi(\rho)\notag\\
  &\quad+\int_{\eta>\rho}
  (q(\eta)-q(\rho))\varphi(\eta)
  \lambda(\rho,\dd\eta),\\
\label{eq:MB-continuum}
  [M,\B]\varphi(\rho)
  &=-a(\rho)q'(\rho)\varphi(\rho)\notag\\
  &\quad+\int_{\eta>\rho}
  \RH{\rho}{\eta}(\eta-\rho)\varphi(\eta)
  \lambda(\rho,\dd\eta).
\end{align}
The Rankine--Hugoniot identity
\(\RH{\rho}{\eta}(\eta-\rho)=q(\eta)-q(\rho)\) proves
\eqref{eq:commutator-identity}. Combining it with
\eqref{eq:stationary-lax} proves \eqref{eq:spectral-lax}. On the compact
density interval, \(M\) and \(N\) are bounded multiplication operators; the
domain issue concerns \(\A\), \(\B\), their compositions, and the common core.
There is no continuum characteristic determinant asserted here.

\subsection{Finite-state check}
\label{subsec:discrete-density-formulation}

The continuum equations in Section \ref{sec:solution-approach} describe a
distribution over all admissible density states. For an optional numerical
check, restrict attention to an ordered grid
\(0\leq\rho_1<\cdots<\rho_N\leq\rhoj\). This reduced model is not a CTM
discretization and does not replace the traffic outputs in the main text. Its
purpose is only to verify that the proposed rate dynamics preserve
nonnegative front-transition rates and an exact matrix invariant.

The construction specializes the finite Markov shock flow of
\citet{menon2011} to increasingly ordered traffic densities. Let
\(A_{ij}\) be the rate per unit distance of a transition from \(\rho_i\) to
\(\rho_j\). Upward fronts give
\begin{equation}
\label{eq:finite-generator-cone}
  A_{ij}\geq0\ (i<j),\qquad A_{ij}=0\ (i>j),
  \qquad A\mathbf 1=0.
\end{equation}
The matrix \(B\) weights each rate by its front speed:
\begin{equation}
\label{eq:G-discrete}
  G_{ij}=
  \begin{cases}
    -\RH{\rho_i}{\rho_j},&i\neq j,\\
    -q'(\rho_i),&i=j,
  \end{cases}
\end{equation}
\begin{equation}
\label{eq:B-discrete}
  B_{ij}=G_{ij}A_{ij}\quad(i\neq j),
  \qquad B_{ii}=-\sum_{j\neq i}B_{ij}.
\end{equation}

\begin{proposition}[Nonnegative finite-state rates]
\label{prop:finite-positive-invariance}
For strictly concave \(q\),
\begin{equation}
\label{eq:finite-lax-A}
  \dot A=[A,B]
\end{equation}
preserves the transition-rate conditions in
\eqref{eq:finite-generator-cone}. Thus a rate that is initially physically
admissible cannot become negative under the reduced evolution.
\end{proposition}

The boundary-face proof is given in Appendix
\ref{app:spectral-integrability}. The same reduced system also supplies an
exact conservation check. Set
\begin{equation}
\label{eq:finite-MN}
  M=\diag(\rho_1,\ldots,\rho_N),
  \qquad
  N=\diag(q(\rho_1),\ldots,q(\rho_N)).
\end{equation}

\begin{theorem}[Finite-state rate consistency]
\label{thm:finite-isospectral}
For every initial matrix satisfying \eqref{eq:finite-generator-cone}, the ODE
\eqref{eq:finite-lax-A} has a unique global solution satisfying
\begin{equation}
\label{eq:finite-commutator}
  [A,N]-[M,B]=0
\end{equation}
and
\begin{equation}
\label{eq:finite-spectral-lax}
  \frac{\dd}{\dd t}(A-\mu M)
  =[A-\mu M,B+\mu N].
\end{equation}
Consequently, for each fixed \(\mu\), the eigenvalues of \(A(t)-\mu M\)
remain unchanged. This invariant checks the numerical rate evolution; it is
not a new traffic observable.
\end{theorem}

Because \(A-\mu M\) is upper triangular, its invariant polynomial is
\begin{equation}
\label{eq:spectral-invariant}
  \det(A-\mu M-zI)
  =\prod_{i=1}^N\bigl(A_{ii}-\mu\rho_i-z\bigr).
\end{equation}
It records the total rate of leaving each density state, but not how that rate
is allocated among destination states. The finite model is therefore a
consistency benchmark, not a unique traffic-state fingerprint. Proofs and the
limits of the upward-only reduction are collected in Appendix
\ref{app:spectral-integrability}.

\subsection{Finite cone and global flow}

We next show that the finite generator remains in its admissible
upper-triangular cone and hence exists for all time. The argument is the
density-order reversal of the lower-triangular cone invariance and global
flow, specialized below to the row-generator convention of this paper.

Proposition \ref{prop:finite-positive-invariance} proves that the vector field
is tangent to the row-sum and triangular constraints and points inward on
every off-diagonal boundary face. Explicitly, when \(A_{ij}=0\), \(i<j\),
\begin{align}
\label{eq:boundary-vector-field}
  \dot A_{ij}
  &=\sum_{i<k<j}(G_{kj}-G_{ik})A_{ik}A_{kj} \notag\\
  &=\sum_{i<k<j}
  \bigl(\RH{\rho_i}{\rho_k}-\RH{\rho_k}{\rho_j}\bigr)
  A_{ik}A_{kj}\geq0
\end{align}
by Lemma \ref{lem:lax-concave}. Because \(A\mathbf1=B\mathbf1=0\), row sums
are preserved, and a commutator of upper-triangular matrices remains upper
triangular. In addition,
\begin{equation}
\label{eq:finite-diagonal-constant}
  \dot A_{ii}=[A,B]_{ii}=0
\end{equation}
because \(A\) and \(B\) are upper triangular. Thus each fixed row satisfies
\[
  \sum_{j>i}A_{ij}(t)=-A_{ii}(0).
\]
All off-diagonal entries remain nonnegative and bounded by the fixed row exit
rate. The polynomial vector field is therefore bounded on the product of
these compact row simplices, precluding finite-time blow-up. This proves the
global-existence part of Theorem \ref{thm:finite-isospectral}.

\subsection{Finite isospectrality}

This final step verifies the finite spectral equation and converts it into a
similarity flow. The commutator construction is the upper-triangular traffic
reflection of the finite shock-clustering flow; once the Lax equation is
established, similarity and spectral invariance are standard
finite-dimensional consequences. The triangular-spectrum and absorption
limitations stated afterward are specific scope conclusions of this paper.

For \(i\neq j\),
\begin{align*}
  [A,N]_{ij}
  &=A_{ij}(q(\rho_j)-q(\rho_i)),\\
  [M,B]_{ij}
  &=(\rho_i-\rho_j)B_{ij}\\
  &=(\rho_i-\rho_j)
    (-\RH{\rho_i}{\rho_j}A_{ij})\\
  &=A_{ij}(q(\rho_j)-q(\rho_i)).
\end{align*}
Both commutators have zero diagonal, proving
\eqref{eq:finite-commutator}. With
\(L_\mu=A-\mu M\) and \(P_\mu=B+\mu N\), this gives
\(\dot L_\mu=[L_\mu,P_\mu]\).

The standard fundamental-matrix argument for a finite-dimensional Lax flow
now gives the similarity representation. Let \(S_\mu(t)\) solve
\begin{equation}
\label{eq:similarity-fundamental-matrix}
  \dot S_\mu=-P_\mu S_\mu,
  \qquad S_\mu(0)=I.
\end{equation}
Then
\[
  \frac{\dd}{\dd t}
  \bigl(S_\mu^{-1}L_\mu S_\mu\bigr)=0,
\]
so
\begin{equation}
\label{eq:finite-similarity}
  L_\mu(t)=S_\mu(t)L_\mu(0)S_\mu(t)^{-1}.
\end{equation}
This proves finite isospectrality without any operator-domain qualification.

Because \(L_\mu\) is upper triangular, its characteristic polynomial is the
product of its diagonal factors, as shown in
\eqref{eq:spectral-invariant}. Equation
\eqref{eq:finite-diagonal-constant} therefore makes the invariance explicit.
The following limitation is identified here for the present traffic-oriented
triangular model: the characteristic polynomial is blind to the allocation
of a fixed row exit rate among different upward transitions. It cannot, by
itself, be a unique fingerprint of shock-amplitude statistics.

Finally, the following absorption observation is specific to the finite
pure-shock restriction used here.

\begin{remark}[Scope of the finite reduction]
\label{rem:finite-pure-shock-scope}
The upward-only chain has an absorbing maximal accessible state. Its invariant
laws are supported on absorbing classes, so it does not supply the strictly
positive stationary marginal used in the continuum \(F=p\lambda\) theory.
For a finite road one may instead evolve an upstream law as
\(p(x)=e^{xA^\top}p(0)\), using the column-density convention
\(\partial_xp=A^\top p\). Adding downward nearest-neighbour transitions to
approximate continuous recovery would destroy the upper-triangular structure;
the reduced model is therefore an exact pure-front regression check, not a general
discretization of the continuum process.
\end{remark}

\section*{Data availability}

The numerical experiments use synthetic data generated by the archived
configurations and scripts described in the reproduction document. Raw NPZ
arrays and processed CSV files are supplied for every table and figure. No
external empirical dataset was used.

\section*{Declaration of generative AI and AI-assisted technologies in the
manuscript preparation process}

OpenAI Codex was used to assist with mathematical and numerical checking,
experiment code, figure construction, LaTeX preparation, and language editing.
The authors are responsible for verifying the calculations, reviewing the
resulting text and figures, and approving the final content.

\bibliographystyle{cas-model2-names}
\bibliography{references}

\end{document}